\documentclass[
]{aa}

\usepackage[varg]{txfonts}
\usepackage{natbib}
\bibpunct{(}{)}{;}{a}{}{,} 
\usepackage[utf8]{inputenc}    

\usepackage[colorlinks=true,
    linkcolor=blue, citecolor=blue, filecolor=blue, urlcolor=blue]{hyperref}

\usepackage[export]{adjustbox}

\usepackage{physics}
\usepackage{siunitx}
\usepackage{physunits}
\DeclareSIUnit\gauss{G}
\usepackage{mathtools}

\usepackage{xcolor}  
\usepackage{txfonts}
\usepackage{amsmath}
\usepackage{lipsum}
\usepackage{graphicx} 
\graphicspath{{./}{fig/}} 

\usepackage{float}
\usepackage{subfig}
\usepackage[rightcaption]{sidecap} 
\sidecaptionvpos{figure}{c}
\usepackage[percent]{overpic}

\newcommand{\bifrost}{\texttt{Bifrost}}
\newcommand{\vapor}{\texttt{Vapor}}
\newcommand{\muram}{\texttt{MURaM}}

\newcommand{\jvec}{\mathbf{j}}
\newcommand{\bvec}{\mathbf{B}}
\newcommand{\fvec}{\mathbf{f}}
\newcommand{\alphan}{\alpha_\mathrm{norm}}
\newcommand{\zmin}{z_\mathrm{CZ}}
\newcommand{\corks}{\texttt{corks}}
\newcommand{\cork}{\texttt{cork}}

\newcommand{\sza}{0.3}
\newcommand{\szb}{0.33}

\newcommand{\szH}{.25}

\let\linenumbers\nolinenumbers\nolinenumbers

\begin{document}

    
    \title{Flux rope formation through flux cancellation of sheared coronal arcades in a 3D convectively-driven MHD simulation}
    \titlerunning{Flux rope formation through flux cancellation of sheared coronal arcades}

    \author{
        S.V.Furuseth\thanks{\email{sondrevf@uio.no}}\inst{1,2}
        \and
        G.Aulanier\inst{1,2,3}
        }

   \institute{
        Rosseland Centre for Solar Physics, University of Oslo, PO Box 1029 Blindern, 0315 Oslo, Norway
        \and 
        Institute of Theoretical Astrophysics, University of Oslo, PO Box 1029 Blindern, 0315 Oslo, Norway
         \and
        Sorbonne Université, 
        Observatoire de Paris-PSL, \'Ecole Polytechnique, IP Paris, CNRS, 
        Laboratoire de Physique des Plasmas, Paris, France
        }

   \date{\today}
\abstract
   {
    Space weather and its potential negative consequences for life on Earth has received increasing scientific attention in recent decades. Particularly predicting CME onset has become important from a security perspective. To predict CMEs, one must first understand the dynamics leading to pre-eruptive magnetic field configurations -- which in many theories include a flux rope. 
   }
   {
    In this study, we investigate the realistic formation of coronal flux ropes above the solar photosphere. 
    The aim is to find if and how flux ropes can form there, and how the formation is related to flux cancellation at the photosphere.
    Previously, such formation has been shown in smooth boundary-driven line-tied simulations and in idealized non-convective and symmetric flux-emergence simulations.
   }
   {
    We run a convective non-symmetric 3D radiative MHD simulation with the code \bifrost{}. Within the simulation box of ${\SI{24}{\mega\meter}\times\SI{24}{\mega\meter}}$ horizontal extent, a linear force-free field with sheared coronal arcades is slowly inserted. Following the insertion, the self-consistent stochastic plasma flows of the convection zone drive several small-scale flux cancellations and magnetic reconnection, without external influence. 
    Lagrangian markers called \corks{} are used to track the dynamic evolution of the magnetic field.
   }
   {
    Over a period of $\SI{2.5}{\hour}$, a flux rope is generated with photospheric footpoints separated by up to $\SI{12}{\mega\meter}$. The flux rope is gradually formed through several individual events, such as slipping reconnection, U-loop emergence, and thick-photosphere tether-cutting reconnection. 
   }
   {
    Flux ropes, which can lead to CMEs, can be formed in the solar atmosphere solely driven by convection and flux cancellations at the photosphere. However, not all flux cancellations contribute to the build-up of the flux rope, and some coronal reconnection events that do are not clearly related to flux cancellation. The formation process of flux ropes from coronal sheared arcades driven by convection is therefore more complex than in the original smooth flux cancellation model. But the end result is qualitatively the same. Flux cancellation works. A flux rope is formed.
   }


   \maketitle

\section{Introduction}
\label{sec_intro}

Solar eruptions are extreme manifestations of the Sun's magnetic activity, explosively releasing parts of the vast amount of energy contained within the solar interior out into the heliosphere, including Earth~\citep{schrijver_events_2012}. Strong eruptions, exemplified by the 1859 Carrington event, can severely affect technology and life on Earth and in its orbit, making space weather an important field of study for the sake of mankind. The most powerful events display bright flares, storms of solar energetic particles, and coronal mass ejections (CMEs). They can erupt from both active regions and long filament or prominence channels, due to a sudden destabilization of force-free magnetic fields in the corona. 

To understand and eventually predict solar eruptions, \citet{Patsourakos_CMEdiscussion_2020} stresses the need for a ``clear understanding of the nature of the pre-eruptive magnetic field configurations of [CMEs]''. However, the mechanisms that bring the magnetic fields and plasma to such a pre-eruptive state are not fully understood, in spite of a plethora of theories~\citep{aulanier_physical_2014,Patsourakos_CMEdiscussion_2020}. Many of these theories include a flux rope in the pre-eruptive magnetic field, but even that is contended by the magnetic breakout model~\citep{antiochos_breakout_1999}. 
Flux ropes are also a common magnetic geometry in filaments that do not erupt~\citep{Mackay_2010_prominence}.
In the present study, the focus is on analyzing in detail the mechanism in which flux cancellation leads untwisted arcades to form a twisted coronal flux rope. In doing so, we aim to increase the understanding of pre-eruptive magnetic fields, which in turn can lead to better prediction of eruptions.

There are several competing theories for the onset of eruptions related to the destabilization of prominence carrying magnetic flux ropes. Furthermore, there are multiple theories for the generation of these flux ropes. These theories are difficult to prove or disprove through observations alone, since flux ropes and their formation are not directly observable. They are only indirectly observable, through, e.g., the corresponding cold and denser prominences or through field extrapolations from photospheric magnetograms~\citep{nakagawa_lfff_1972,Wiegelmann_extrapolation_2008}, and through coronal cavities~\citep{Tandberg-Hanssen_1995, gibson_2018}. Therefore, it is important to supplement theories and observations with numerical simulations.

The suggested mechanisms for flux rope formation can in short be split in two: 
(i) Pre-twisted flux ropes emerging from the convection zone through the photosphere into the solar atmosphere, as observed by~\citet{tanaka_emergence_1001,Lites_emergence_1995,Leka_emergence_1996};
(ii) Flux ropes forming above the photosphere, due to flux cancellation in the photosphere, driven by plasma convection of the field line footpoints, motivated by observations~\citep{harvey_cancellation_1973, martin_cancellation_1979, wang_75hr_1989, Livi_cancellation_1989,green_cancellation_2011}.
Both theories have been studied extensively through numerical simulations.

The emergence of pre-twisted flux ropes has been simulated with various simplifications. 
Some simulations model the photosphere as the upper boundary, studying the rise of a pre-twisted buoyant flux rope inside an initially stratified convection zone, first in 1D and than in 2D~\citep{emonet_emergence_1998,Jouve2013}.
Then kinematic emergence simulations brought the flux rope into the corona by modeling the rise of the flux rope through the convection zone kinematically, whereupon the physics in the atmosphere was described by a full MHD model~\citep{fan_emergence_2004,amari_emergence_2004,amari_emergence_2005}. 
Also, the convection zone was modeled by the full MHD equations, both in 2D~\citep{archontis_emergence_2007} and 3D~\citep{Fan_2001, archontis_flux_2008}, albeit being initially stratified. The flux ropes do not fully emerge in these simulations, but are tied to the photosphere due to the heavy plasma that they do not manage to carry into the atmosphere. Additionally, as the convection zone is initially stratified but not convective, they do not fully model the mechanisms of the Sun. 
\citet{hotta_sunspot_2020} has recently run simulations of an emerging flux rope in a convectively unstable and relaxed convection zone, showing it can be done, achieving the goal of forming sunspots.

How flux ropes can form in the atmosphere by flux cancellation was theorized in the seminal paper by \citet{van_ballegooijen_formation_1989}. A continued shearing of overlying coronal arcades, combined with a priori convectively-driven converging flows toward a straight polarity inversion line (PIL), would cause reconnection as the opposing fluxes would cancel, gradually creating a twisted flux rope.
This mechanism has typically been studied numerically in 3D by considering the photosphere as the lower boundary, where the magnetic field line footpoints are line-tied. The flux rope formation and dynamics is then driven by prescribed convection and/or diffusion of the field line footpoints at the boundary, in the manner described by~\citet{van_ballegooijen_formation_1989}. This approach has been successfully repeated several times over the last two decades, using smooth surface flux distributions and flux cancellation all along well-defined PILs \citep{amari_cme1_2003,amari_cme2_2003,Mackay_corona1_2006,Mackay_corona2_2006,aulanier_torus_2010,Jiang_eruption_2024,Xing_CME_rise_2024,Xing_CME_fp_2024}. 

A more critical problem of the line-tied flux-cancellation models, where it breaks physics, is that the magnetic flux diffuses as a scalar at the photospheric boundary. That leads to a diminishing flux until it disappears by diffusion~\citep[Fig 5.]{Aulanier_2010_linetying}. In reality, these field lines reconnect and continue down below the photosphere where they are redistributed and reenter the magnetic structure. By not properly modeling the reconnection of field lines, which is at the root of the mechanism for forming flux ropes, there are valid concerns whether the flux ropes generated are physical or simply artifacts of the model and the prescribed driving at the boundary. Therefore we need a proper treatment of flux cancellation to assess whether it can produce flux ropes in the corona, or not.

While observations alone cannot explain all details of how flux ropes are formed, they do show several attributes of the real process that the models should replicate. Real flux cancellation does not occur along a straight PIL, as in several of the simulations described above. In observations, the flux cancellation is fragmented between patchy flux concentrations along the PIL, even in $\delta$-spots, between opposite polarity sunspots isolated from other flux, and in decaying active regions~\citep{Schrijver_cme_2011,vanDriel_2003,wang_2013}. 
Flux ropes do also typically not form in a single cancellation event, but are rather built up through several events that contribute to the same rope or filament~\citep{wang_2007,wang_2013}.
Both of these are also true for flux ropes that erupt. In the observation by~\citet{green_cancellation_2011}, a bipole with an initially smooth PIL connected by slightly sheared arcades, turned into an eruptive sigmoid through several cancellation episodes between increasingly patchy polarities.

In this study, we simulate flux rope formation through flux cancellation of sheared coronal arcades, driven by stochastic convection from a convectively unstable convection zone. This counters many of the artificial approximations of previous flux-cancellation simulations. We do so by adding a linear force-free field (LFFF) in the form of sheared coronal arcades to a relaxed Quiet Sun (QS) simulation in \bifrost{}~\citep{gudiksen2011}, which combines into an enhanced network (EN) with a patchy PIL. Then, we let the stochastic convection drive the evolution of the atmospheric magnetic fields. We find that the result is a stable twisted flux rope formed in the solar atmosphere, but that unlike in the smooth flux cancellation model, the flux rope formation results from a variety of local events, including previously undocumented convectively-driven thick-photosphere tether-cutting reconnection.

\section{Numerical method}

\subsection{The \bifrost{} code}
 
The simulations discussed in this work has been run with the parallel radiative-MHD code \bifrost{}~\citep{gudiksen2011}. The code is designed to study a box in the Sun, starting in the upper convection zone and extending through the photosphere into the corona. We will refer to these vertical boundaries as the lower and coronal boundaries, respectively. The complex dynamics within this box in the Sun are modeled as realistically as currently possible by modeling several different physical processes in a modular approach. The implemented physics used here include radiative transfer, Spitzer thermal conductivity along magnetic field lines using the wave method discussed by~\citet{rempel2016}, hydrogen kept in local thermal equilibrium (LTE), and numerical hyper-diffusivity that recently have been proven to model the resistivity realistically, even at poor resolutions~\citep{faerder_comparative_2023,faerder_comparative_2024}.

There are various vertical boundary conditions (BC) implemented in \bifrost{}, including constant, symmetric, antisymmetric, hydrostatic, and divergence-free magnetic field extrapolations. These again can be treated either as standard time-derivative equations or by the method of characteristics. By using divergence-free magnetic field extrapolations, the code suppresses numerical growth of nonphysical ${\div{\textbf{B}}}$ and thereby reducing the need for iteratively canceling it to suppress nonphysical magnetic monopoles. 
The horizontal BCs are typically periodic in \bifrost{} simulations. 
Other types of horizontal BCs are possible in \bifrost{}, but are typically not used. Open BCs lead to loss of convective power and hard wall BCs lead to strong constraints on the magnetic field and plasma motion close to the boundary. To get a compact simulation box with minimal artifacts, periodic horizontal BCs are preferred.

The staggered grid used is 3D. It is uniform in the horizontal ${(x,y)}$~plane and nonuniform in the vertical $z$~direction, in such a way that it is better resolved closer to the photosphere and chromosphere than high up in the corona. In \bifrost{}, $z$ increases along the line of sight (LOS), meaning it is positive in the convection zone and negative in the corona. The spatial derivatives are 6th order and the time stepping is done using the explicit 3rd order Hyman predictor-corrector scheme~\citep{hyman_1979}. The state of the simulation is written to file every ${\SI{10}{\second}}$ (solar time), hereafter referred to as `snapshots'. That includes for each grid cell the density, 3 momentum components, internal energy, and 3 magnetic field components, 8 in total. One can additionally store other auxiliary variables, such as the temperature, that alternatively can be calculated a posteriori from the 8 independent variables.

\subsection{Initial conditions: A modified QS simulation}
\label{sec_met_QS}

The simulation presented here starts at a copied state from a QS simulation that had been relaxed and run for ${\SI{8000}{\second}=\text{133m20s}}$, twice as long as necessary to typically reach a QS state in a \bifrost{} simulation. That the simulation had indeed reached a QS state, was determined by considering the average signed and unsigned flux at the photosphere, reported below, and checking for the absence of organization of the flux into poles, similar to the state in Fig.~\ref{fig_ramp_z0_1000}. The solar times referred to in this paper are relative to this state, i.e., the copy was made and the solar time was reset to $\SI{0}{\second}$. This original QS simulation has a grid consisting of $768^3$ grid cells. 
The horizontal cross-section of the simulation is square with side lengths ${L_x=L_y=\SI{24}{\mega\meter}}$, divided uniformly such that the horizontal resolution is $\SI{31.25}{\kilo\meter}$. The vertical direction extends~$\SI{2.5}{\mega\meter}$ into the convection zone, referred to as the lower boundary in this article, and $\SI{-14.3}{\mega\meter}$ into the corona, referred to as the coronal boundary. The vertical grid is not split uniformly, the resolution is higher close to the photosphere than in the corona.

Up until the copy was made, the QS simulation was subject to magnetic flux feeding from the convection zone when the plasma velocity through the lower boundary was directed into the simulation box. This was done to replace the magnetic flux that was convected out and to compensate for the lack of small scale dynamo. The injected flux was horizontal, 
first with a strength of ${B_y=\SI{45}{\gauss}}$ for ${\SI{4000}{\second}}$, followed by ${B_y=\SI{110}{\gauss}}$ for another ${\SI{4000}{\second}}$, until the time of the copy. As a result, at the time of the copy of the QS simulation, the middle signed horizontal flux was measured to be ${\langle B_y(z=\SI{2.5}{\mega\meter})\rangle\approx\SI{100}{\gauss}}$ at the lower boundary and about $\SI{10}{\gauss}$ at the photosphere. This was due to the flux feeding. The vertical field was balanced with a middle signed flux of ${\langle B_z\rangle\approx\SI{0}{\gauss}}$, as it should be for a QS simulation, and a middle unsigned flux at the photosphere of ${\langle\abs{B_z(z=0)}\rangle\approx\SI{20}{\gauss}}$.

The copied state of the QS simulation was modified before restarting it. The resolution of the original simulation was first halved in each spatial dimension to conserve numerical resources, using the \texttt{backstaff}\footnote{https://github.com/lars-frogner/Backstaff} tool~\citep{frogner_2024}. This tool was designed to ``provide a fast, reliable and flexible framework for computations on Bifrost simulation data'', such as interpolations to different grids, tracing of field lines, and synthesizing of optically thin spectral lines.
Then the coronal boundary was extended from ${-\SI{14.3}{\mega\meter}}$ to ${-\SI{28.0}{\mega\meter}}$ above the photosphere by a hydrostatic extrapolation and adding of extra vertical grid cells. This  successfully alleviated various numerical problems close to the coronal boundary, further elaborated in App.~\ref{app_Simulation}. The new staggered grid consists of ${m_x\times m_y \times m_z = 384\times 384\times472}$ cells. That gives a horizontal resolution ${\Delta x = \Delta y = \SI{62.5}{\kilo\meter}}$, while the vertical resolution varies between a minimum ${\Delta z = \SI{24}{\kilo\meter}}$ in the chromosphere and transition region and a maximum ${\Delta z = \SI{155}{\kilo\meter}}$ in the corona. Since restarting it, there has been no magnetic flux fed through the vertical boundaries.

The horizontal BC were periodic. The vertical boundary ghost zones are filled by extrapolating from inside the main domain. The magnetic field BC were set to be symmetric around the boundary in the horizontal components, while the vertical component was used to make the field divergence free. The density is extrapolated via the hydrostatic condition in the coronal ghost zones and logarithmically extrapolated in the lower boundary ghost zones. The internal energy is constant in the coronal ghost zones and extrapolated in the lower boundary ghost zones. To keep the lower boundary inside the convection zone at a certain energy level, it continuously edged the average density and internal energy toward some constant values. The momentum is kept symmetric around the border in both the coronal and lower boundary ghost zones.

Some trial-and-error was necessary to make a best-possible numerical model for this physical effect, more discussion on this experimentation is presented in App.~\ref{app_Simulation}.

\subsection{Linear force-free magnetic fields (LFFF) }

According to~\citet{van_ballegooijen_formation_1989}, flux ropes can be generated by first shearing potential magnetic arcades along the PIL whereupon the shear is increased by convective flows toward the PIL from both polarities. Here, we inserted already sheared arcades in the form of an LFFF.
Such magnetic fields $\textbf{B}$, and corresponding currents $\textbf{j}$, has zero Lorentz force, thereby satisfying 
\begin{align}
    \div\bvec           &= 0    , \label{eq_divb}\\
    \mu_0\jvec          &= \curl\bvec = \alpha \bvec   \label{eq_j},
\end{align}
where $\alpha$ is a shearing scalar, being constant for an LFFF, and $\mu_0$ is the magnetic vacuum permeability.
By taking the curl of Eq.~\eqref{eq_j}, and using Eq.~\eqref{eq_divb}, one gets the Helmholtz differential equation
\begin{equation}
    \nabla^2 \bvec = -\alpha^2 \bvec. 
    \label{eq_helmholtz}
\end{equation}

In Cartesian coordinates, the solution of Eq.~\eqref{eq_helmholtz} in a box can be expressed as periodic harmonics in the $x$ and $y$ directions that are exponentially decaying along $z$ into the atmosphere, (here toward negative $z$) as expressed by~\citet{nakagawa_lfff_1972,demoulin_lfff_1989}. 
When removing the dependence on the $y$-coordinate, the magnetic field components  become
\begin{align}
	B_x &= -\sum_{n=1}^{n_h}  
	\frac{B_n l_n}{k_{nx}} \cos(k_{nx} x)\exp[l_n (z-\zmin)],
    \label{eq_lfff_bx}
	\\
	B_y &= +\sum_{n=1}^{n_h}  
	\frac{B_n \alpha}{k_{nx}} \cos(k_{nx} x)\exp[l_n (z-\zmin)],
    \label{eq_lfff_by}
	\\
	B_z &= -\sum_{n=1}^{n_h}  
	B_n \sin(k_{nx} x) \exp[l_n (z-\zmin)],
    \label{eq_lfff_bz}
\end{align}
where $\zmin$ is a constant reference depth, here set to the depth of the lower boundary, $B_n$ is the amplitude of the $n$th harmonic, and 
\begin{align}
	k_{nx} &= \frac{2\pi}{L_x}n ,  \\
	l_n &= \sqrt{k_{nx}^2-\alpha^2} = \frac{2\pi}{L_x} \sqrt{n^2-\alphan^2} ~,
\end{align}
are the corresponding wave number and exponential decay rate, respectively.
Furthermore, $\alphan$ is the normalized shearing parameter, which has a maximum real value of 1. It has been assumed that the horizontal interval is ${x\in[0,L_x]}$. In the limit of ${n_h\rightarrow1}$ or ${z \ll \zmin}$ (far above the lower boundary) the shearing angle between $\bvec$ and the $x$~axis at the PIL is given by 
\begin{equation}
	\lim\limits_{n_h\rightarrow1}\tan(\theta) = \frac{B_y}{B_x} = \frac{\alpha}{l_1} = \frac{1}{\sqrt{1-\alphan^2}},
	\label{eq_tantheta}
\end{equation}
such that, e.g., ${\alphan^2=0.5}$ corresponds to ${\theta=45\deg}$, which just happens to be the shearing constant used in the simulation presented in this paper.
With additional harmonics  ${(n_h>1)}$ the shearing angle is lower closer to $\zmin$. That is caused by the higher harmonics, for which $l_n/\alpha$ is higher, before their field strength fall off quicker with height due to the factor ${\exp[l_n (z-\zmin)]}$ in Eqs.~(\ref{eq_lfff_bx}-\ref{eq_lfff_bz}), as discussed by~\citet{demoulin_lfff_1989}.

Due to the orthogonality of the different harmonics, one can generate different LFFF by appropriately selecting the amplitudes $B_n$. The goal here is to make it dipolar and focused around the central PIL at ${x_\mathrm{PIL}=L_x/2}$ as opposed to around the PIL at the $x$~boundary. 
To achieve the second goal, one can set two requirements. First, assuming there are more than one harmonic, require the horizontal derivative of $B_z$ of ever higher orders to be 0 at ${z=\zmin}$ and ${x=0}$, making $B_z$ as small as possible in the close vicinity
\begin{equation}
	\left.\pdv[p]{B_z}{x}\right|_{x=0}  \propto  
	\begin{cases}
    		\sum\limits_{n}^{n_h}
                n^p B_n = 0 &,~p~\text{odd}~\text{and}~n_h>1 \\
			\sin(0)=0 &,~p~\text{even}.
	\end{cases}
	\label{eq_dBzdx0}
\end{equation}
Secondly, make the field as strongly dipolar as possible at ${x=L_x/2}$, by making the horizontal derivatives of the different vertical harmonics to have the same sign
\begin{equation}
	\left.\pdv{B_{zn}}{x}\right|_{x=\tfrac{L_x}{2}}  = B_n \cos(\pi n) = B_n (-1)^n > 0,
	\label{eq_Bn_sign}
\end{equation}
which requires every other amplitude to be of opposite sign. It turned out, in all the cases we tested, that this is achieved automatically by solving Eq.~\eqref{eq_dBzdx0}. 

To find $n_h$ magnetic amplitudes, one requires as many linearly independent equations. The first degree of freedom is the overall amplitude of $\bvec$, which can be set by ${B_1=1}$. All the amplitudes can later be scaled by a constant to get a different overall amplitude.
To get the remaining ${n_h-1}$ equations, one must solve Eq.~\eqref{eq_dBzdx0} up to order ${p=2(n_h-1)-1}$, e.g. for ${n_h=3}$, one must go to the derivative of order ${p=3}$ to get 2 equations more.
Coefficients for the first numbers of harmonics are given in Table~\ref{table_Bn}. Except for $B_1$, the other $B_n$ depend on $n_h$, and are of opposite sign, as required by Eq.~\eqref{eq_Bn_sign}.

\begin{table}[t]
	\caption{Harmonic amplitudes for a focused dipolar field.}              
	\label{table_Bn}      
	\centering                        
	\begin{tabular}{c|ccccc}          
		\hline\hline                  
		$n_h$  & $B_1$   & $B_2$   & $B_3$   & $B_4$ & $B_5$\\ 
		\hline                                   
        1        &  1 &   			 & 			 & &   \\
        2       &  1 &  $-\dfrac{1}{2}$  & 			&  & \\[8pt]
        3       &  1 &  $-\dfrac{4}{5}$ & $ \dfrac{1}{5}$ &   & \\[8pt]
        4       &  1 &  $-1$ 	 & $ \dfrac{3}{7} $ & $-\dfrac{1}{14}$ &  \\[8pt]
        5       &  1 &  $-\dfrac{8}{7}$ 	 & $\dfrac{9}{14}$ & $-\dfrac{4}{21}$ &$\dfrac{1}{42}$  \\[6pt]
		\hline
	\end{tabular}
\end{table}

The plan is to add the LFFF to an existing QS \bifrost{} simulation. In general, one introduces new Lorentz forces by adding a magnetic field to another, even if the added field is force-free. To illustrate this, imagine a configuration with a background magnetic field $\bvec_0$ with a force density $\fvec_0$. When you add an LFFF denoted $\bvec_1$, this can introduce new force terms since
\begin{equation}
\begin{split}
\fvec   &= \jvec \times \bvec \\
        &= (\jvec_0+\jvec_1) \times (\bvec_0 + \bvec_1) \\
        &= \underbrace{\jvec_0\times\bvec_0}_{\fvec_0} 
        + \underbrace{\left( \jvec_0\times\bvec_1 + \jvec_1\times\bvec_0 \right)}_{ \text{new force terms, $\fvec_{01} + \fvec_{10}$}}
        + \underbrace{\jvec_1\times\bvec_1}_{\fvec_1=0}  .
\end{split}
\label{eq_newforce}
\end{equation}
Note that the additional force terms ${\fvec_{01}+\fvec_{10}}$ are proportional to the field strength of the added LFFF. To minimize the consequential adjustment, the LFFF should be added incrementally. We detail how we do this in our experiment in Sect.~\ref{sec_exp_lfff}.

\subsection{Tracking field lines with \texttt{corks} in \texttt{Vapor}}

In addition to the physics modules that make the simulations of \bifrost{} realistic, there are also modules implemented to aid the analysis. One such module of great relevance to this study is the \corks{} module~\citep{zacharias_cork_2018,druett_cork_2022}. This module was employed in a similar fashion by~\citet{robinson_incoherent_2022,robinson_quiet_2023}. In short, \corks{} are Lagrangian point markers that are convected by the velocity-field in the simulation without affecting the physics.  
These markers track the plasma motion at the numerical time resolution of the simulation, not at the often much larger time resolution between output snapshots. Hence, the \corks{} give an accurate description of the simulation evolution, otherwise lost by limiting the output cadence, due to disk storage concerns, to once every ${\SI{10}{\second}}$ of solar time in this case.

The \corks{} were inserted at a specific time, 1 \cork{} per grid cell. To avoid over- or under-representation of any areas, the \corks{} where pruned (removed) if there were more than 2 per grid-cell and added if there were 0 in a grid-cell. For each snapshot, an identification number and the 3 spatial coordinates were stored per \cork{}.
Simple arithmetic shows that $N$ grid cells will approach $2N$ corks, making the storage requirement comparable to that of the 8 physical parameters of the plasma itself.

To accurately visualize the magnetic field lines in 3D, we have used the Visualization and Analysis Platform for Ocean, Atmosphere, and Solar Researchers (\vapor{})~\citep{vapor2019}. This software can visualize 3D parameters in a multitude of ways, such as showing the magnetic field lines integrated in 3D on top of a 2D magnetogram illustrating $B_z$ at ${z=0}$. In particular interest for this study, \vapor{} can select random seed points from which to start the integration of magnetic field lines in 3D, biased by another parameter. We often select such seed points at a specific time by biasing the algorithm to find large values of 
\begin{equation}
	\frac{\abs{\textbf{j}}}{\mu_0\abs{\textbf{B}}} = 
		\frac{\abs{\curl{\bvec}}}{\abs{\textbf{B}}},
        \label{eq_joverb}
\end{equation}
which is a normalized measure of the current density or complexity of the magnetic field, almost identical to the absolute value of the shearing parameter $\alpha$ in Eq.~\eqref{eq_j}.
We selected \corks{} close to these seed points and found where those \corks{} had been or would be at different times to track and analyze the evolution of the magnetic field lines and connectivity in particular.

Tracking individual field lines is a key component of the analysis presented in Sect.~\ref{sec_res_full}. While the \corks{} module proved utterly necessary in this endeavor, it has its limitations, some of which are further elaborated in App.~\ref{app_tracking}.

\section{Numerical experiment}

The \bifrost{} simulation presented and analyzed in this paper has been run as following: 
(i) Relax a modified QS simulation for 1000 snapshots ($\text{166m40s}$), starting at a reference time ${t_0=\SI{0}{\second}}$;
(ii) Ramp up the LFFF for 120 snapshots until solar time $\text{186m40s}$;
(iii) Initialize the \corks{} at solar time $\text{203m30s}$;
(iv) End the simulation at solar time $\text{368m00s}$. 

\begin{figure}[t]
    \centering
    \includegraphics[width=0.8\linewidth]{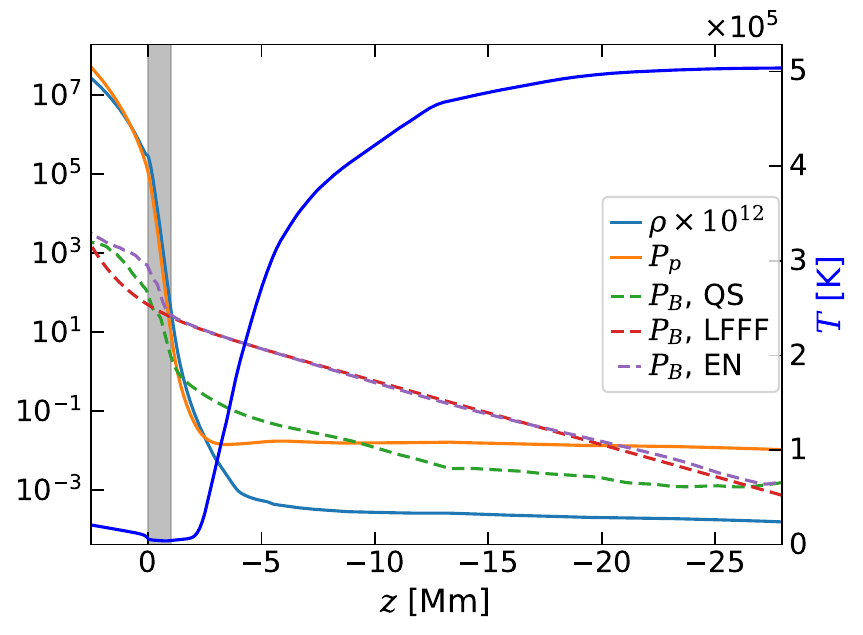}
    \caption{Horizontal averages of the \bifrost{} simulation. 
    The magnetic pressures ($P_B$) are given in $\si{\dyne\per\centi\meter^2}$ for the QS simulation before the ramp at snapshot 1000 (green dashed), for the time-independent LFFF (red dashed), and for the simulation after the ramp at snapshot 1120 (purple dashed).
    The pressure (orange), the density (light blue), and the temperature (dark blue) of the atmosphere are plotted at snapshot 1000.
    The density is given in $\si{\gram\per \centi\meter^3}$, multiplied with a factor $10^{12}$ to fit on the same axis as the magnetic pressures, while the temperature is given on the right axis.
    The gray shaded area (${z\in[0,-1]~\si{\mega\meter}}$) represents the ``thick'' photosphere, up to the altitude of the temperature minimum, i.e. the height where the density stratification is the strongest (see Sect.~\ref{sec_exp_thick} for more details).
    }
    \label{fig_ramp_1d}
\end{figure}

\begin{figure*}[t]
	\centering
    \subfloat[QS (snapshot 1000, ${t\!=\!\text{166m40s}}$)]{
        \includegraphics[width=0.245\linewidth,trim=2mm 0 3mm 0, clip]{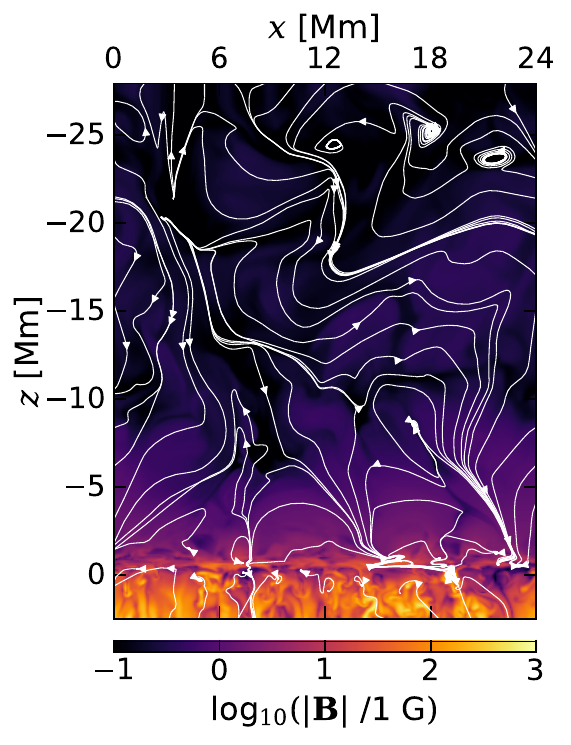}
        \label{fig_ramp_QS}
    }%
    \subfloat[LFFF]{
        \includegraphics[width=0.245\linewidth,trim=2mm 0 3mm 0, clip]{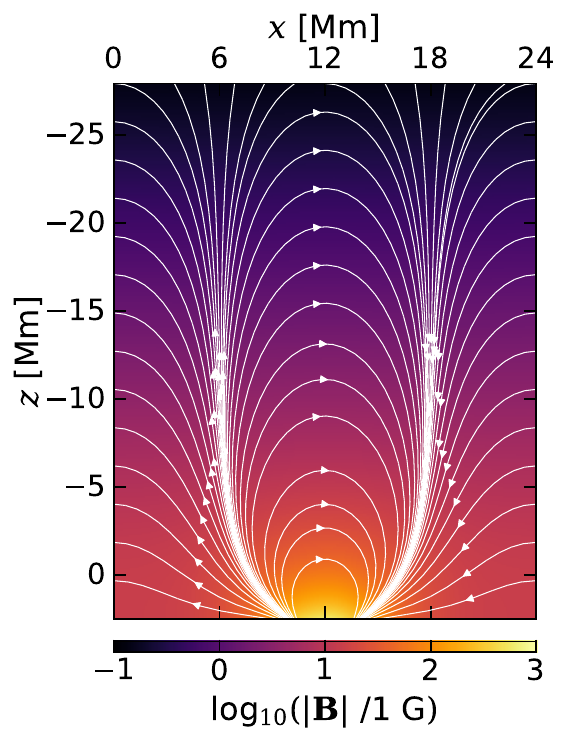}
        \label{fig_ramp_lfff}
    }
    \subfloat[QS + LFFF]{
        \includegraphics[width=0.245\linewidth,trim=2mm 0 3mm 0, clip]{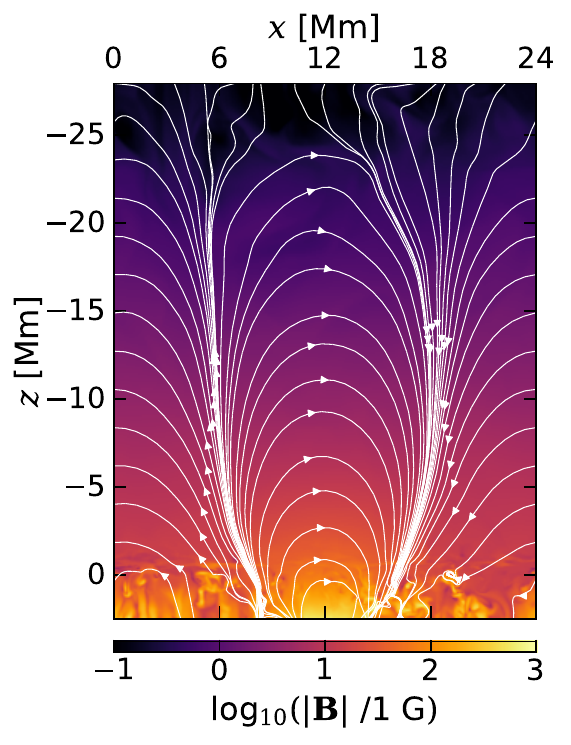}
        \label{fig_ramp_qslfff}
    }%
    \subfloat[EN (snapshot 1120, ${t\!=\!\text{186m40s}}$)]{
        \includegraphics[width=0.245\linewidth,trim=2mm 0 3mm 0, clip]{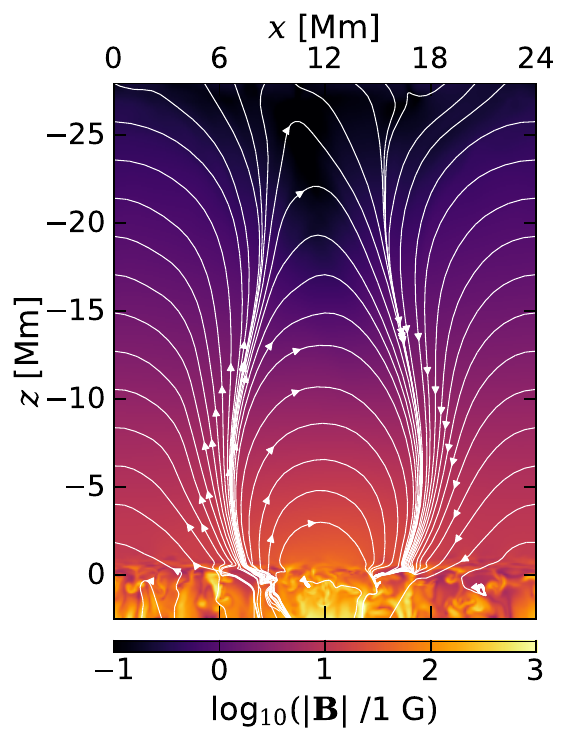}
        \label{fig_ramp_en}        
    }%
    \caption{ 2D cross sections of the magnetic field at ${y=\SI{10}{\mega\meter}}$. The streamlines show field lines, but their density does not signify the field strength. The field strength is shown by the color.
            (a) is the QS field right after the relaxation, (b) is the LFFF, (c) is the sum of the two proceeding, while (d) shows the EN right after the ${\SI{20}{\minute}}$ long ramp of the LFFF into the QS simulation.
    }
    \label{fig_ramp_Y10}
\end{figure*}

\begin{SCfigure*}[.3]
\begin{wide}
    \centering
    \renewcommand{\sza}{.37}
    \subfloat[QS (snapshot 1000)]{
        \includegraphics[width=\sza\textwidth,trim=0mm 0 0mm 0, clip]{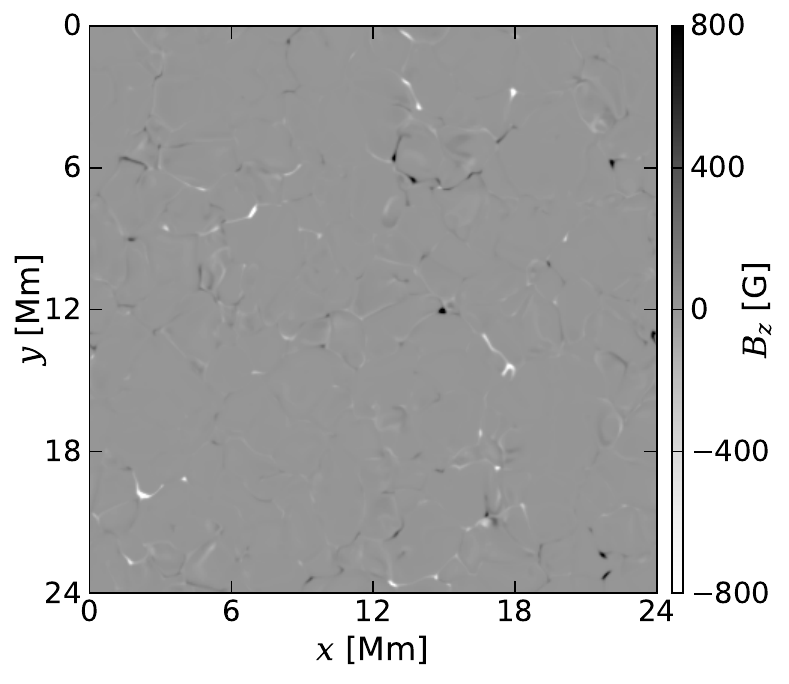}\label{fig_ramp_z0_1000}
	}
	\subfloat[EN (snapshot 1120)]{
    \includegraphics[width=\sza\textwidth,trim=0mm 0 0mm 0, clip]{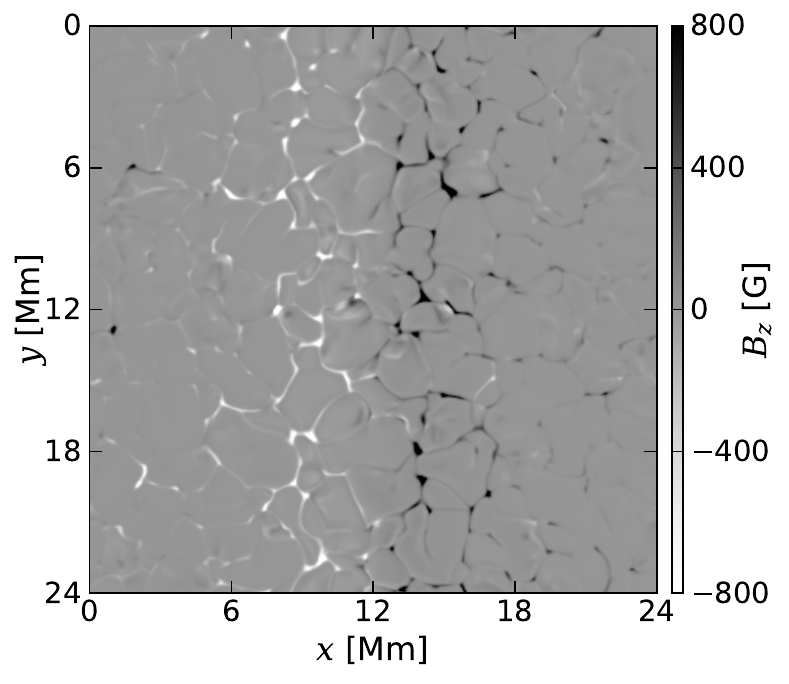}\label{fig_ramp_z0_1120}
	\hspace{4mm}}
    \caption{
        Magnetograms at the photosphere (${z=0}$) in the \bifrost{} simulation. (a) is just after the relaxation of the QS. (b) is just after the ramp of the LFFF. The white polarity (north pole) signifies flux coming out of the solar surface, being negative here as $z$ is positive along the LOS. The black polarity (south pole) signifies flux entering the solar surface.
        }
    \label{fig_ramp_z0}
\end{wide}
\end{SCfigure*}

\subsection{Relaxing the modified QS simulation}
\label{eq_exp_QS}

We started the simulation as a relaxed QS simulation to not have a smooth idealized photosphere with a straight PIL, but rather one that was more comparable to real-Sun observations.
After having reduced the resolution and extended the corona, as described in Sect.~\ref{sec_met_QS}, it was necessary to first relax the modified initial condition. As discussed in App.~\ref{app_Simulation}, it was found more than sufficient to relax it for ${\SI{10000}{\second}=\text{166m40s}}$. 
In Fig.~\ref{fig_ramp_1d}, we display the horizontal averages of the temperature, density, plasma pressure, and magnetic pressure in snapshot 1000 after this relaxation.

A vertical 2D cut of the magnetic field is shown in Fig.~\ref{fig_ramp_QS} and a horizontal 2D cut of the magnetogram at the photosphere is shown in Fig.~\ref{fig_ramp_z0_1000}.
After this relaxation without flux feeding through the bottom boundary, the vertical field was still balanced, while the mean signed flux in the $y$ direction had dropped to below $\SI{1}{\gauss}$, both at the lower boundary and at the photosphere. Hence, the simulation was ready as a blank slate for being injected with an LFFF.

\subsection{Ramping of the LFFF}
\label{sec_exp_lfff}

The LFFF added to the simulation was chosen to fulfill several requirements. These requirements represent our hypothesis for a field with a high likelihood of eventually forming a flux rope through several episodes of flux cancellation and magnetic reconnection as in~\citet{van_ballegooijen_formation_1989}, but driven by the stochastic convection. 
The shape was to be dipolar in $B_z$ at the lower boundary, sheared along the PIL and be stronger across the PIL in the middle of the simulation box than at the $x$~boundary. The strength was to dominate the preexisting magnetic and plasma pressure in the atmosphere, but not dominate the plasma pressure in the convection zone.
Through a thorough analysis we decided to define the shape of our LFFF by ${\alphan^2=0.5}$, to make the shearing angle ${\theta\leq \SI{45}{\degree}}$, and ${n_h=30}$, to make the field sufficiently focused around the central PIL, as illustrated in Fig.~\ref{fig_ramp_lfff}. After finding the relative strengths of the coefficients, the overall scaling of the field was decided  by setting the strongest vertical field at the photosphere to ${\max[B_z(z=0)]=\SI{50}{\gauss}}$. This value was chosen to ensure that there would be coronal arcades from which the flux rope could be built after the insertion. Simultaneously, the arcades should not be present in the convective layer, because the convection zone was to self-consistently drive the motion. This is achieved, as can be seen in Fig.~\ref{fig_ramp_1d}, where the magnetic pressure of the LFFF does dominate the QS magnetic and plasma pressure in most of the atmosphere, but not in the convection zone.

Immediately adding the designed LFFF to the QS simulation, gave the magnetic field illustrated in Fig.~\ref{fig_ramp_qslfff}. The LFFF dominates the original magnetic field in most of the box, except for in the convection zone and highest up in the corona.
Adding the field this abruptly perturbed the QS simulation too much, often making it crash after significant plasma flows. The crashes typically occurred at the coronal boundary when calculating the Spitzer conductivity. Even without the crashes, the induced plasma flows were deemed an artifact of the sudden change of magnetic field and therefore undesirable in a realistic modeling of this mechanism.

To reduce the reactions to the injected field, the LFFF in the presented simulation was slowly introduced in $120$ equal parts every $\SI{10}{\second}$, totaling up to $\SI{20}{\minute}$ of solar time. This ramp of the LFFF was conducted from snapshot~1000 to snapshot~1120, including a $\SI{10}{\second}$ cool-down period after the last injection. The hyper-diffusivity was kept slightly larger during the ramp and an additional $\SI{1000}{\second}$, to diffuse small-scale effects that caused crashes after the injection. This was acceptable as we were interested in long-term large-scale effects. The magnetic pressure of this EN after the ramp is quite close to that of the LFFF in most of the atmosphere, as seen in Fig.~\ref{fig_ramp_1d}. A vertical 2D cut of this magnetic field after the ramp is shown in Fig.~\ref{fig_ramp_en}, being appreciably different from Fig.~\ref{fig_ramp_qslfff}, particularly higher up in the corona.

The change of the simulation caused by adding the LFFF is clearly seen from the QS magnetogram in Fig.~\ref{fig_ramp_z0_1000} to the EN magnetogram in Fig.~\ref{fig_ramp_z0_1120}. After the ramp, there is a clear separation in white and black poles. The stochastic and self-consistent plasma convection have moved the field lines to the intergranular lanes, where they have been concentrated and consequently increased the field strength locally.
While the maximum field strength of the smooth LFFF was $\SI{50}{\gauss}$, these poles have much higher maximum field strengths. One thing lacking from this plot, combined with Fig.~\ref{fig_ramp_en}, is that the field lines are also sheared. In Fig.~\ref{fig_ramp_z0_1120}, the field lines would cross the PIL at approximately ${\SI{45}{\degree}}$ upwards to the right. This is exemplified by the field lines in Fig.~\ref{fig_frT_1}.

\subsection{Stochastic driving by self-consistent convection}
\label{sec_exp_convection}

During and after the ramp of the LFFF, the self-consistent \bifrost{} simulation controls what happens to the magnetic field. Hence, there is no prescribed convection pattern, and the surface flows develop freely and self-consistently, following the laws of physics. Since the plasma beta ${\beta_p=P_p/P_B\gg1}$ in the convection zone, the plasma dictates the motion of the `photospheric footpoints' of the coronal parts of the magnetic field line. During the ramp, the convection moves the magnetic flux into the granular lanes, as indicated by the end of the ramp in Fig.~\ref{fig_ramp_z0}. After the ramp, the self-consistent convection continues to stochastically make positive and negative polarities converge and cancel in the photosphere. This can be seen as the mutual cancellation of black and white polarities in Fig.~\ref{fig_ramp_z0} (see additional movie made available online for long-term evolution of the magnetogram). According to~\citet{van_ballegooijen_formation_1989}, these cancellations can cause reconnection and flux rope formation.

\subsection{Reconnection in the thick photosphere}
\label{sec_exp_thick}

Before describing what happened in the simulation, we find it instrumental to stress that the photosphere is not an infinitesimal boundary or plane. We define the photosphere as the extended region located between ${z=0}$ and the height of the temperature minimum, ${z(T_\mathrm{min})}$, where the density stratification is the strongest with the smallest pressure scale-height. The width of this layer is typically approximated in the literature as $\SI{0.5}{\mega\meter}$~\citep{priest1984}. However, the thickness of the photosphere is not a constant, it varies in space and time. 

In the simulation presented in this article, the altitude of the temperature minimum is ${z(T_\mathrm{min})=\SI[parse-numbers=false]{0.73 \pm 0.24}{\mega\meter}}$ after ramping of the LFFF. There are even transient excursions up to $\SI{2}{\mega\meter}$. To be conservative, we define the average thick photosphere to extend up to $\SI{1}{\mega\meter}$ above ${z=0}$. We illustrate the width of this extended thick photosphere as the gray area in Fig.~\ref{fig_ramp_1d}. It is within this region that we see much of the atmospheric reconnection stochastically driven by the self-consistent convection.

\section{Flux rope generation by flux cancellation}
\label{sec_res_full}

The primary goal of this work was to know whether or not a twisted flux rope would form in the atmosphere of a stochastic self-consistent realistic radiative MHD simulation. We actually find that it does. 

\subsection{High-level summary of the flux rope generation}

\begin{figure*}[!t]
    \centering 
    \renewcommand{\sza}{0.32} 

    \subfloat{\includegraphics[width=\sza\textwidth,trim= 3cm 14mm 17mm 0, clip,valign=t]{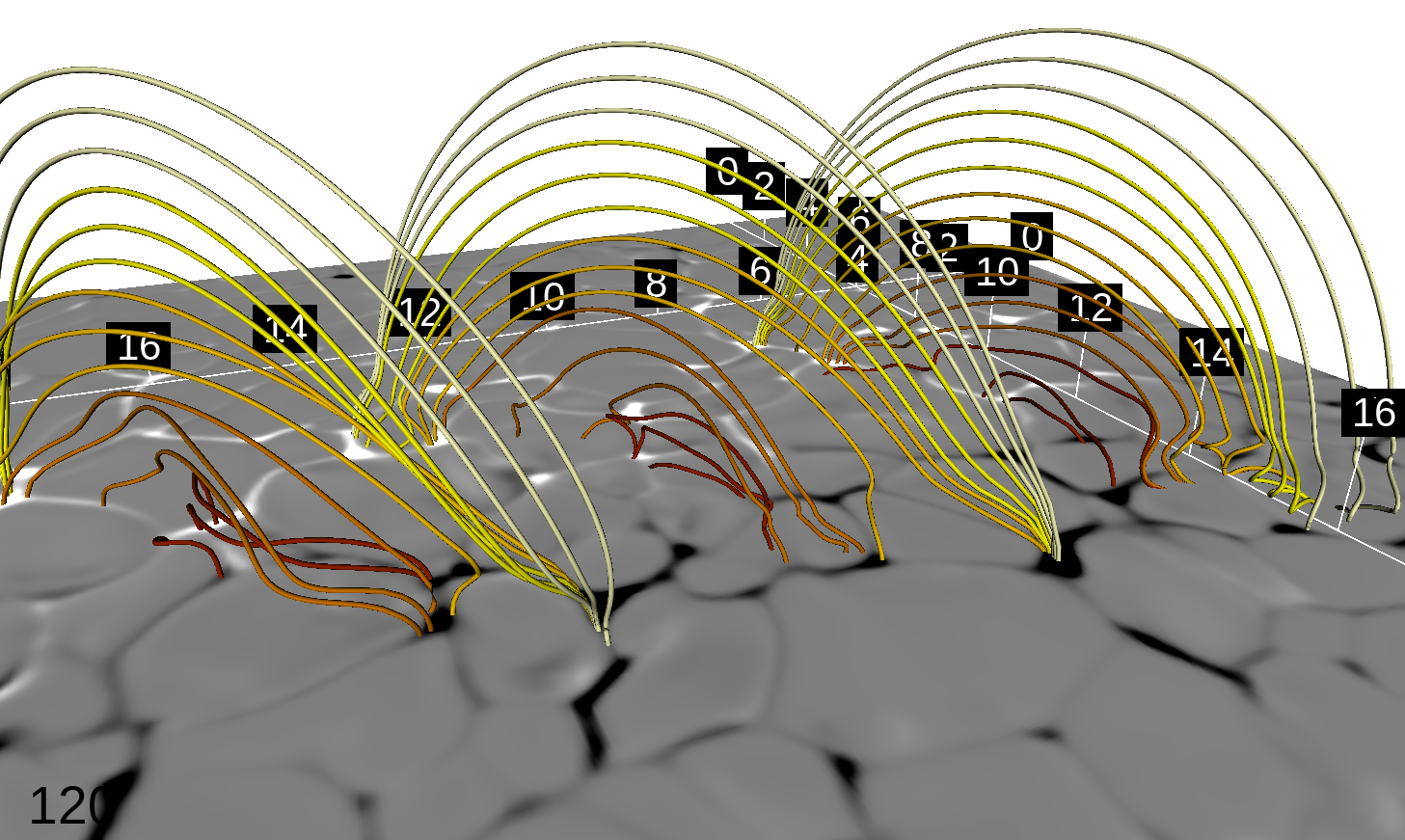}}
    \hfill\!%
    \subfloat{\includegraphics[width=\sza\textwidth,trim= 3cm 14mm 17mm 0, clip,valign=t]{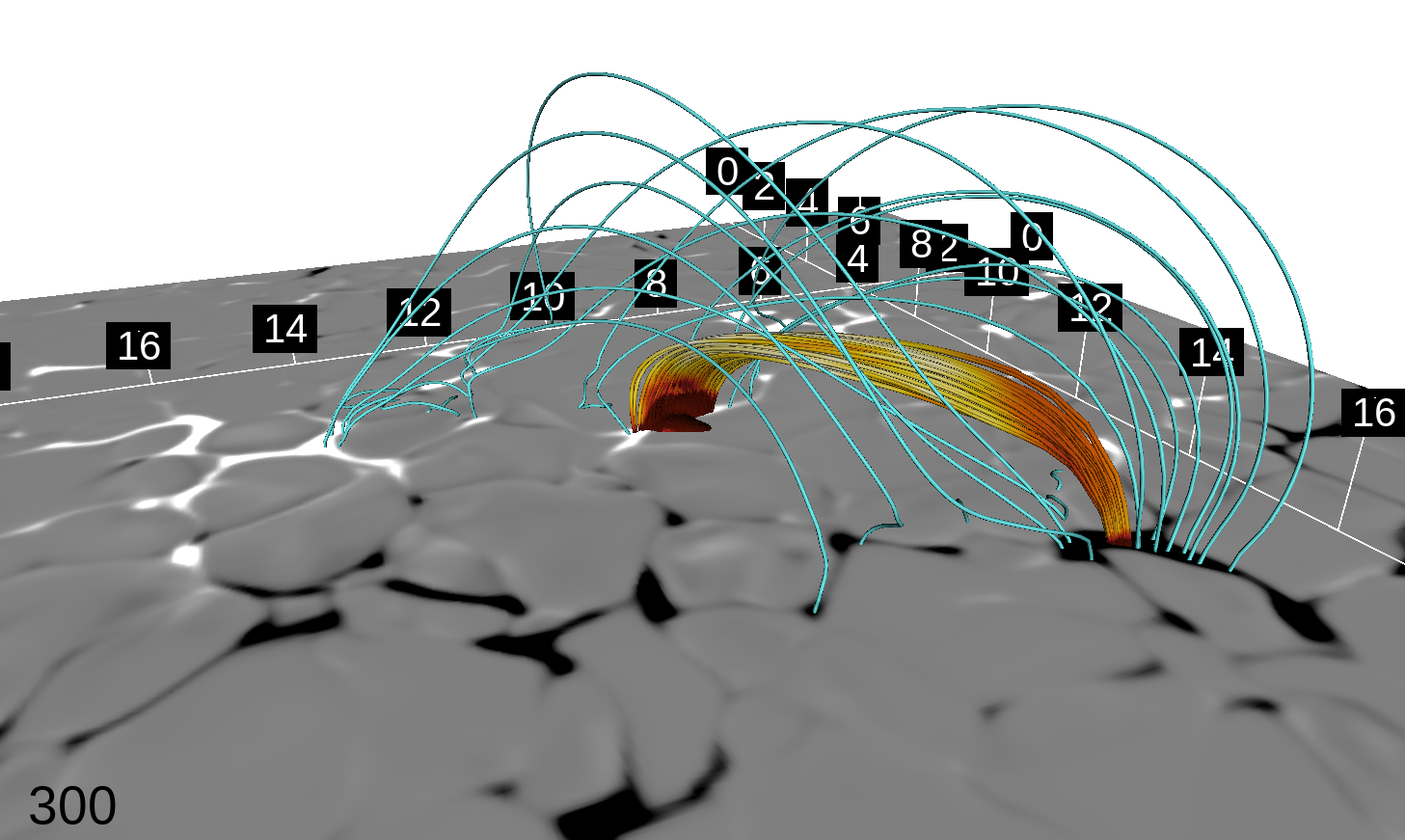}}
    \hfill\!%
    \subfloat{\includegraphics[width=\sza\textwidth,trim= 3cm 14mm 17mm 0, clip,valign=t]{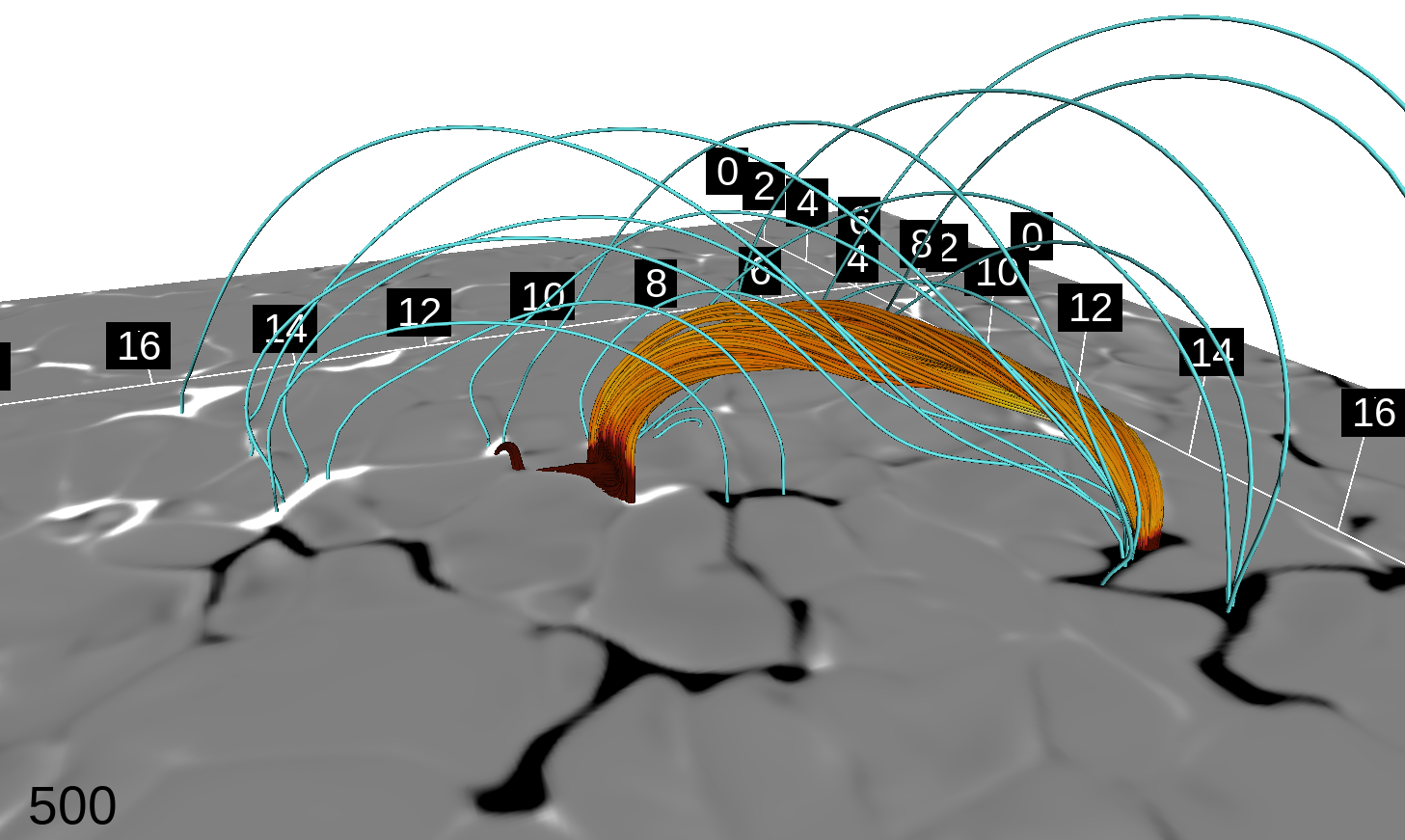}}
    \vspace{-6pt}\\
    \setcounter{subfigure}{0}

    \subfloat[Snapshot 1120, $t=\text{186m40s}$]{\includegraphics[width=\sza\textwidth,trim= 7cm 0 7.7cm 0, clip]{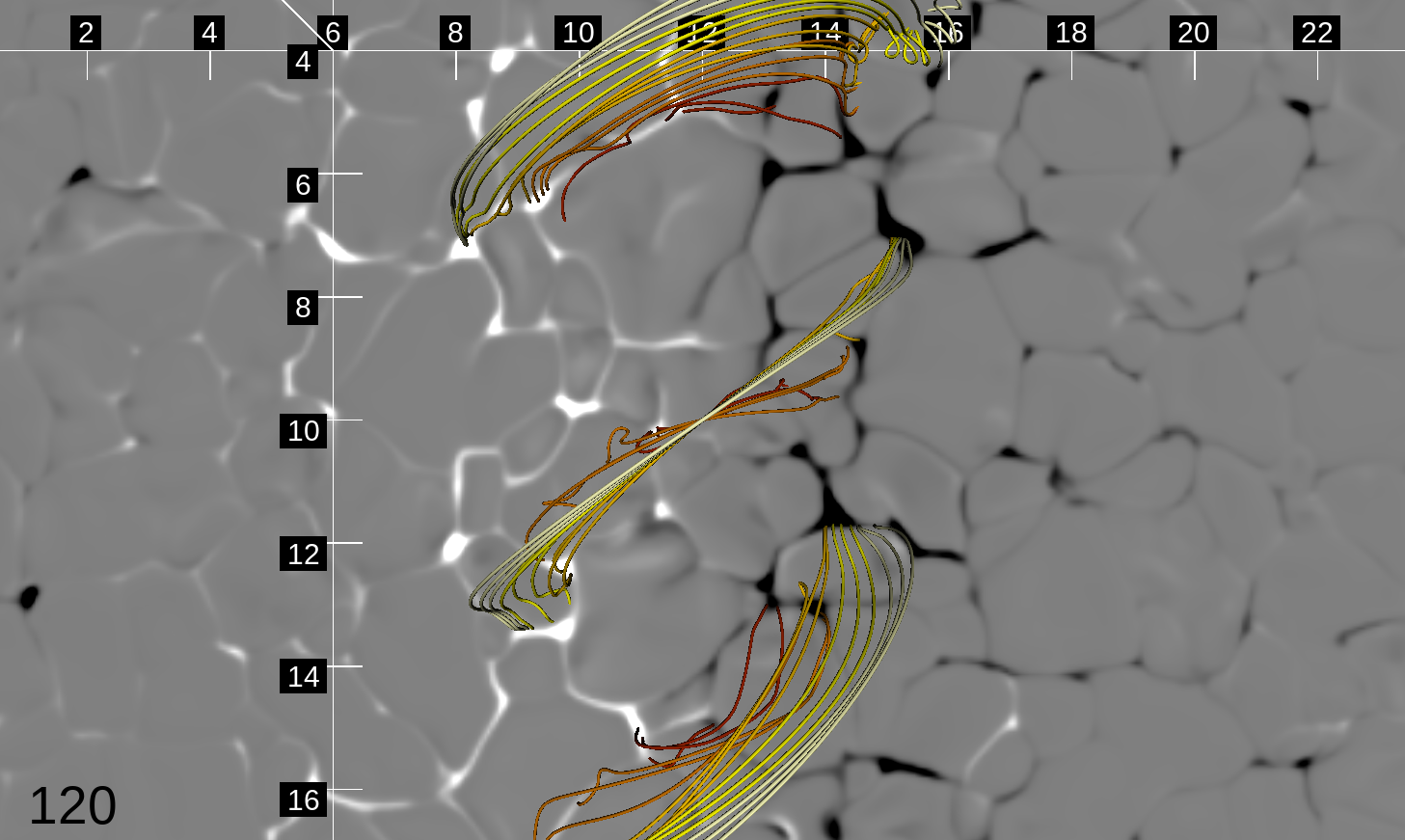}\label{fig_frT_1}}
    \hfill\!%
    \subfloat[Snapshot 1300, $t=\text{216m40s}$]{\includegraphics[width=\sza\textwidth,trim= 7cm 0 7.7cm 0, clip]{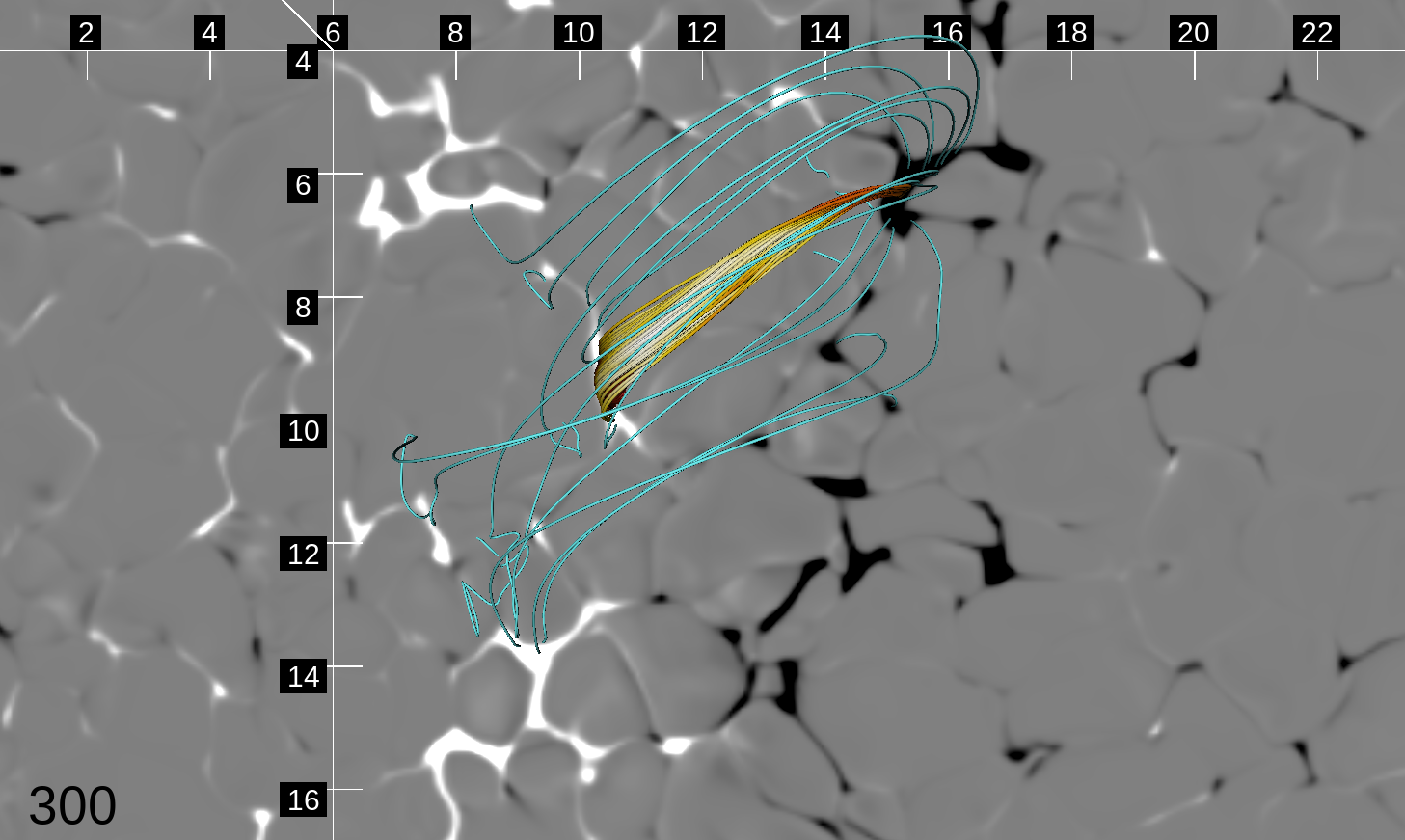}\label{fig_frT_2}}
    \hfill\!%
    \subfloat[Snapshot 1500, $t=\text{250m00s}$]{\includegraphics[width=\sza\textwidth,trim= 7cm 0 7.7cm 0, clip]{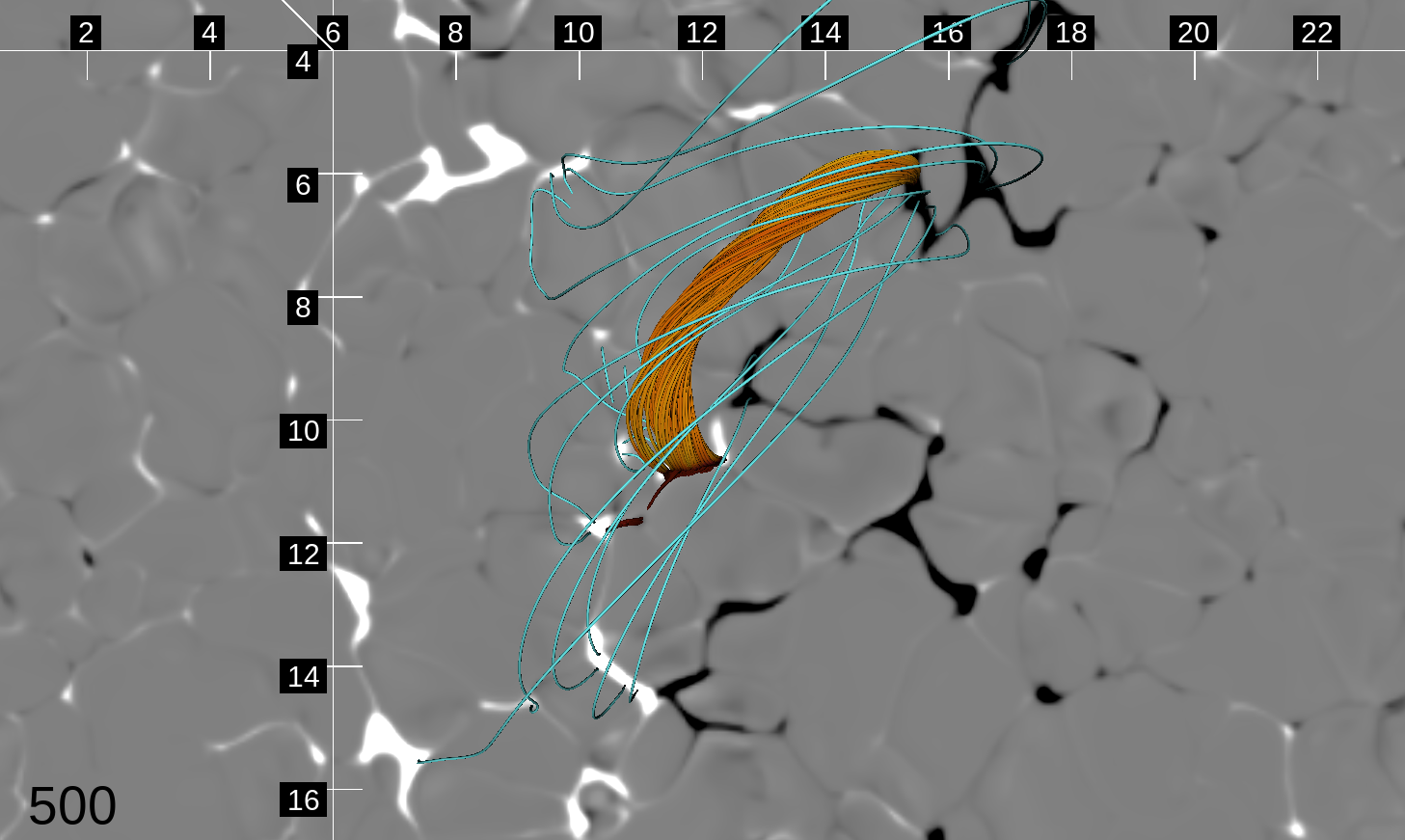}\label{fig_frT_3}}
    \\

    \subfloat{%
        \begin{overpic}[width=\sza\textwidth,trim= 3cm 14mm 17mm 0, clip]{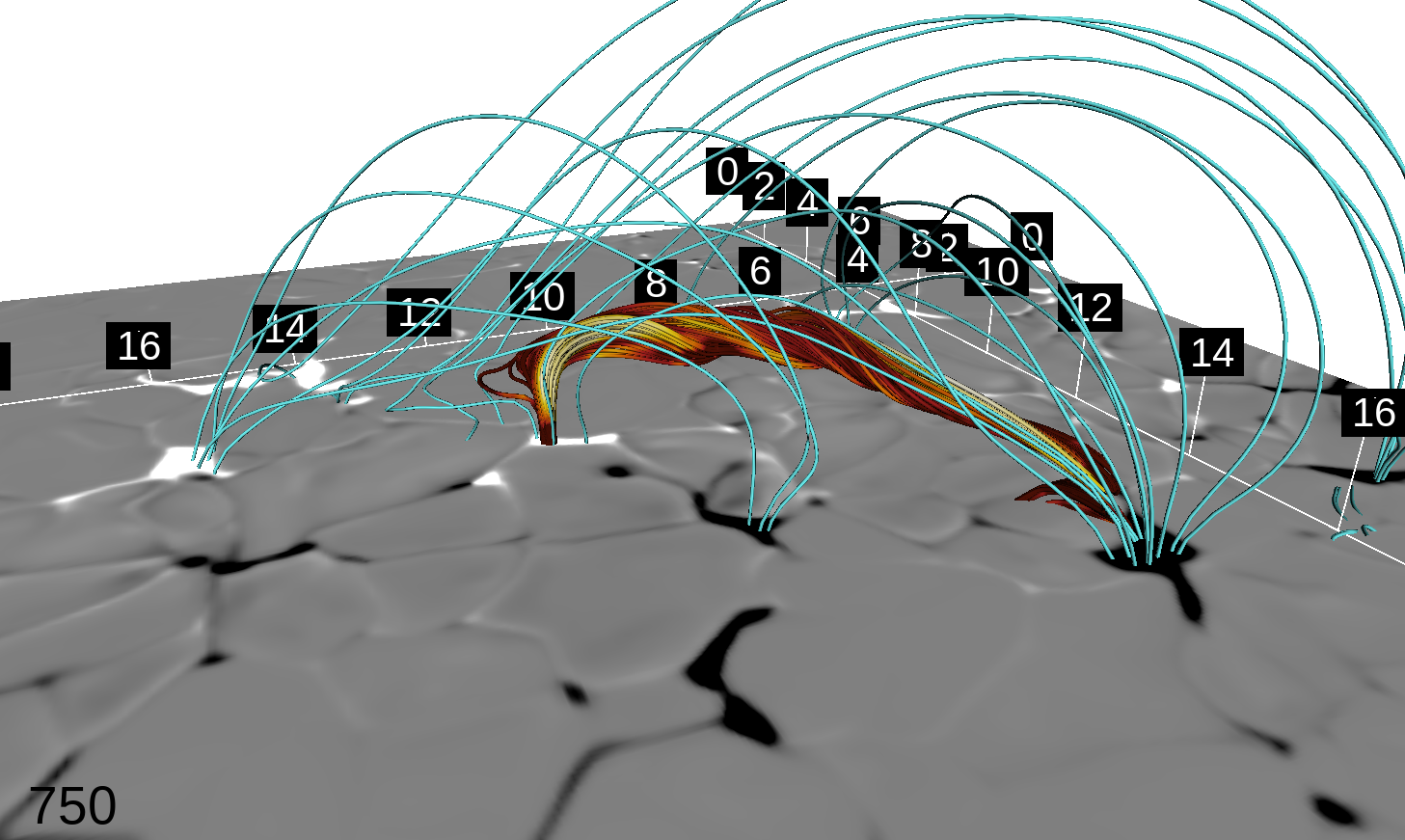}%
        \put (30,22) {\normalsize\textcolor{white}{\textbf{W1}}}%
        \put (1,20) {\normalsize\textcolor{white}{\textbf{W2}}}%
        \put (78,12) {\normalsize\textbf{B1}}%
        \put (55,17) {\normalsize\textbf{B2}}%
        \end{overpic}}
    \hfill\!%
    \subfloat{\includegraphics[width=\sza\textwidth,trim= 3cm 14mm 17mm 0, clip]{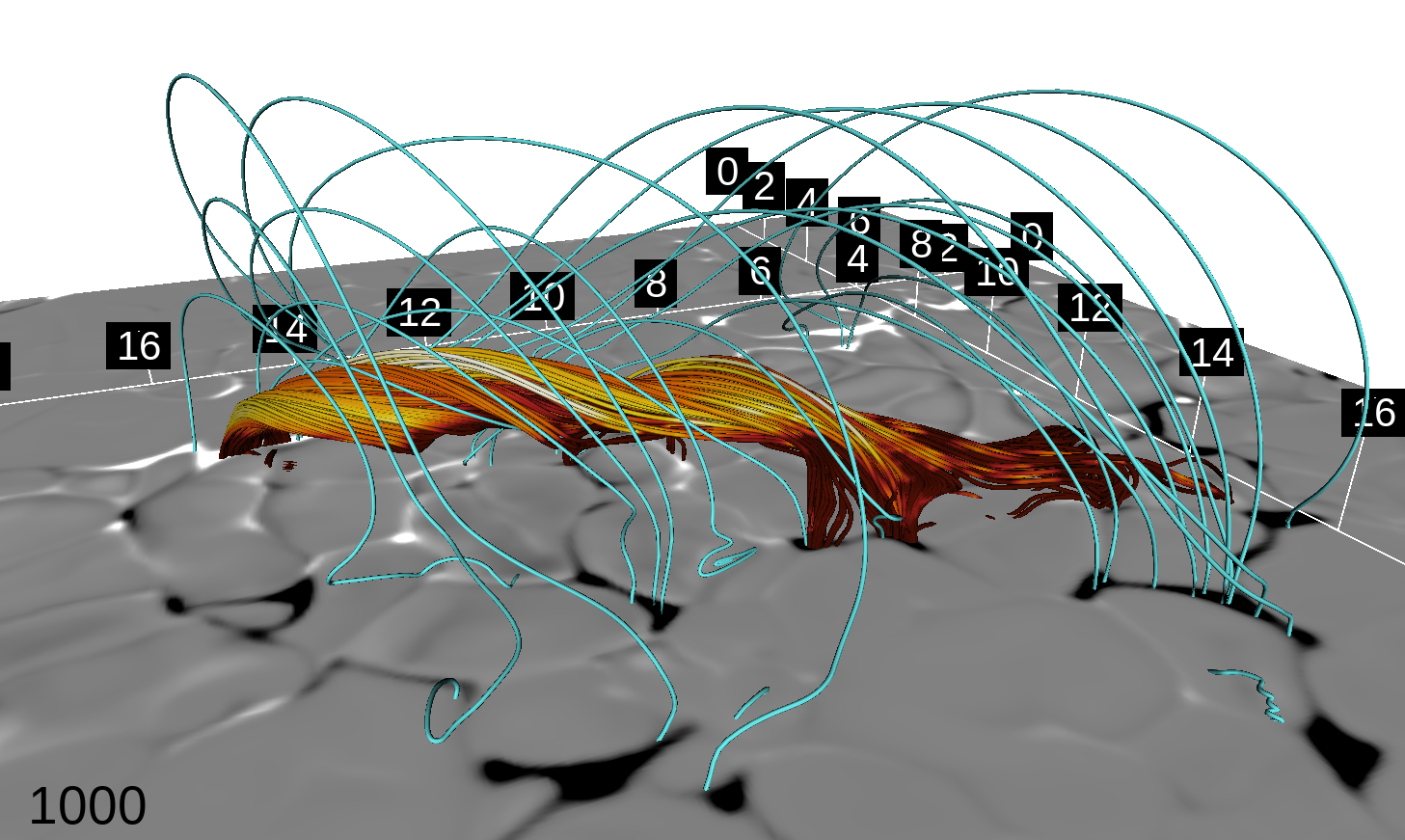}}
    \hfill\!%
    \subfloat{\includegraphics[width=\sza\textwidth,trim= 3cm 14mm 17mm 0, clip]{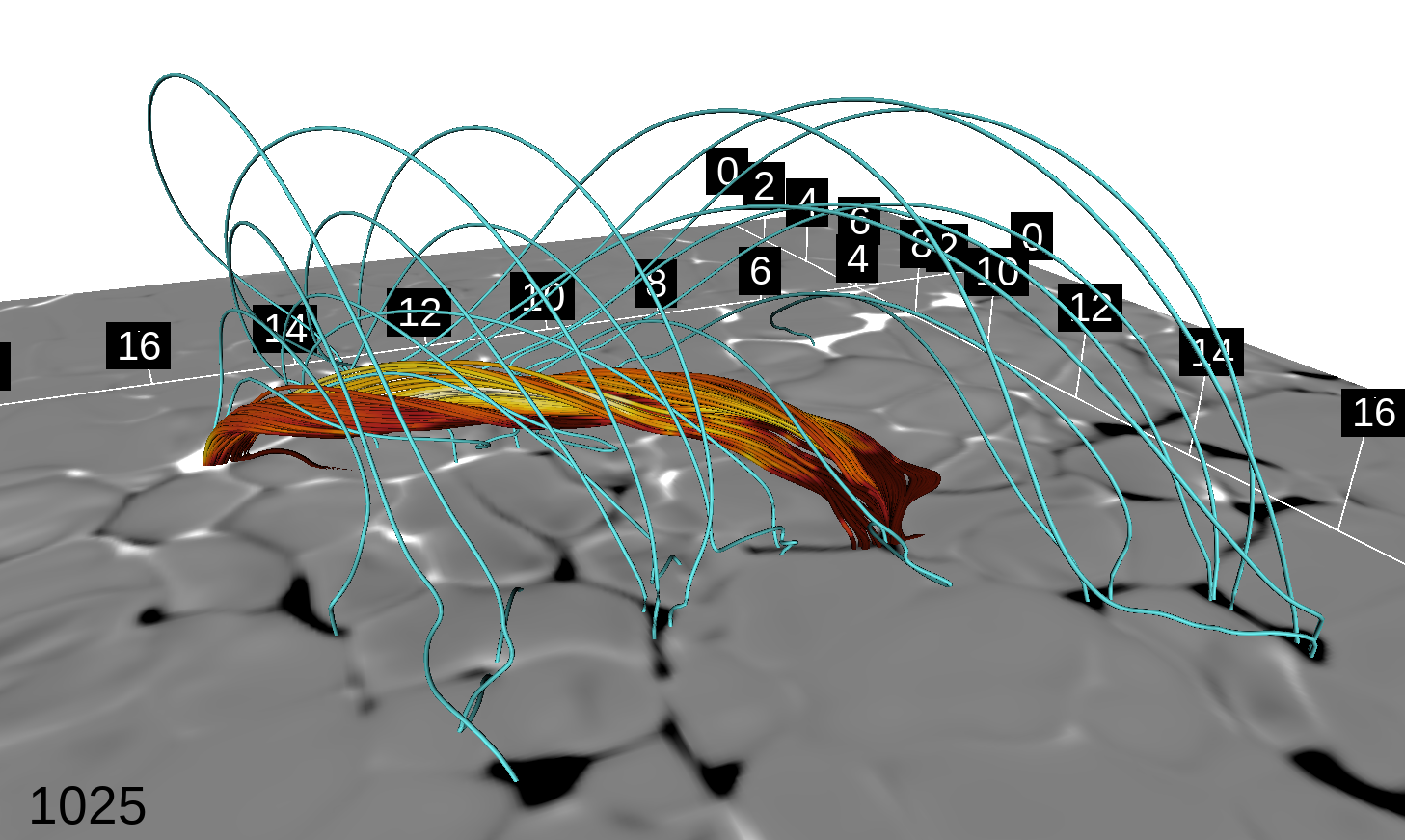}}
    \vspace{-6pt}\\

    \setcounter{subfigure}{3}
    
    \subfloat[Snapshot 1750, $t=\text{291m40s}$]{%
        \begin{overpic}[width=\sza\textwidth,trim= 7cm 0 7.7cm 0, clip]{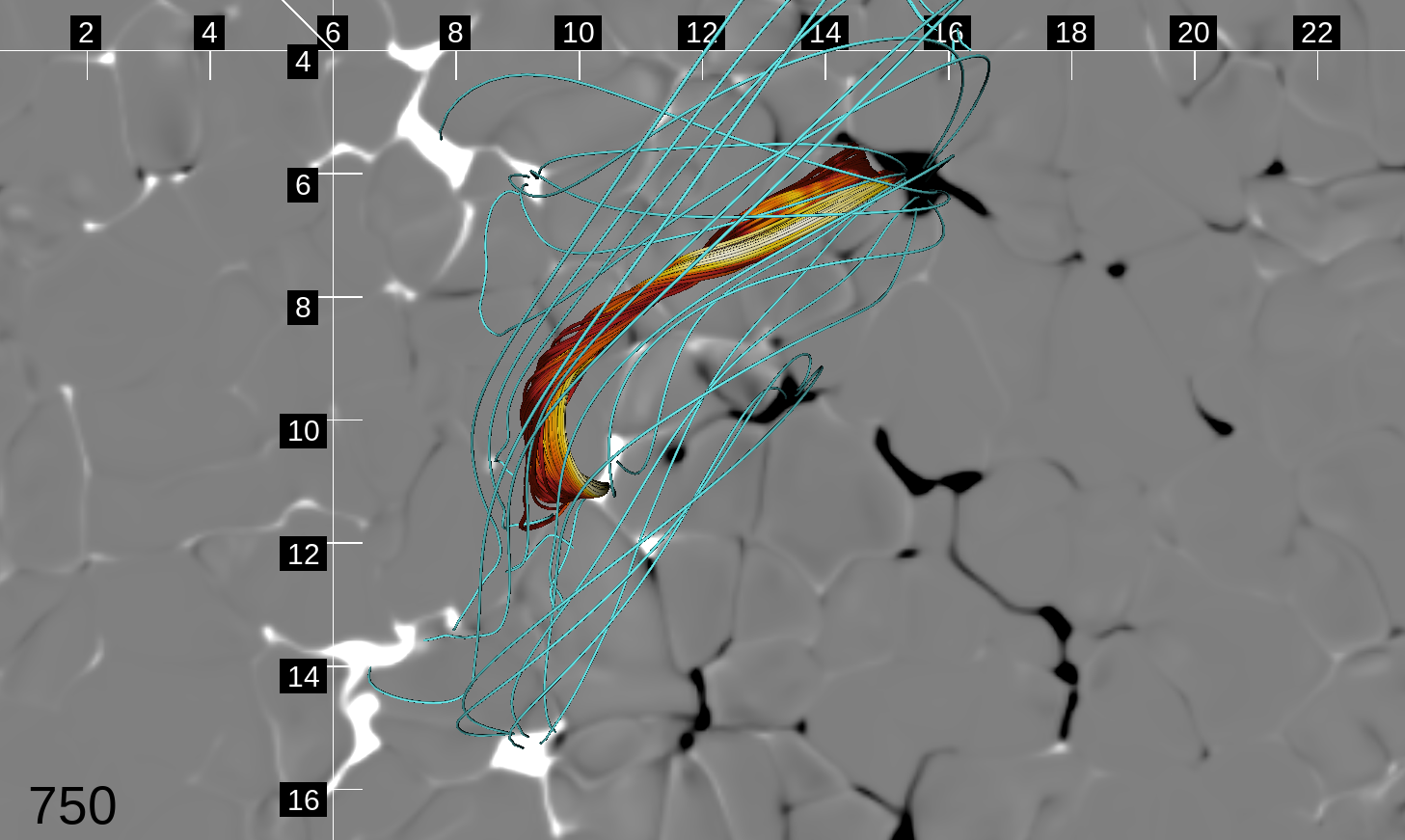}%
        \put (30,34) {\normalsize\textcolor{white}{\textbf{W1}}}%
        \put (22,2) {\normalsize\textcolor{white}{\textbf{W2}}}%
        \put (60,45) {\normalsize\textbf{B2}}%
        \put (82,78) {\normalsize\textbf{B1}}%
        \end{overpic}\label{fig_frT_4}}
    \hfill\!%
    \subfloat[Snapshot 2000, $t=\text{333m20s}$]{\includegraphics[width=\sza\textwidth,trim= 7cm 0 7.7cm 0, clip]{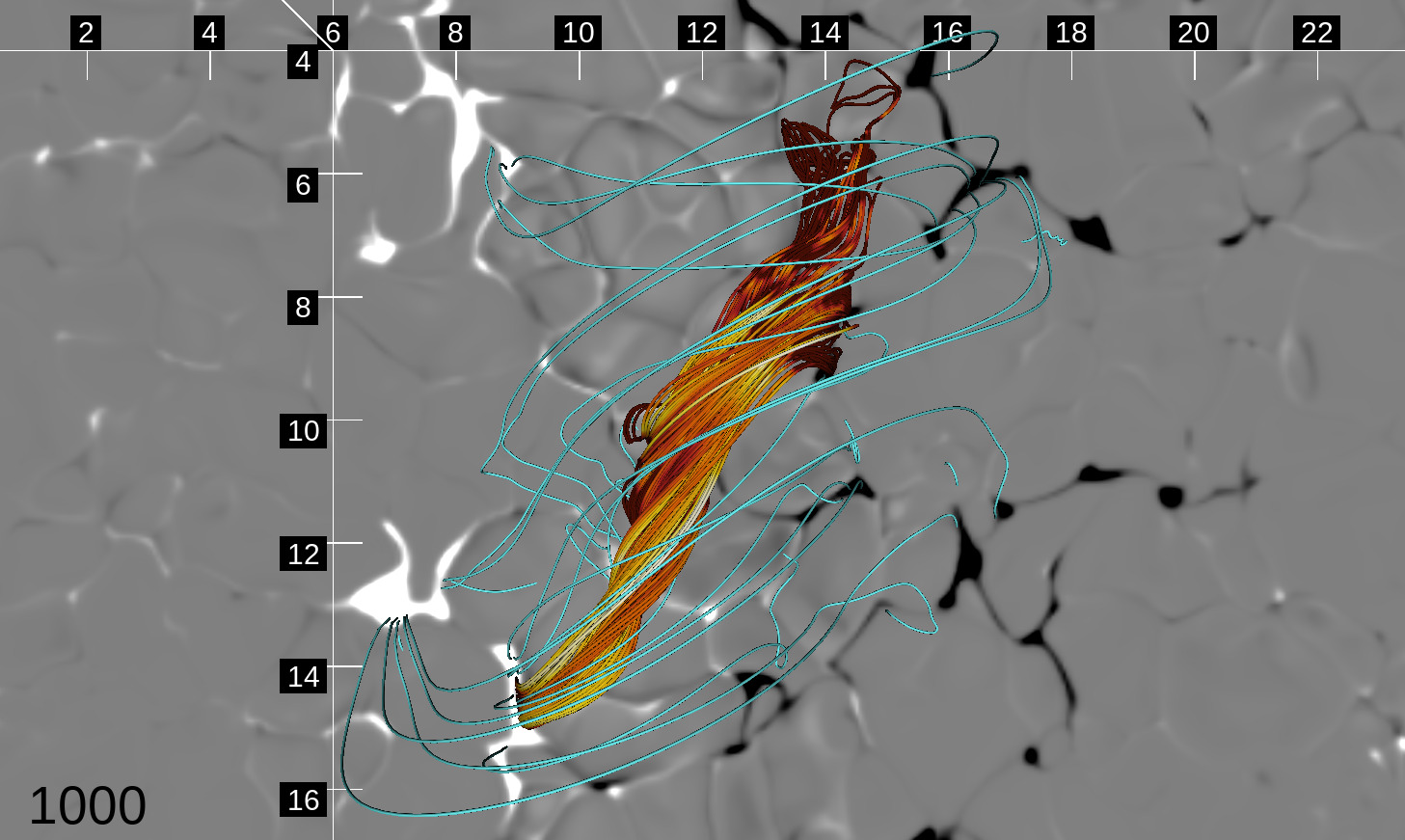}\label{fig_frT_5}}
    \hfill\!%
    \subfloat[Snapshot 2025, $t=\text{337m30s}$]{\includegraphics[width=\sza\textwidth,trim= 7cm 0 7.7cm 0, clip]{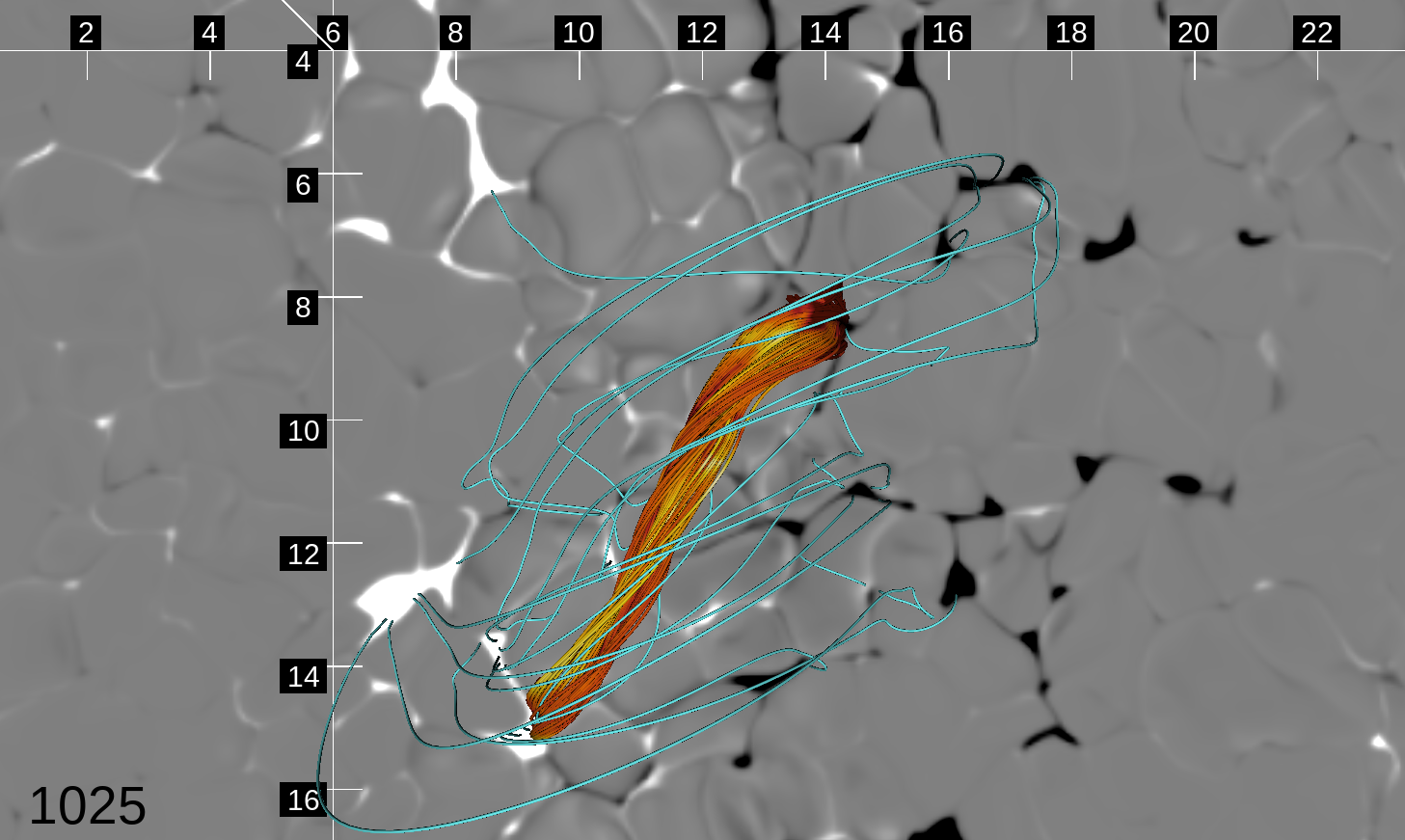}\label{fig_frT_6}}
      
    \caption{
        Gradual formation and evolution of a flux rope in the \bifrost{} simulation. 
        Each subfigure contains two panels, an angled view above a vertical top view of the same set of field lines. The angled views are identical in all subfigures, oriented such that the $x$-axis increases down to the right and the $y$-axis increases down to the left. The numbers on the axes are in units of $\si{\mega\meter}$.
        (a) is right after the end of ramp of the LFFF, and differs from the rest by only showing field lines seeded at ${x=\SI{12}{\mega\meter}}$, ${y\in\{5,10,15\}~\si{\mega\meter}}$, and ${z\in (0,-5)~\si{\mega\meter}}$ in gradually brighter colors.
        (b)-(f) show field lines representative for the flux rope in yellow-red colors, darker colors are shown for larger values of ${\abs{J}/\abs{B}}$, and overlying arcades in cyan. (d) also shows the four poles W1, W2, B1, and B2 referred to in the text.
        }
    \label{fig_frT}
\end{figure*}

The gradual generation and evolution of such a flux rope, over $\SI{2.5}{\hour}$ of solar time after the end of the ramp of the LFFF, is shown in Fig.~\ref{fig_frT}. Fig.~\ref{fig_frT_1} shows the shearing angle of the LFFF, being $\SI{45}{\degree}$ high up in the atmosphere, but slightly less closer to the photosphere. The arcades are also less smooth closer to the photosphere, where the plasma pressure dominates the magnetic pressure. No appreciable twisting is visible in Fig.~\ref{fig_frT_2}, $\SI{30}{\minute}$ later.
A twisted flux rope is gradually formed through Fig.~\ref{fig_frT_3} and Fig.~\ref{fig_frT_4}, with footpoints in a white polarity (W1) and a black polarity (B1). In Fig.~\ref{fig_frT_4}, there are also cyan field lines going from a white pole (W2) at ${(x,y)=(\SI{9.5}{\mega\meter},\SI{15.5}{\mega\meter})}$ to a black pole (B2) at ${(x,y)=(\SI{13.5}{\mega\meter},\SI{9.5}{\mega\meter})}$.
The flux rope later reconnects with these field lines W2B2 to form a longer flux rope in Fig.~\ref{fig_frT_5}, before B1 starts to disentangle toward Fig.~\ref{fig_frT_6}. The resulting flux rope in Fig.~\ref{fig_frT_5} has a maximum footpoint separation of approximately $\SI{12}{\mega\meter}$. At this time, the top of the flux rope extends vertically up to ${z=-\SI{2}{\mega\meter}}$ into the solar atmosphere.

The secondary goal was to understand how such a flux rope could be formed. In particular, how the forming and twisting of the flux rope were related to the flux cancellations at the photosphere. After the ramp of the LFFF, described in Sect.~\ref{sec_exp_lfff}, the simulation was left to its own devices. Various behaviors were observed, of which we in the following analyze and discuss four distinct processes in the order in which they occur during the cancellation process:
(1) Assumed slipping reconnection, without appreciable flux cancellation at the photosphere, aiding in forming the flux rope 
between Fig.~\ref{fig_frT_2} and~\ref{fig_frT_3};
(2) Flux cancellation linked to U-loop emergence, increasing the twist of the flux rope
between Fig.~\ref{fig_frT_3} and~\ref{fig_frT_4};
(3) Flux cancellation linked to $\Omega$-loop submergence (also between Fig.~\ref{fig_frT_3} and~\ref{fig_frT_4}), located right under the flux rope, but not increasing the twist of the flux rope appreciably;
(4) Thick-photosphere tether cutting (TPTC) reconnection and flux cancellation between a preexisting flux rope and comparably untwisted arcades between Fig.~\ref{fig_frT_4} and~\ref{fig_frT_5}.


\begin{figure*}[!t]
    \renewcommand{\sza}{0.32}
    \renewcommand{\szb}{0.33}
    
    \centering
    \vspace{20pt}
    \subfloat{\includegraphics[width=\sza\textwidth,trim= 21mm 15mm 17mm 0mm, clip]{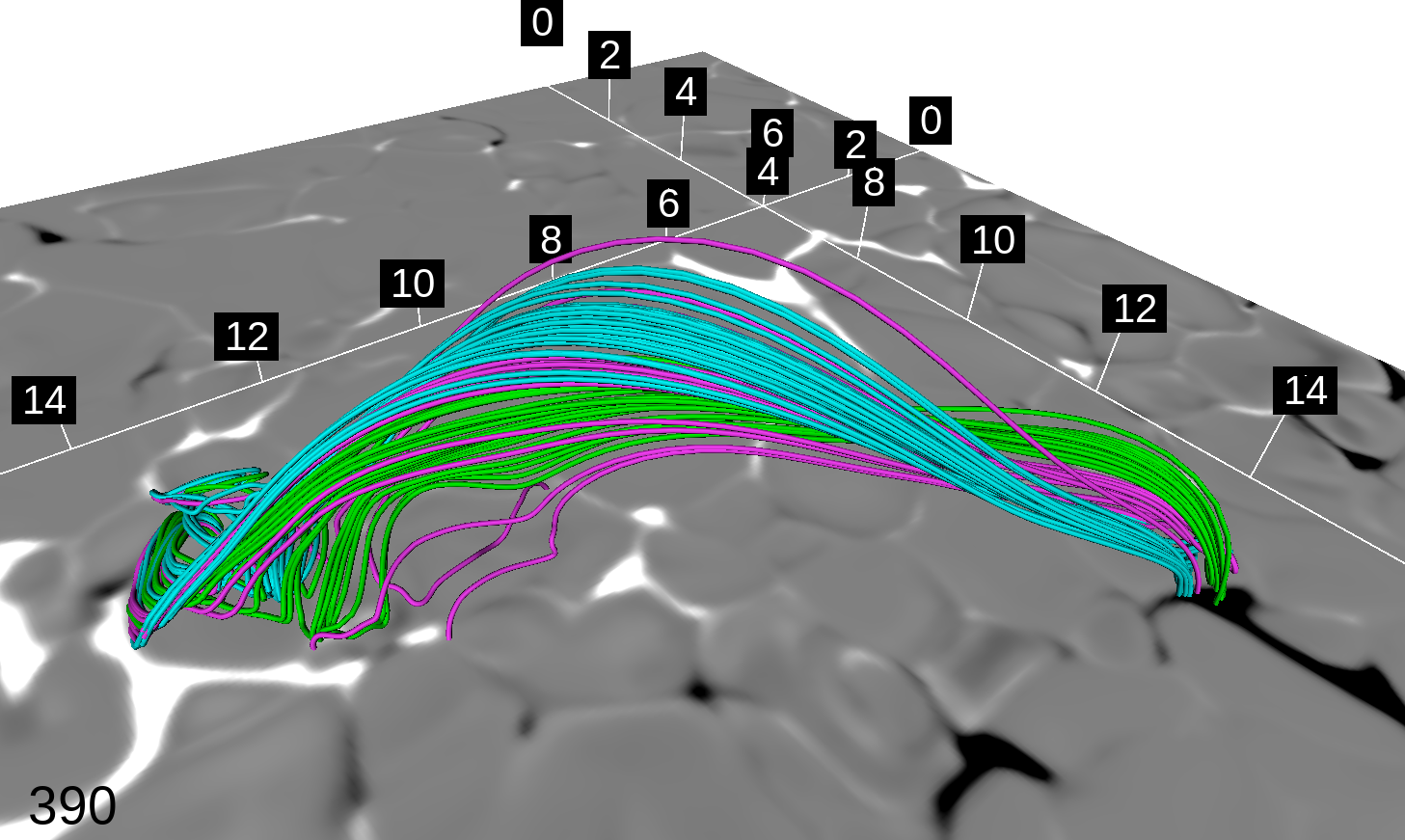}}
    \hfill%
    \subfloat{\includegraphics[width=\sza\textwidth,trim= 21mm 15mm 17mm 0mm, clip]{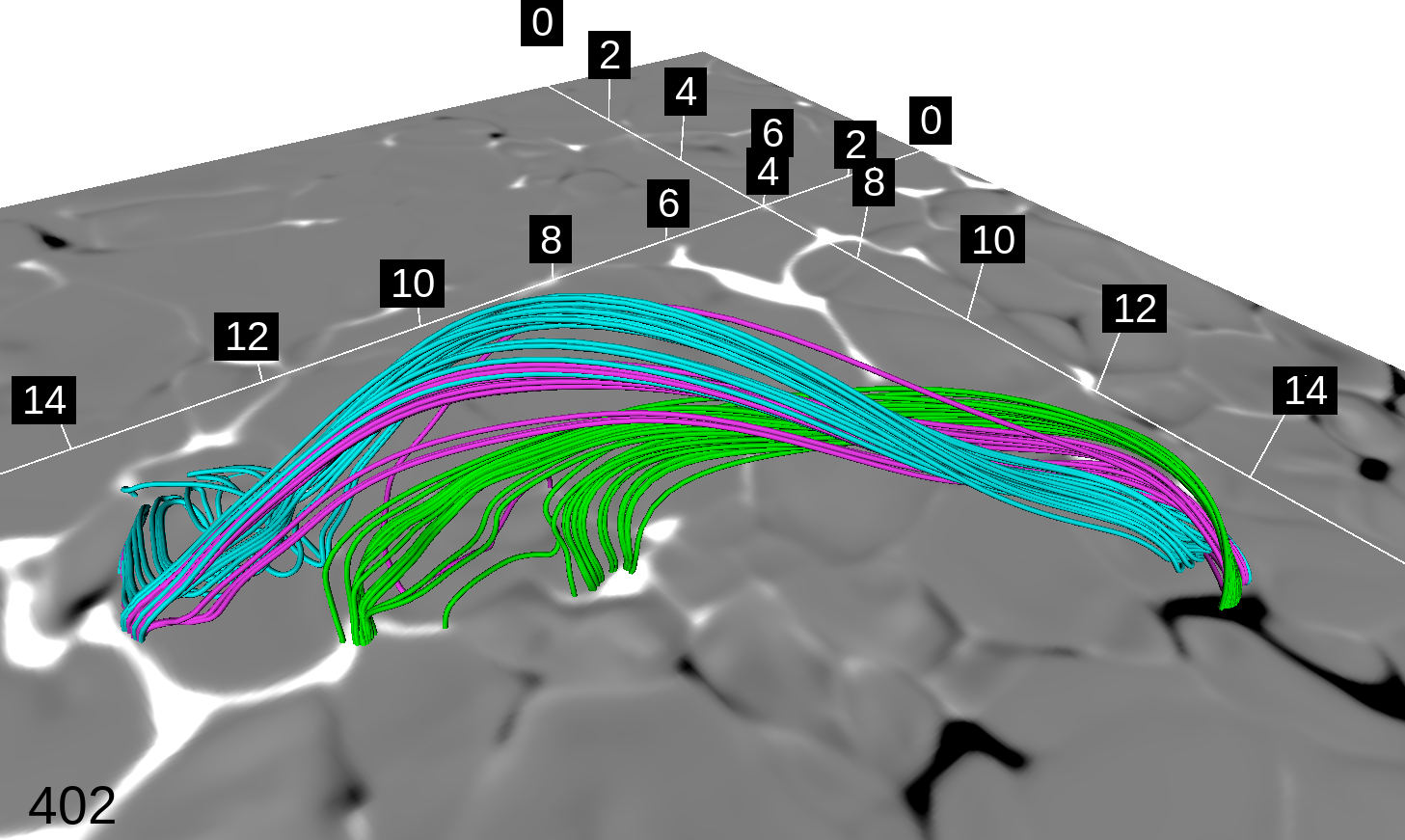}}
    \hfill%
    \subfloat{\includegraphics[width=\sza\textwidth,trim= 21mm 15mm 17mm 0mm, clip]{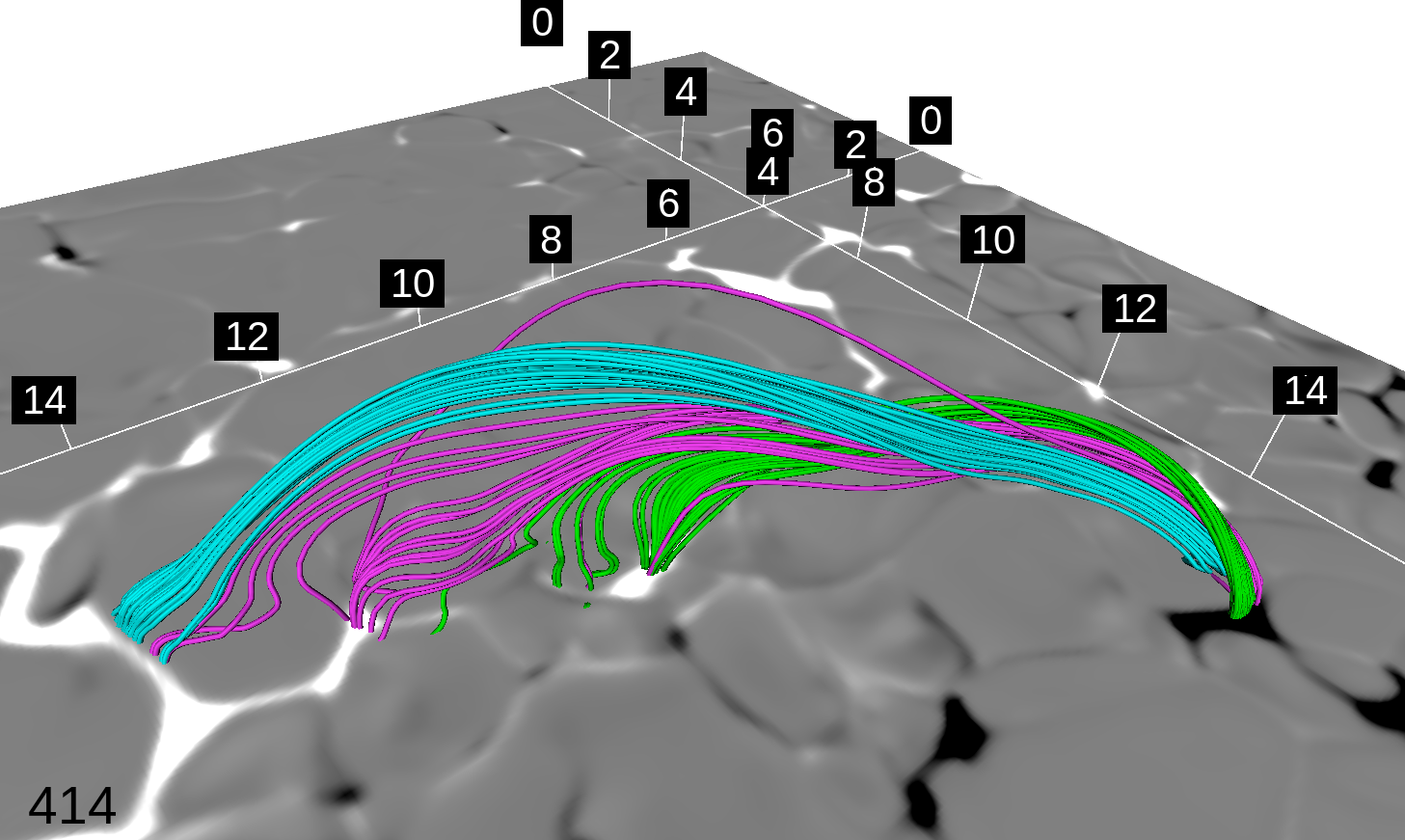}}

    \setcounter{subfigure}{0}

    \subfloat[Snapshot 1390, $t=t_a-\SI{14}{\minute}$]{\includegraphics[width=\szb\textwidth,trim= 0 0 0 0, clip]{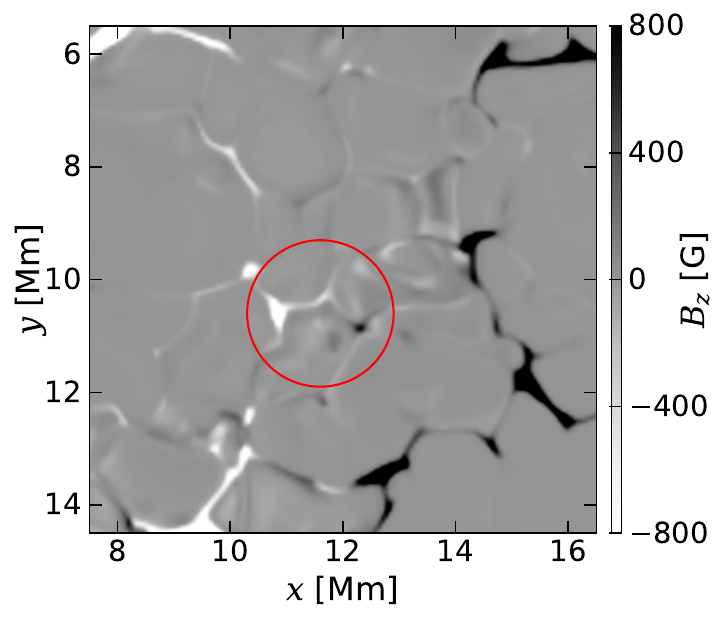}\label{fig_fr1a_1}} 
    \subfloat[Snapshot 1402, $t=t_a-\SI{12}{\minute}$]{\includegraphics[width=\szb\textwidth,trim= 0 0 0 0, clip]{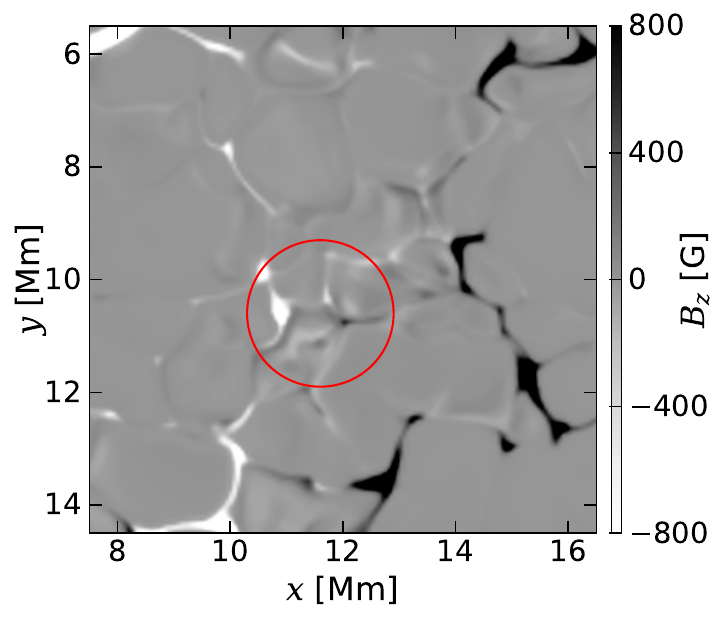}\label{fig_fr1a_2}} 
    \subfloat[Snapshot 1414, $t=t_a-\SI{10}{\minute}$]{\includegraphics[width=\szb\textwidth,trim= 0 0 0 0, clip]{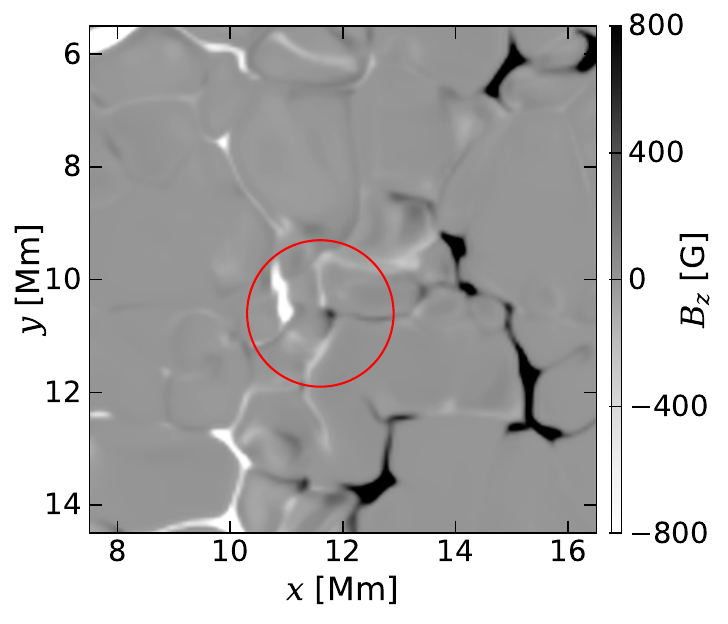}\label{fig_fr1a_3}}
    \\  

    \setcounter{subfigure}{0}

    \vspace{20pt}
    \subfloat{\includegraphics[width=\sza\textwidth,trim= 21mm 15mm 17mm 0mm, clip]{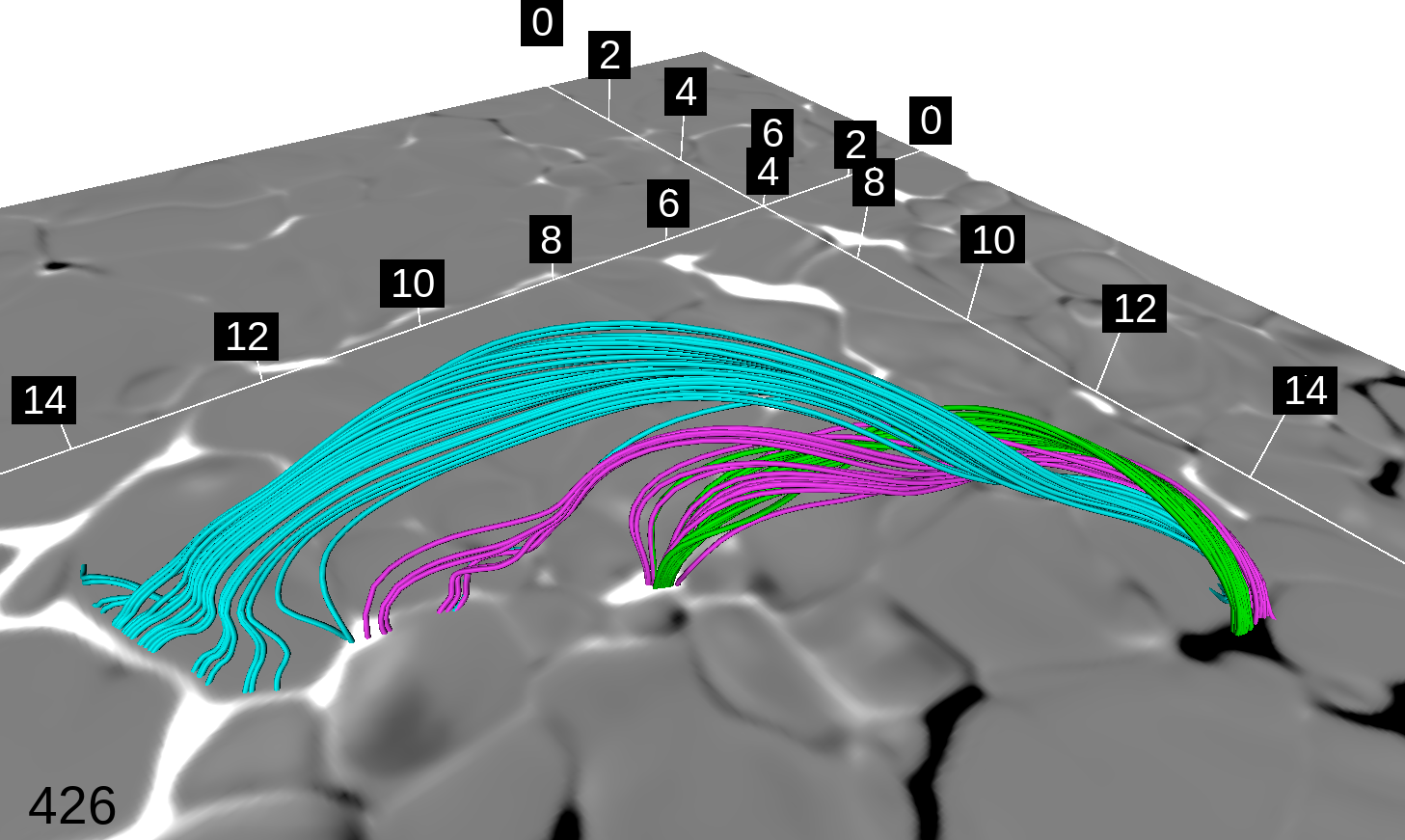}}
    \hfill%
    \subfloat{\includegraphics[width=\sza\textwidth,trim= 21mm 15mm 17mm 0mm, clip]{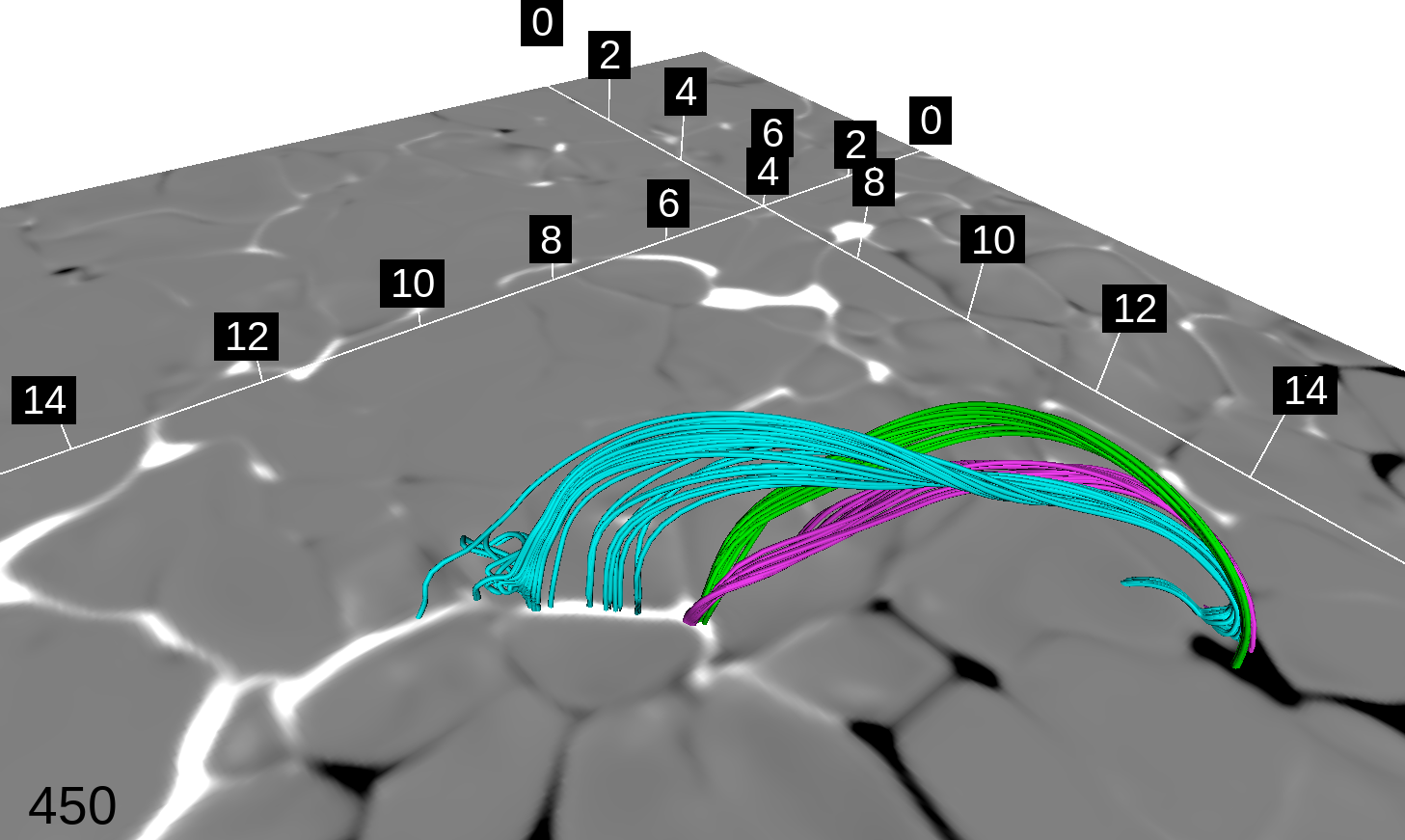}}
    \hfill%
    \subfloat{\includegraphics[width=\sza\textwidth,trim= 21mm 15mm 17mm 0mm, clip]{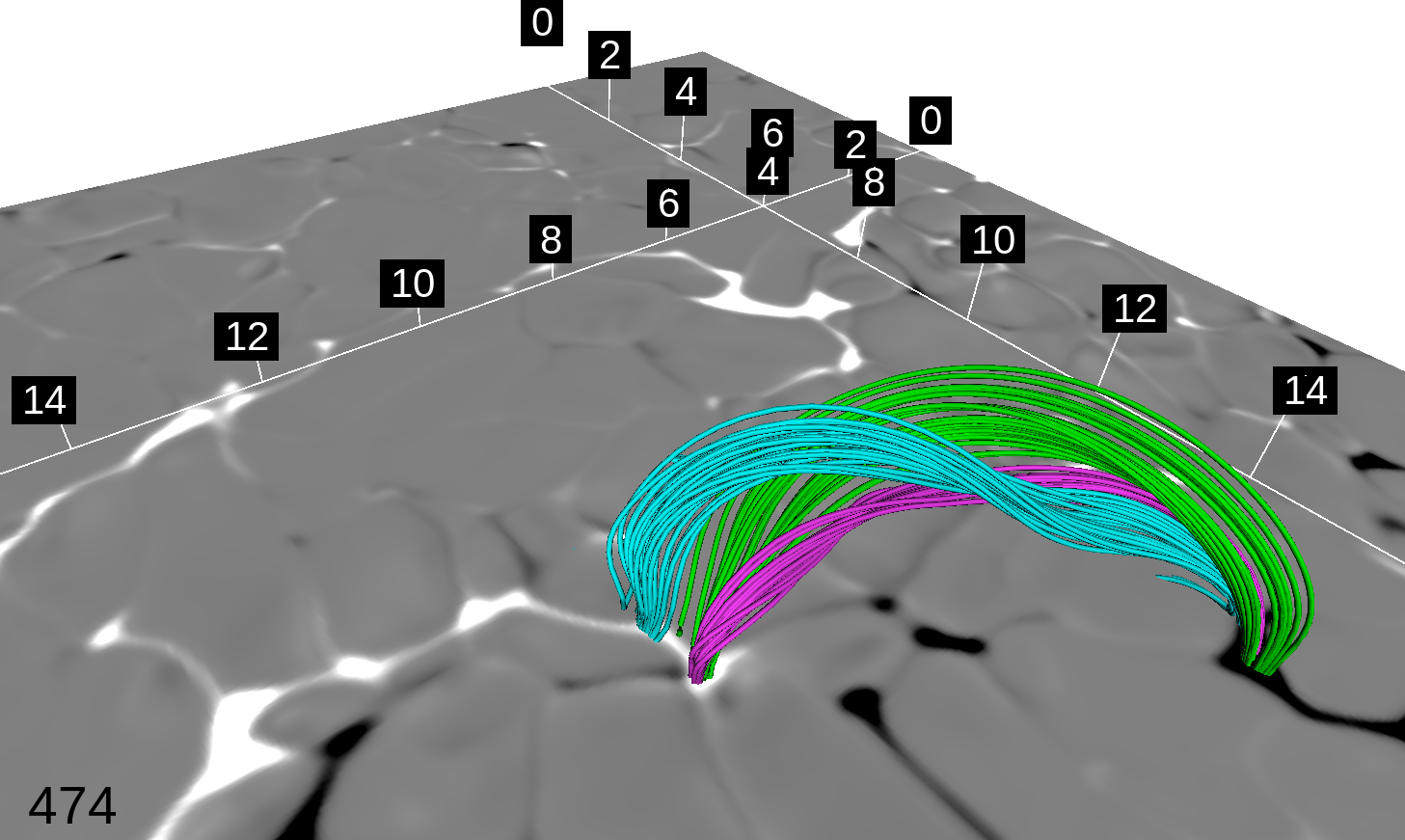}}

    \subfloat[Snapshot 1426, $t=t_a-\SI{8}{\minute}$]{\includegraphics[width=\szb\textwidth,trim= 0 0 0 0, clip]{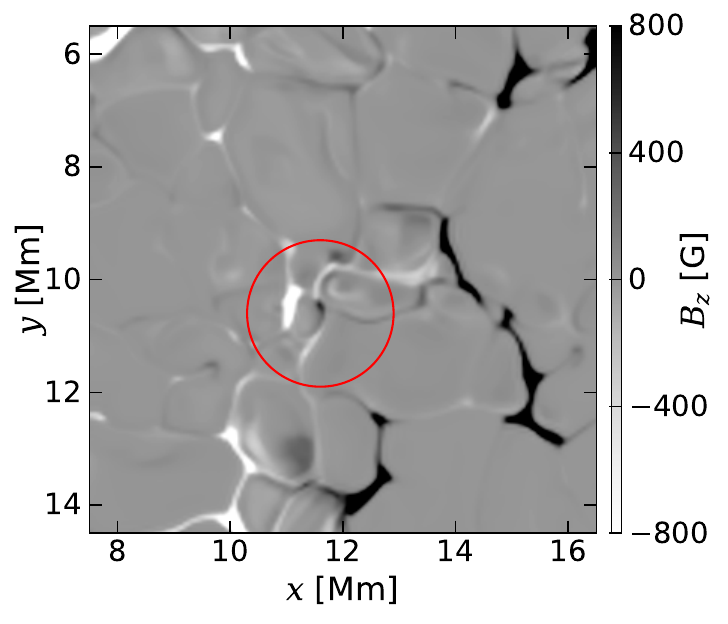}\label{fig_fr1a_4}} 
    \subfloat[Snapshot 1450, $t=t_a-\SI{4}{\minute}$]{\includegraphics[width=\szb\textwidth,trim= 0 0 0 0, clip]{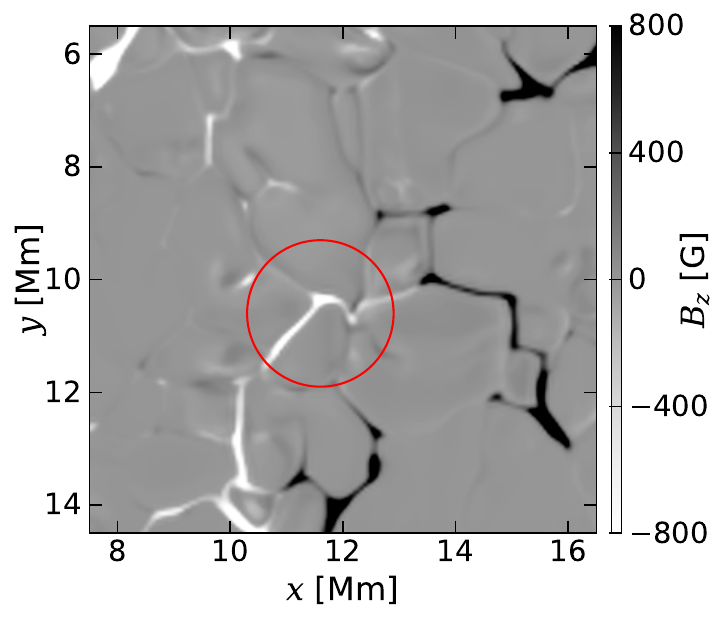}\label{fig_fr1a_5}} 
    \subfloat[Snapshot 1474, $t=t_a=\text{245m40s}$]{\includegraphics[width=\szb\textwidth,trim= 0 0 0 0, clip]{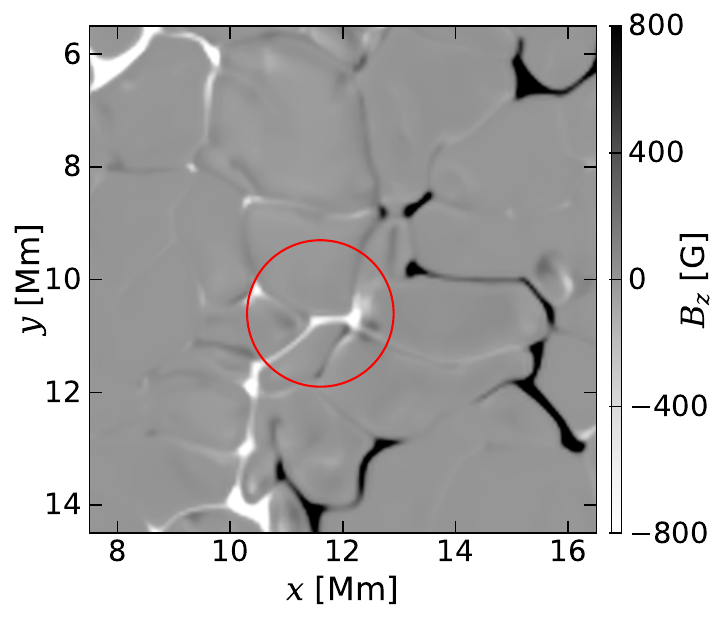}\label{fig_fr1a_6}}
    
    \caption{
        Slipping reconnection in the \bifrost{} simulation. Combined, the subfigures labeled (a)-(f) show the evolution until the time ${t_a=\text{245m40s}}$ when this slipping process is complete. 
        Each subfigure contains a top panel with an angled view of three families of 3D field lines (cyan, green, magenta), tracked by corks, above a 2D magnetogram. The angled view is angled from the bottom right corner of the 2D magnetogram toward the top left corner.
        On the bottom of each subfigure is a panel with only the 2D magnetogram , seen directly from above.
        The red circles in the bottom panels of each subfigure all highlight the same area of interest. 
    }
    \label{fig_fr1a}
\end{figure*}

\begin{figure*}[!t]
    \renewcommand{\sza}{0.32}
    \centering
    \subfloat{\includegraphics[width=\sza\textwidth,trim= 5cm 0 7cm 0, clip]{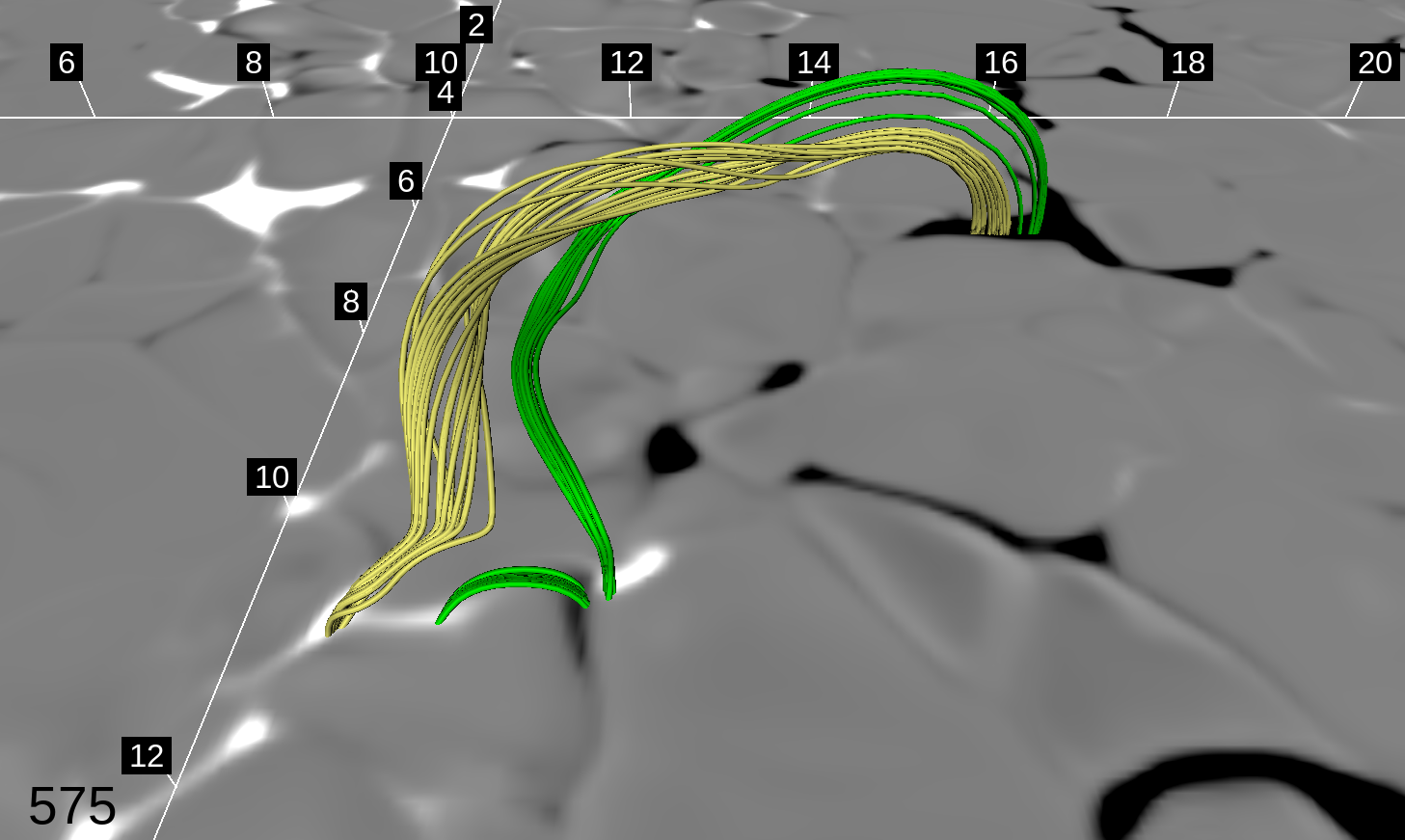}}
    \hfill\!%
    \subfloat{\includegraphics[width=\sza\textwidth,trim= 5cm 0 7cm 0, clip]{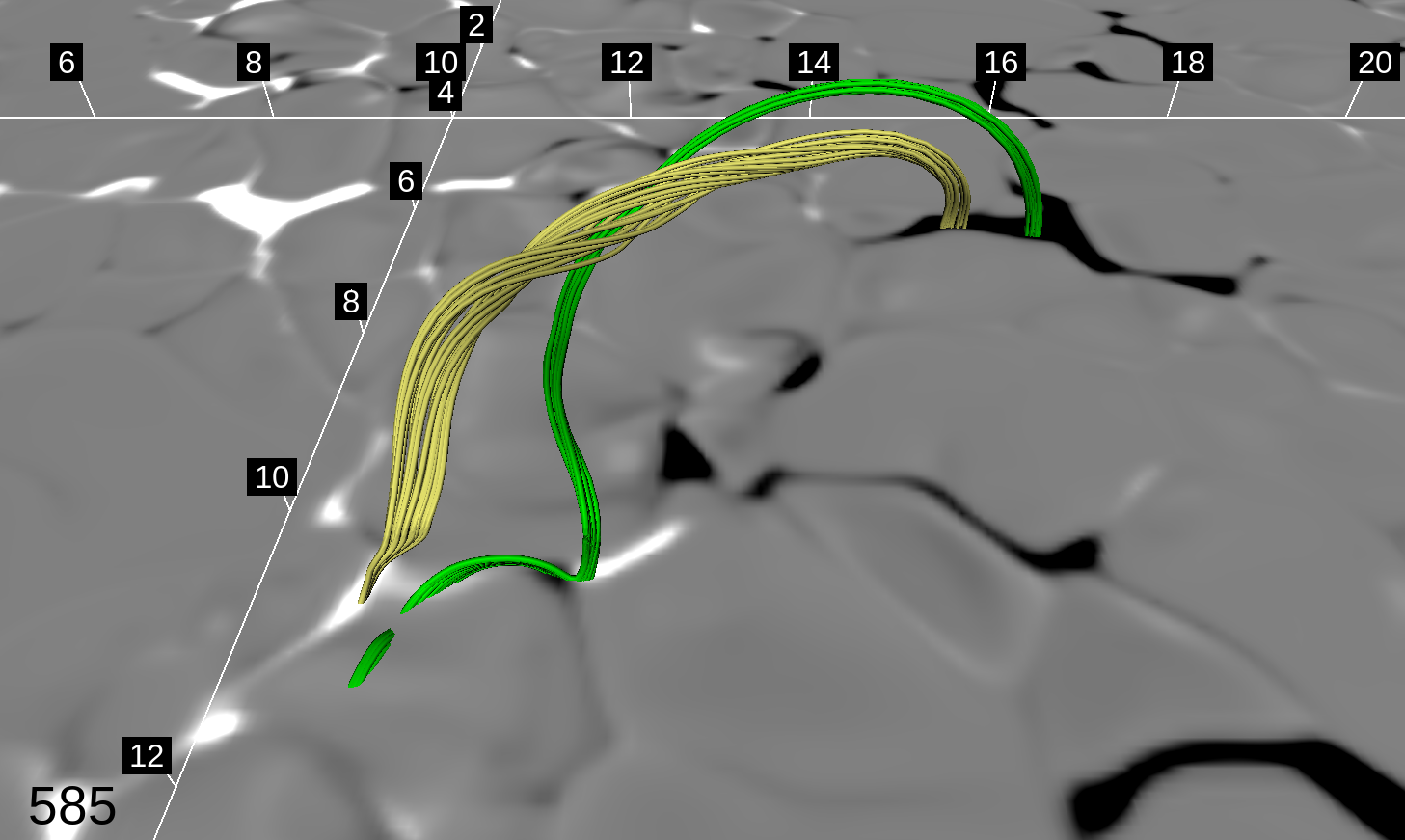}}
    \hfill\!%
    \subfloat{\includegraphics[width=\sza\textwidth,trim= 5cm 0 7cm 0, clip]{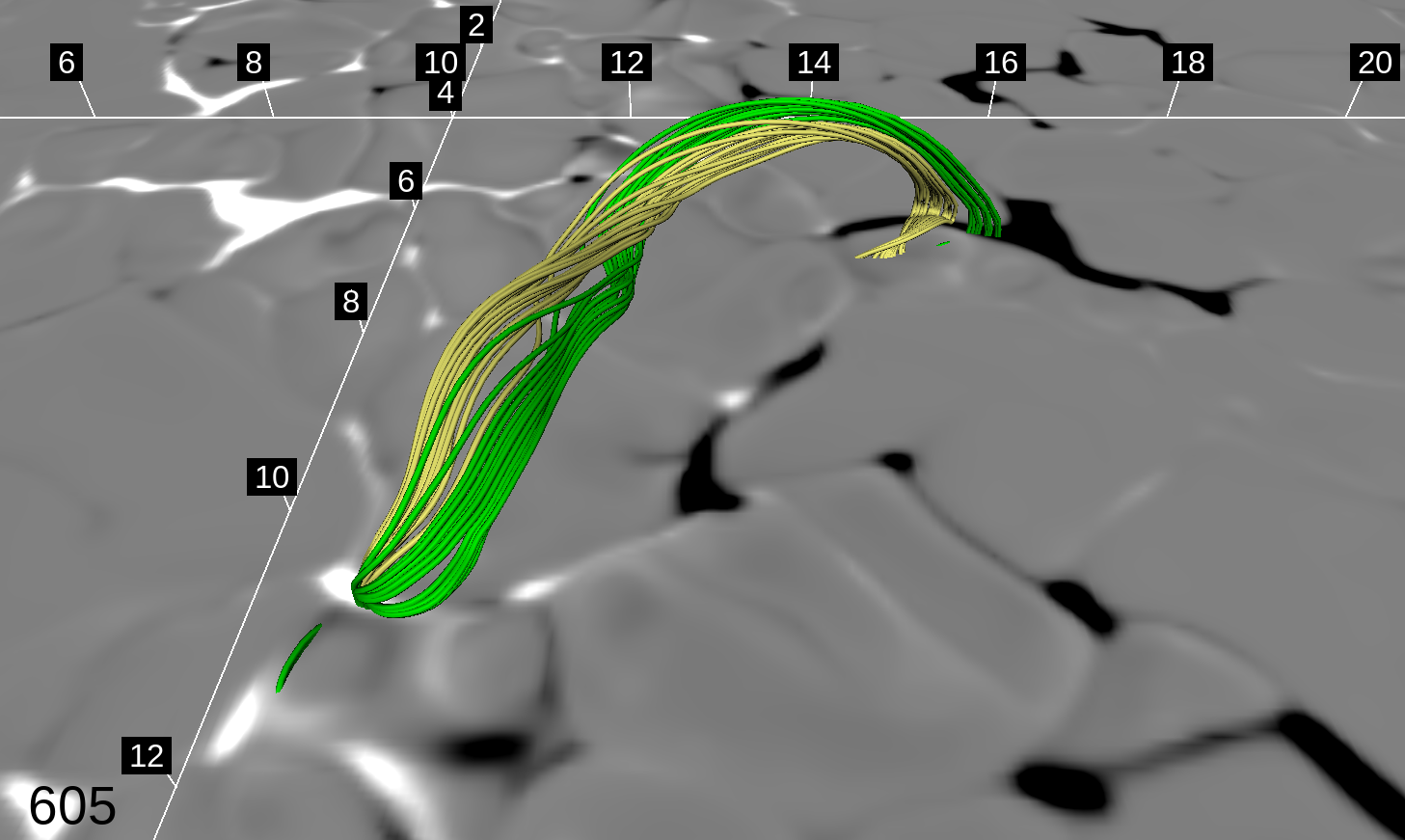}} 
    \vspace{-10pt}\\
    \setcounter{subfigure}{0}
    %
    %
    \subfloat[Snapshot 1575, $t=t_b-\SI{100}{\second}$]{\includegraphics[width=0.33\textwidth]{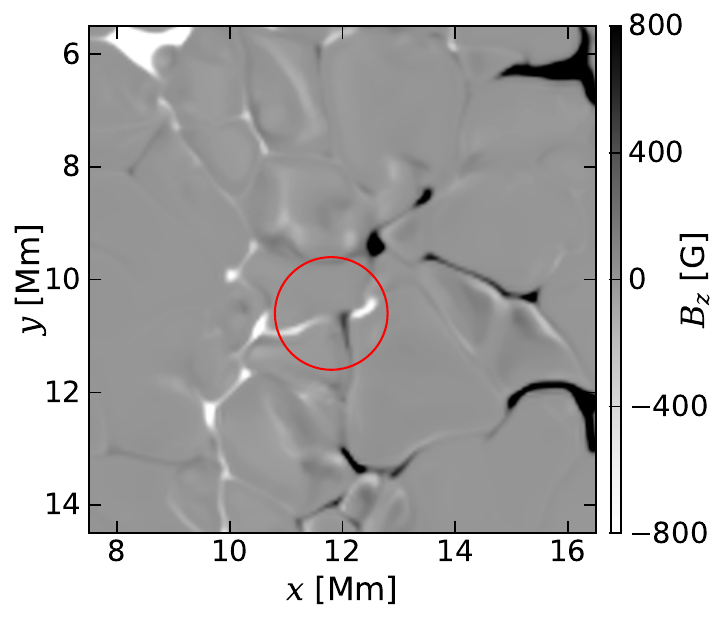}\label{fig_fr1b_1}}
    \hfill\!%
    \subfloat[Snapshot 1585, $t=t_b=\text{264m10s}$]{\includegraphics[width=0.33\textwidth]{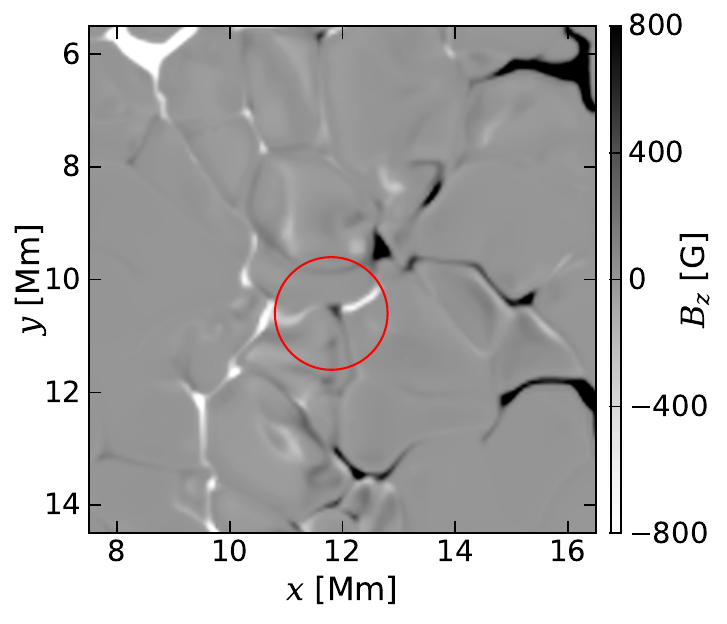}\label{fig_fr1b_2}}
    \hfill\!%
    \subfloat[Snapshot 1605, $t=t_b+\SI{200}{\second}$]{\includegraphics[width=0.33\textwidth]{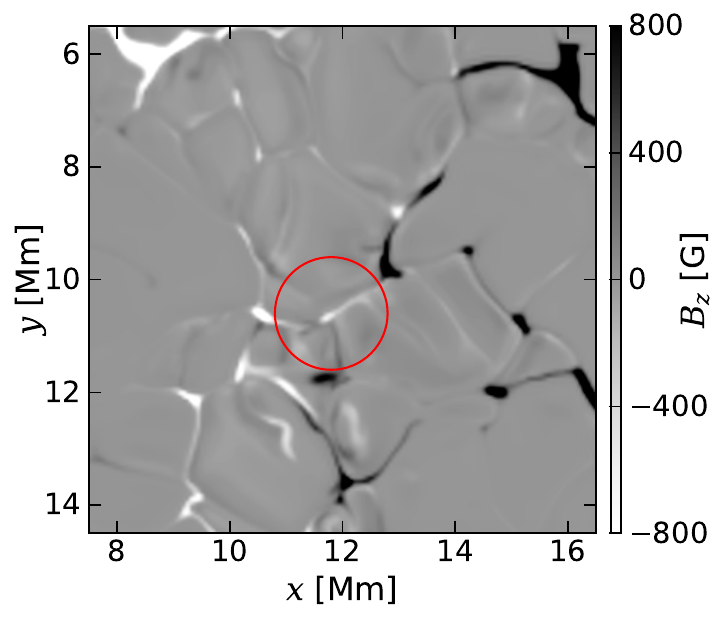}\label{fig_fr1b_3}}\\
    \caption{
        U-loop emergence in the \bifrost{} simulation. 
        From left to right, there are three different times relative to $t_b$, a representative time for the emergence: (a) ${\SI{100}{\second}}$ earlier, (b) $t_b$, (c) ${\SI{200}{\second}}$ later, after the end of the relevant flux cancellation. The upper panels of each subfigure show the 3D field lines of interest in green, tracked by corks, and pale yellow field lines representing the core of the flux rope, above a 2D magnetogram. 
        The view in the upper panels is angled toward lower $y$. The lower panels show the same 2D magnetogram, directly from above. The cancellation of interest occurs within the red circle. 
    }
    \label{fig_fr1b}
\end{figure*}

\begin{figure*}[!t]
    \renewcommand{\sza}{0.32}
    \renewcommand{\szb}{0.33}
    \centering

    \subfloat{\includegraphics[width=\sza\textwidth,trim= 60mm 0cm 0mm 6mm, clip]{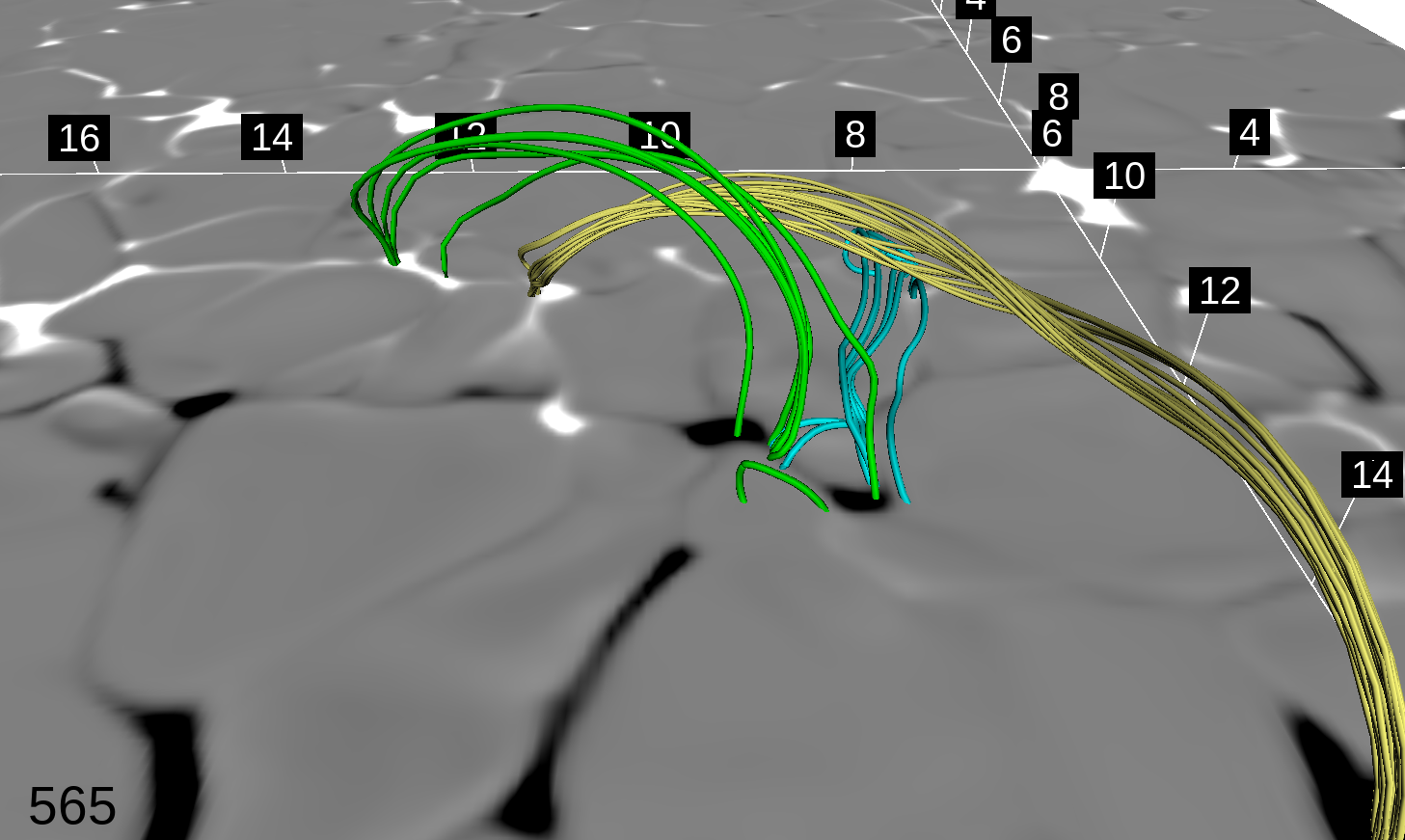}}
    \hfill%
    \subfloat{\includegraphics[width=\sza\textwidth,trim= 60mm 0cm 0mm 6mm, clip]{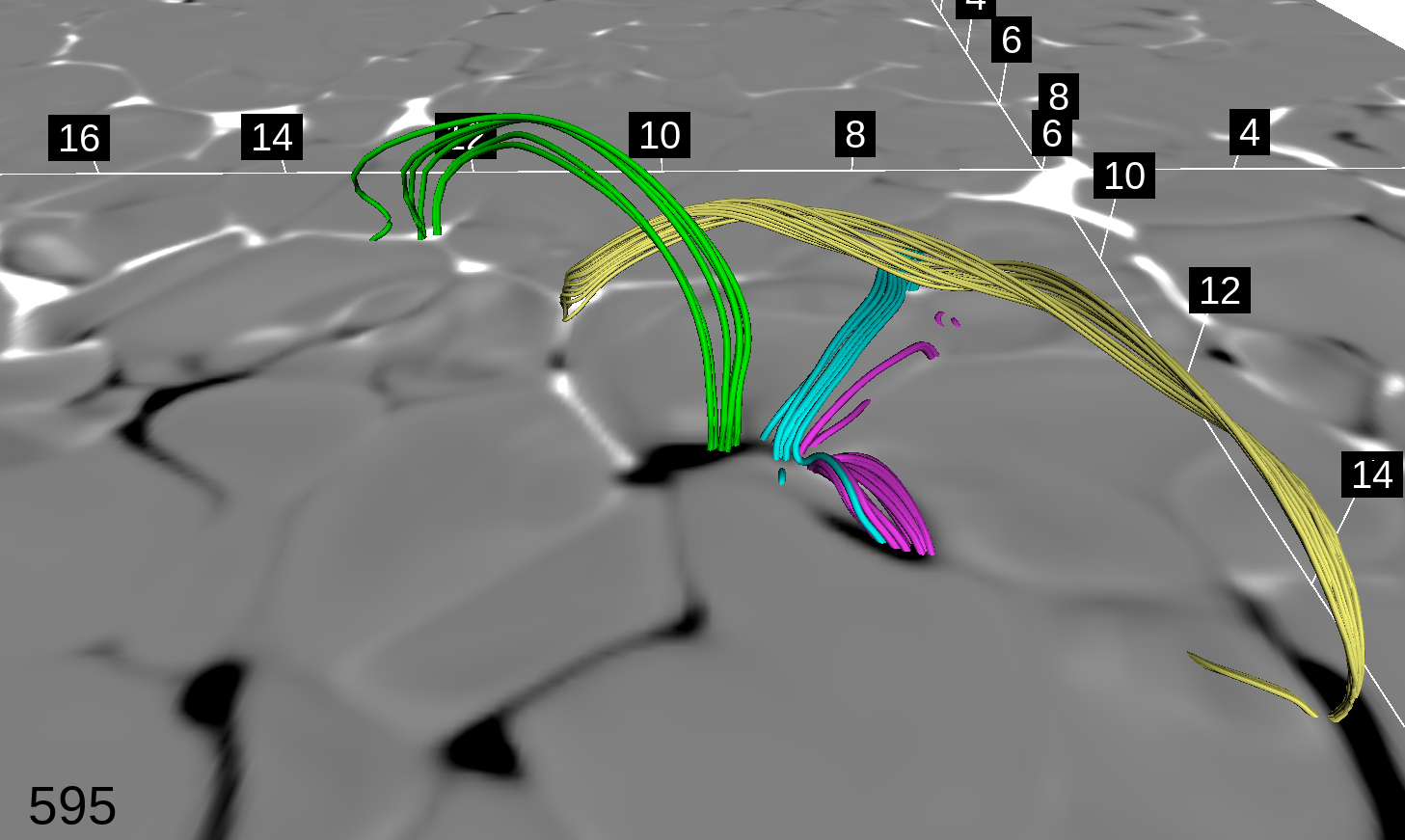}}
    \hfill%
    \subfloat{\includegraphics[width=\sza\textwidth,trim= 60mm 0cm 0mm 6mm, clip]{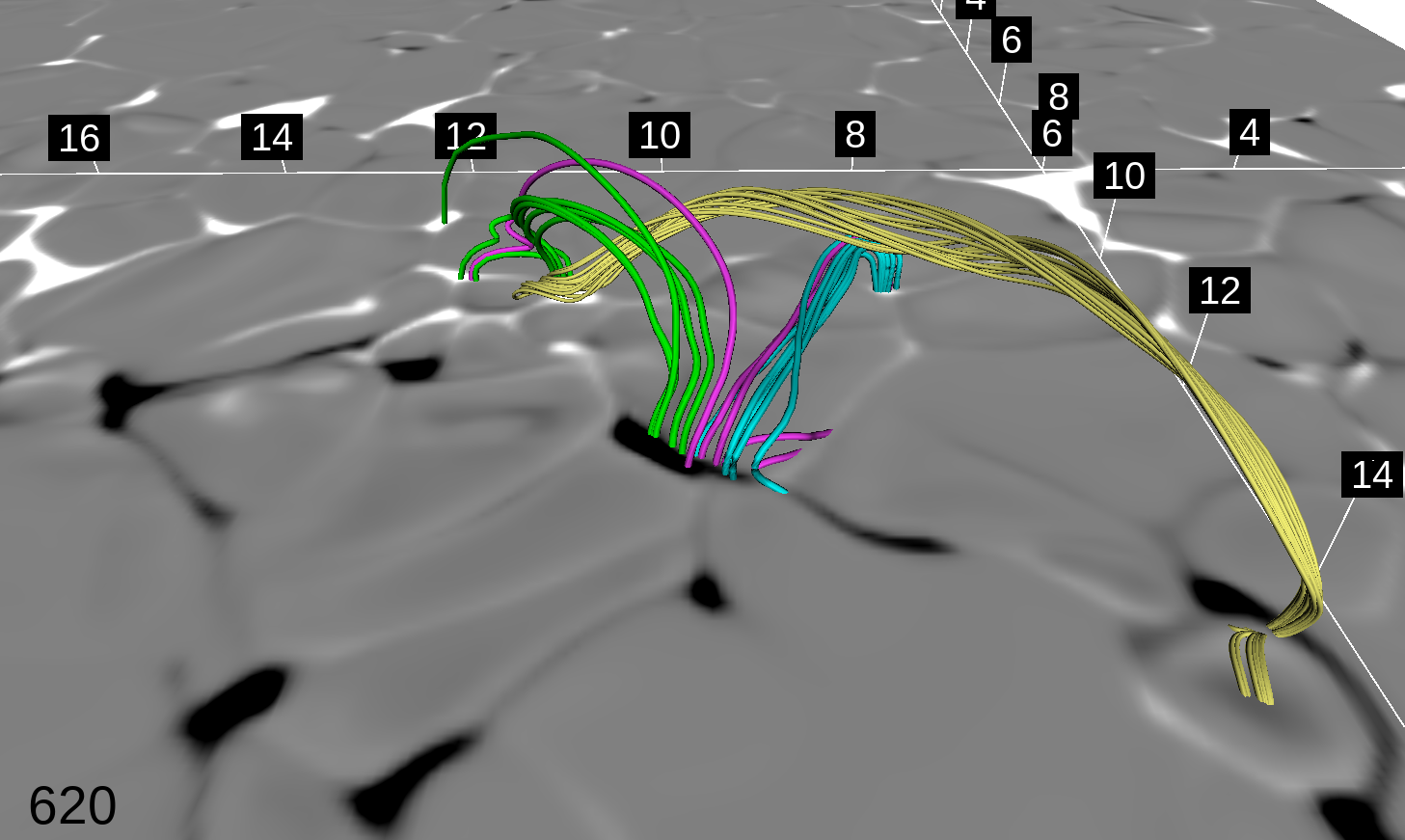}}
    \vspace{-6pt}\\

    \subfloat{\includegraphics[width=\sza\textwidth,trim= 60mm 0 0mm 6mm, clip]{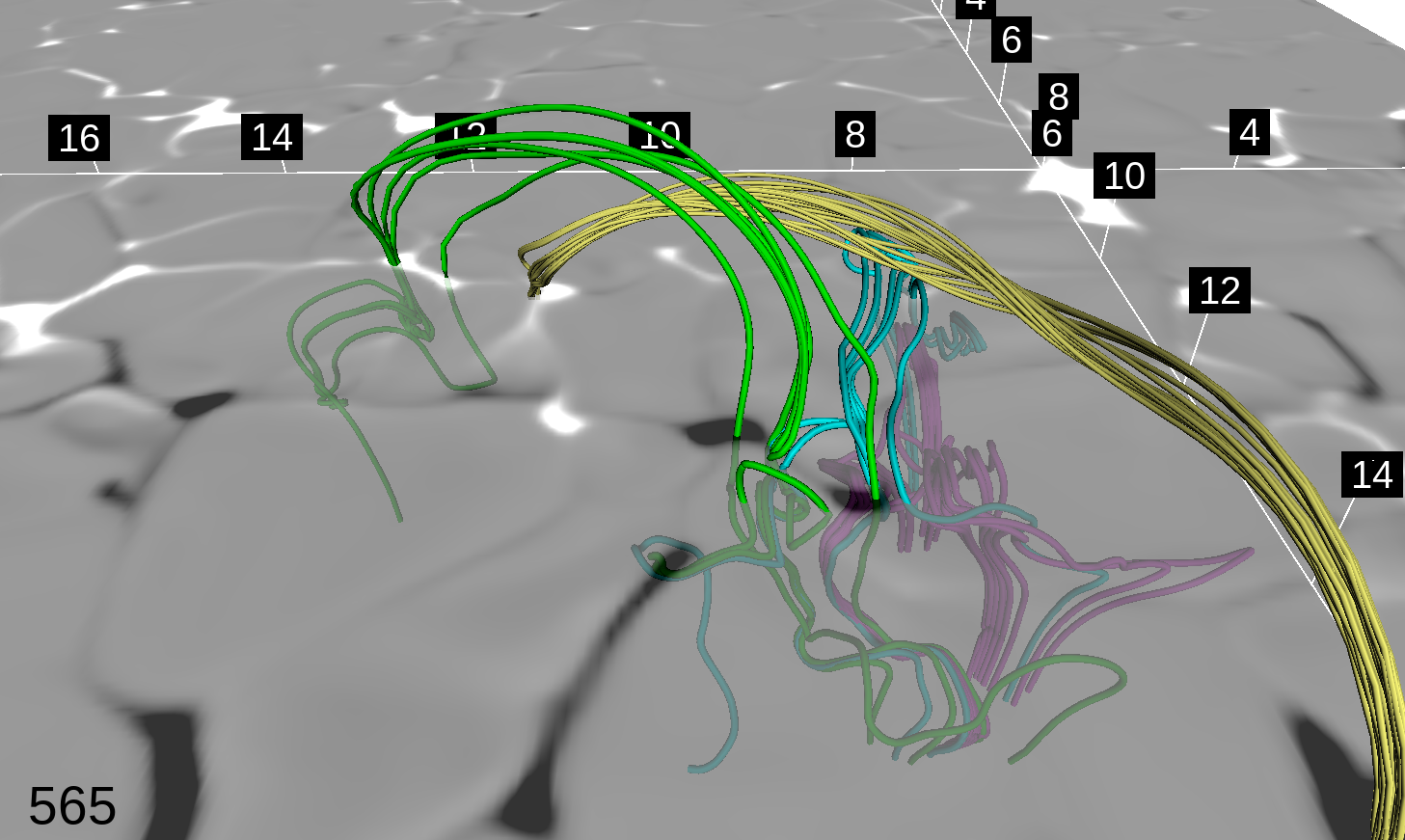}}
    \hfill%
    \subfloat{\includegraphics[width=\sza\textwidth,trim= 60mm 0 0mm 6mm, clip]{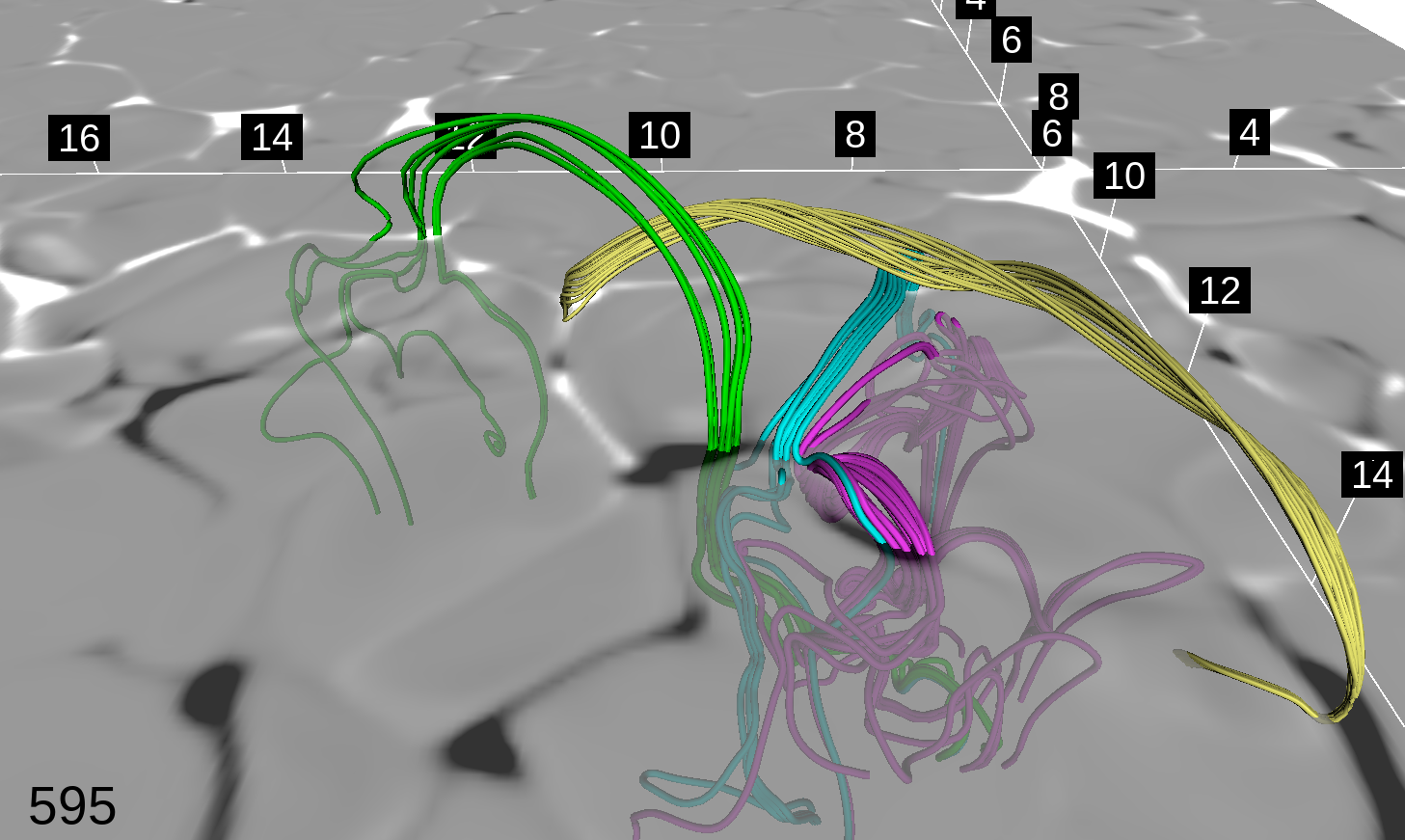}}
    \hfill%
    \subfloat{\includegraphics[width=\sza\textwidth,trim= 60mm 0 0mm 6mm, clip]{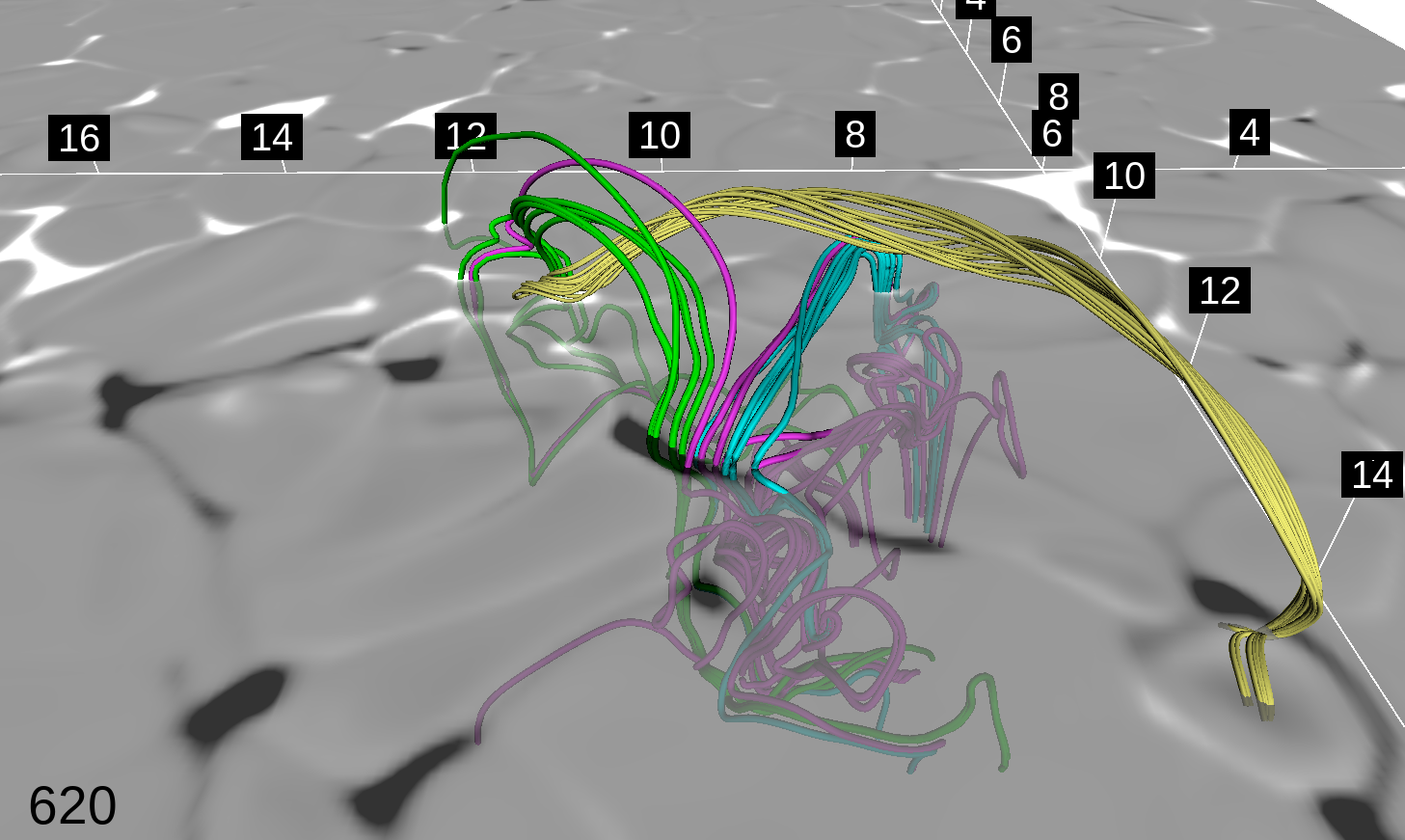}}
    \vspace{-10pt}\\

    \setcounter{subfigure}{0}
    
    \subfloat[Snapshot 1565, $t=t_c-\SI{300}{\second}$]{\includegraphics[width=\szb\textwidth]{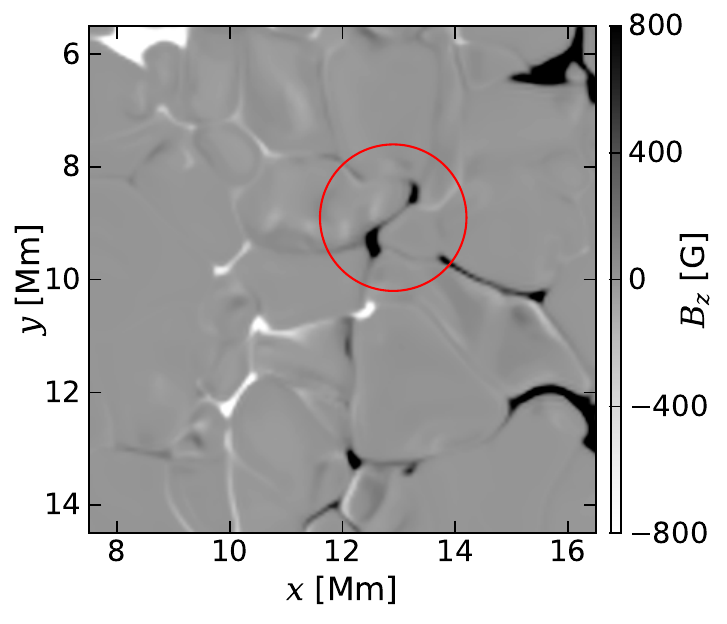}\label{fig_fr1c_1}}\hfill%
    \subfloat[Snapshot 1595, $t=t_c=\text{265m50s}$]{\includegraphics[width=\szb\textwidth]{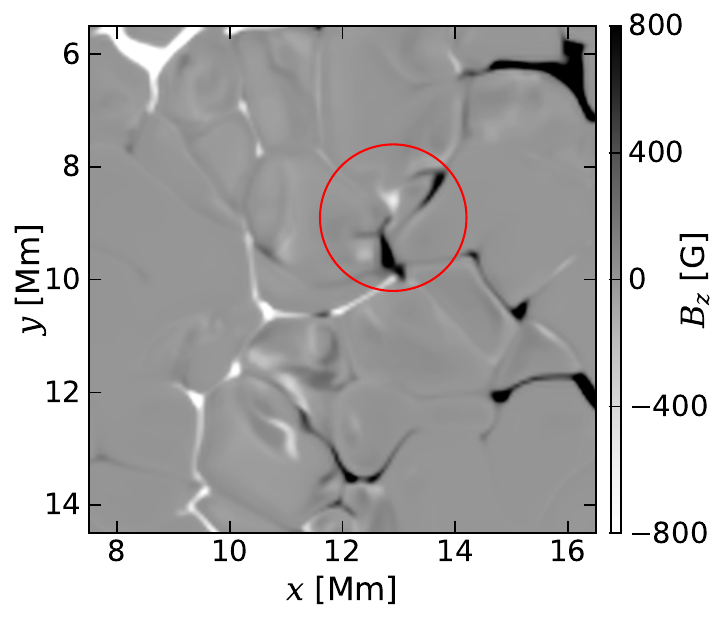}\label{fig_fr1c_2}}\hfill%
    \subfloat[Snapshot 1620, $t=t_c+\SI{250}{\second}$]{\includegraphics[width=\szb\textwidth]{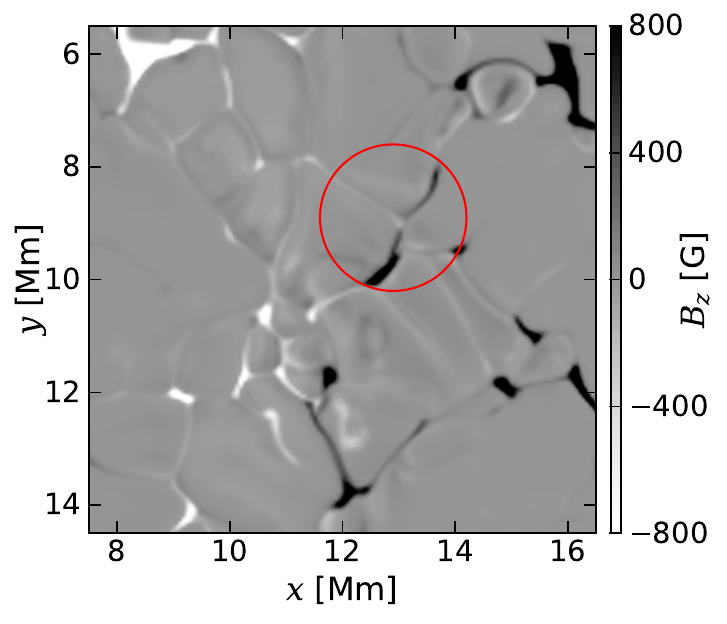}\label{fig_fr1c_3}}\\
    \caption{
        $\Omega$-loop submergence in the \bifrost{} simulation. 
        From left to right, there are three different times relative to $t_c$, a representative time for the submergence: (a) ${\SI{300}{\second}}$ earlier, (b) $t_c$, (c) ${\SI{250}{\second}}$ later, after the end of the flux cancellation.
        The top panel of each subfigure shows three families of 3D field lines of interest in different colors (cyan, green, magenta), tracked by corks, and pale yellow field lines representing the core of the flux rope, above a 2D magnetogram. The middle panel shows the same, but with a partially transparent photospheric magnetogram, to reveal the field lines below the surface. The bottom panel shows only the 2D magnetogram, where the cancellation of interest occurs within the red circle. We stress that the two upper rows are viewed from the right, toward lower $x$, to better visualize the dynamics of interest. 
    }
    \label{fig_fr1c}
\end{figure*}

\begin{figure*}
    \centering 
    \renewcommand{\sza}{.17}
    \renewcommand{\szH}{0.20}
    \setcounter{subfigure}{0}
    %
    \raisebox{.16\height}{
    \subfloat[Snapshot 1750, ${t=\text{291m40s}}$]{%
        \begin{overpic}[height=\sza\textheight,trim= 2.5cm 2cm 15mm 0, clip]{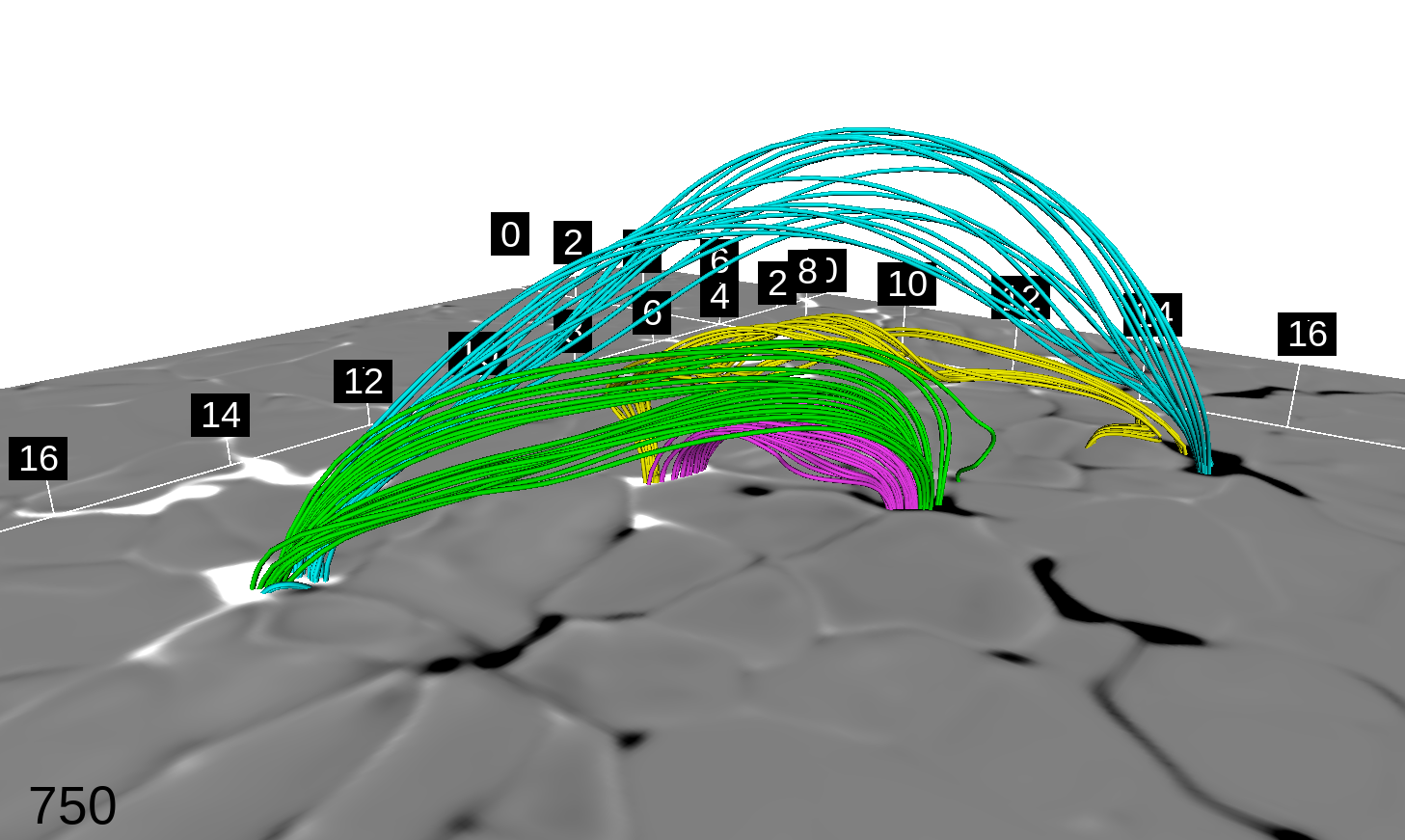}%
        \put (46,18) {\normalsize\textcolor{white}{\textbf{W1}}}%
        \put (14,8.5)  {\normalsize\textcolor{white}{\textbf{W2}}}%
        \put (86,18) {\normalsize\textbf{B1}}%
        \put (62.4,15) {\normalsize\textbf{B2}}%
        \end{overpic}\label{fig_fr2_1}}}%
    \hfill\!%
    \subfloat{%
        \begin{overpic}[height=\szH\textheight,trim= 6.9cm 0 7.8cm 0, clip]{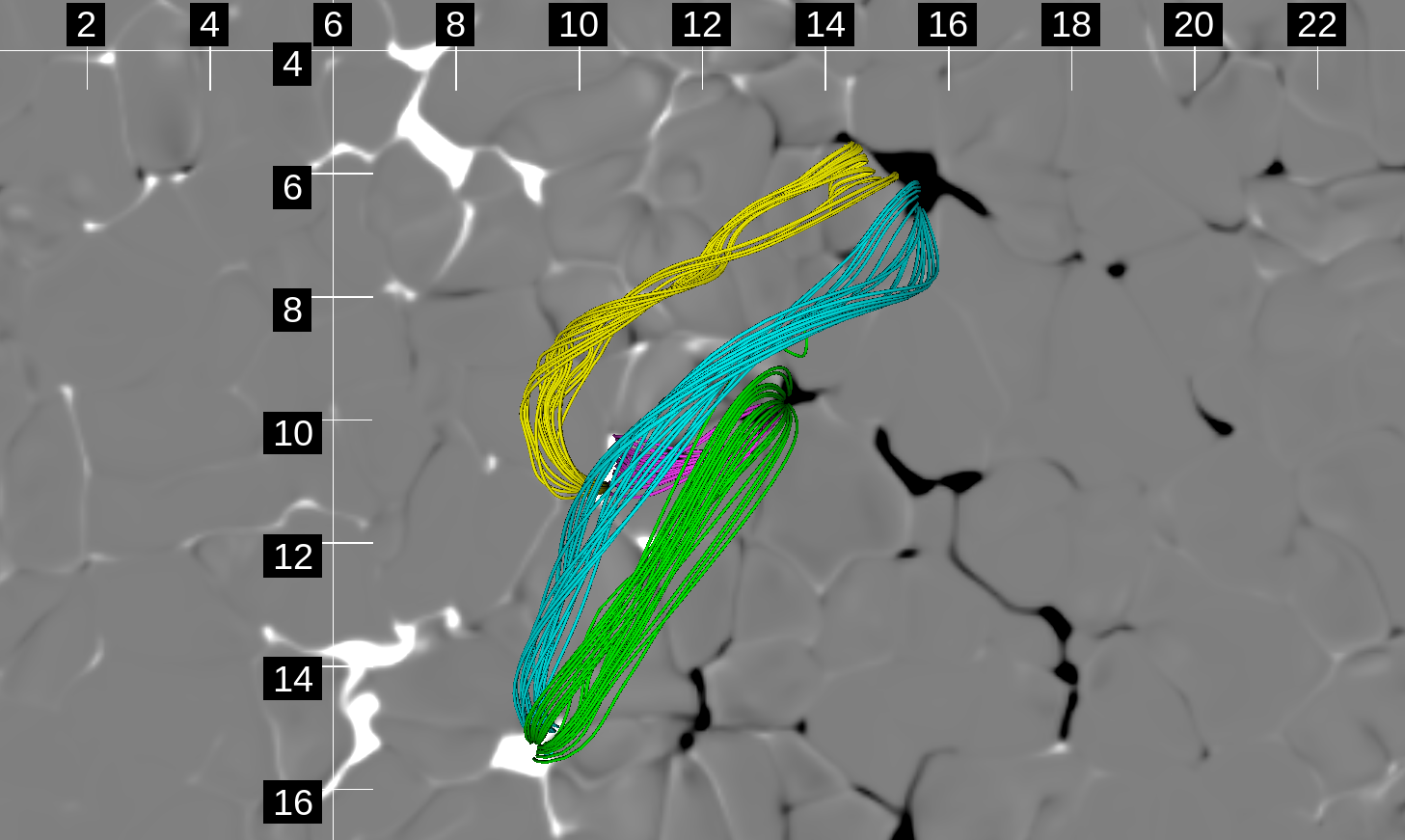}%
        \put (24,34) {\normalsize\textcolor{white}{\textbf{W1}}}%
        \put (22,2) {\normalsize\textcolor{white}{\textbf{W2}}}%
        \put (68,52) {\normalsize\textbf{B2}}%
        \put (73,82) {\normalsize\textbf{B1}}%
        \end{overpic}}%
    \hfill\!%
    \subfloat{\includegraphics[height=\szH\textheight,trim= 0 2mm 0 2mm, clip]{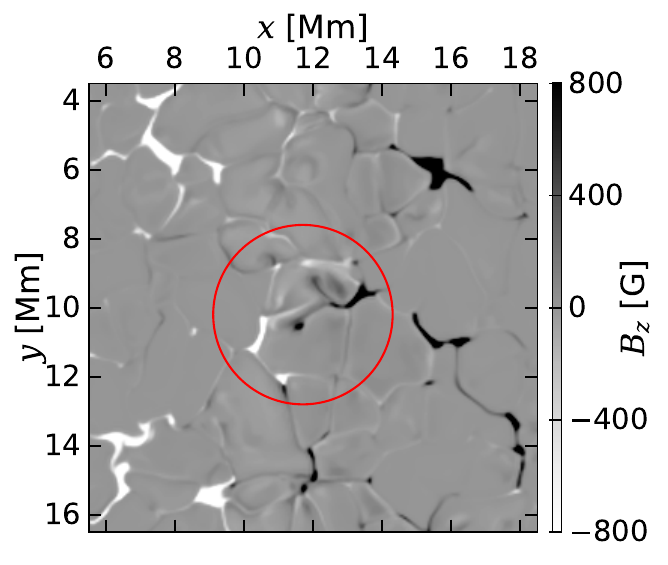}}\\

    \setcounter{subfigure}{1}
    \subfloat[Snapshot 1850, ${t=\text{308m20s}}$]{\includegraphics[height=\sza\textheight,trim= 2.5cm 2cm 15mm 0, clip, valign=t]{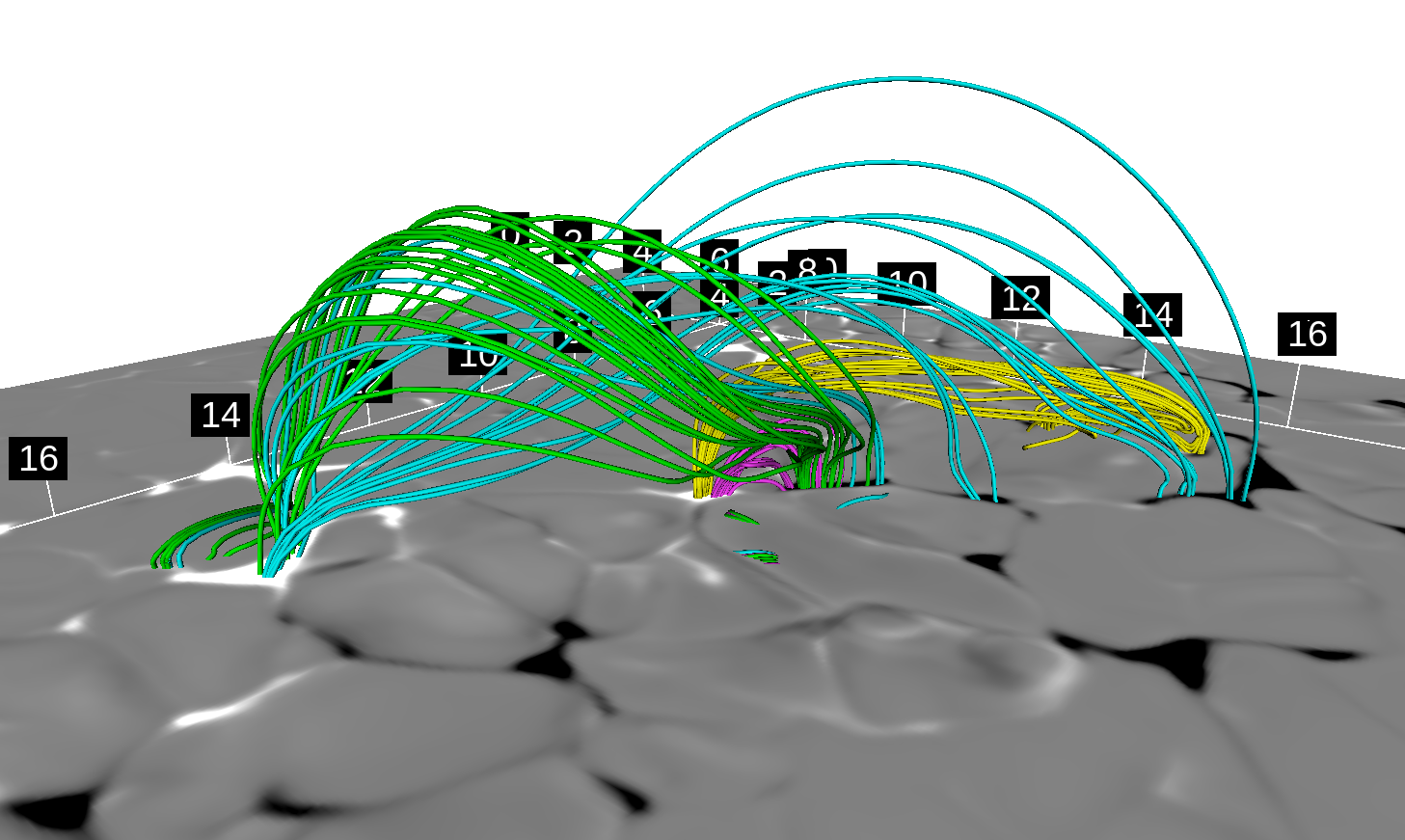}\label{fig_fr2_2}}%
    \hfill\!%
    \subfloat{\includegraphics[height=\szH\textheight,trim= 6.9cm 0 7.8cm 0, clip, valign=t]{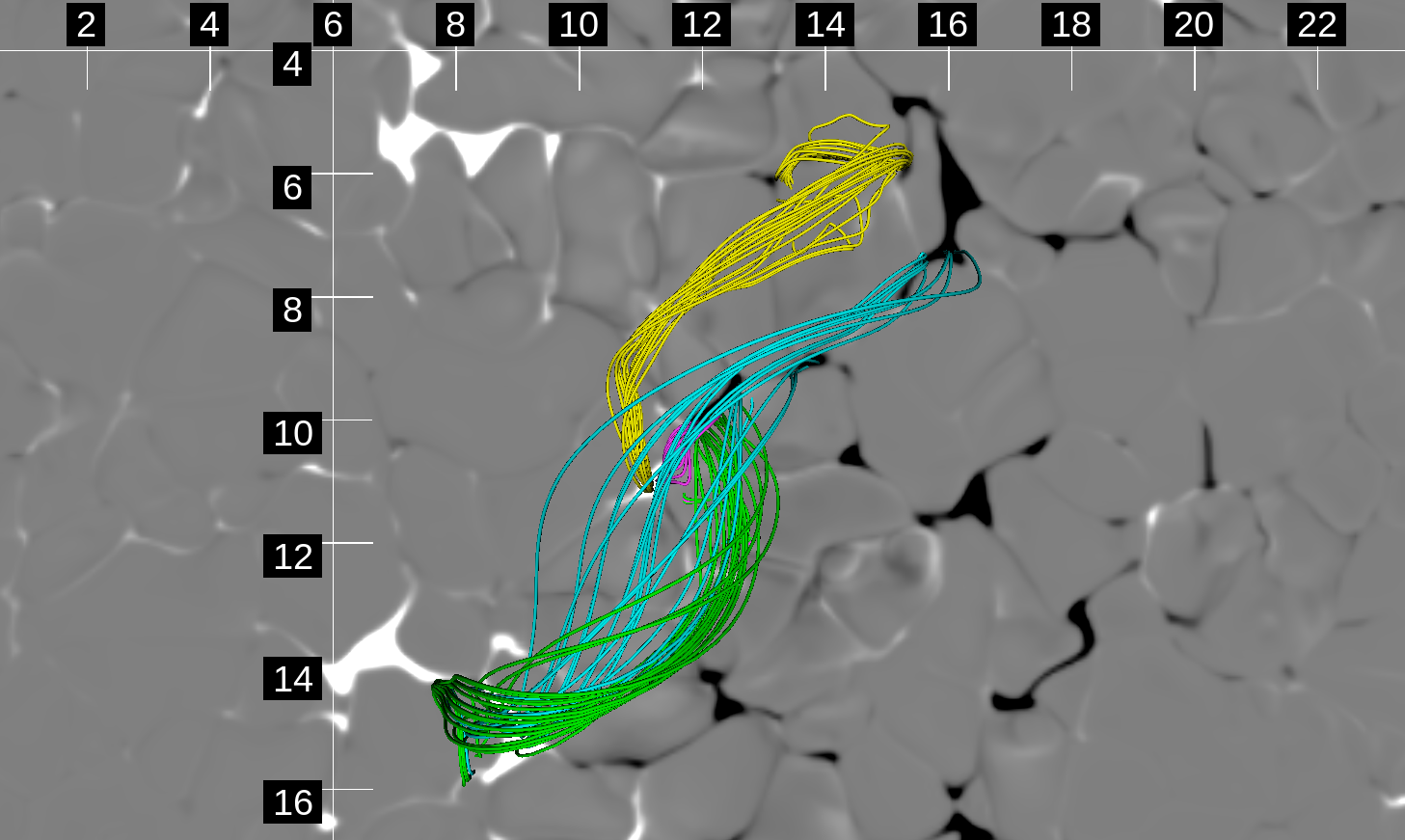}}
    \hfill\!%
    \subfloat{\includegraphics[height=\szH\textheight,trim= 0 2mm 0 2mm, clip,valign=t]{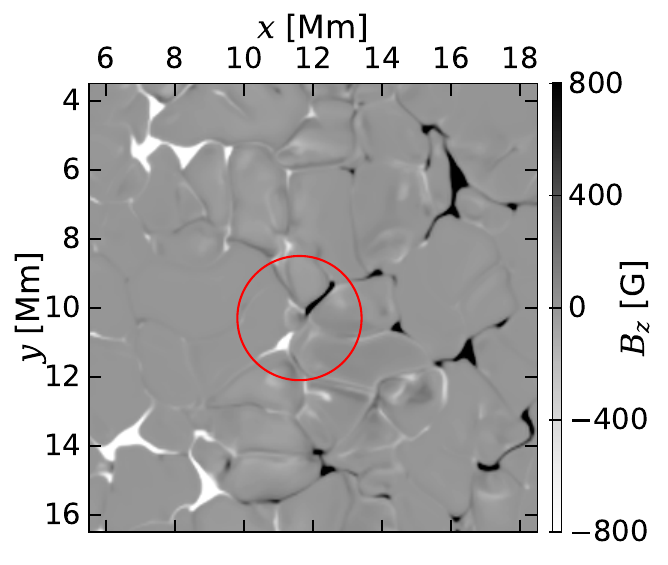}}\\

    \setcounter{subfigure}{2}
    \subfloat[Snapshot 1950, ${t=\text{325m0s}}$]{\includegraphics[height=\sza\textheight,trim= 2.5cm 2cm 15mm 0, clip, valign=t]{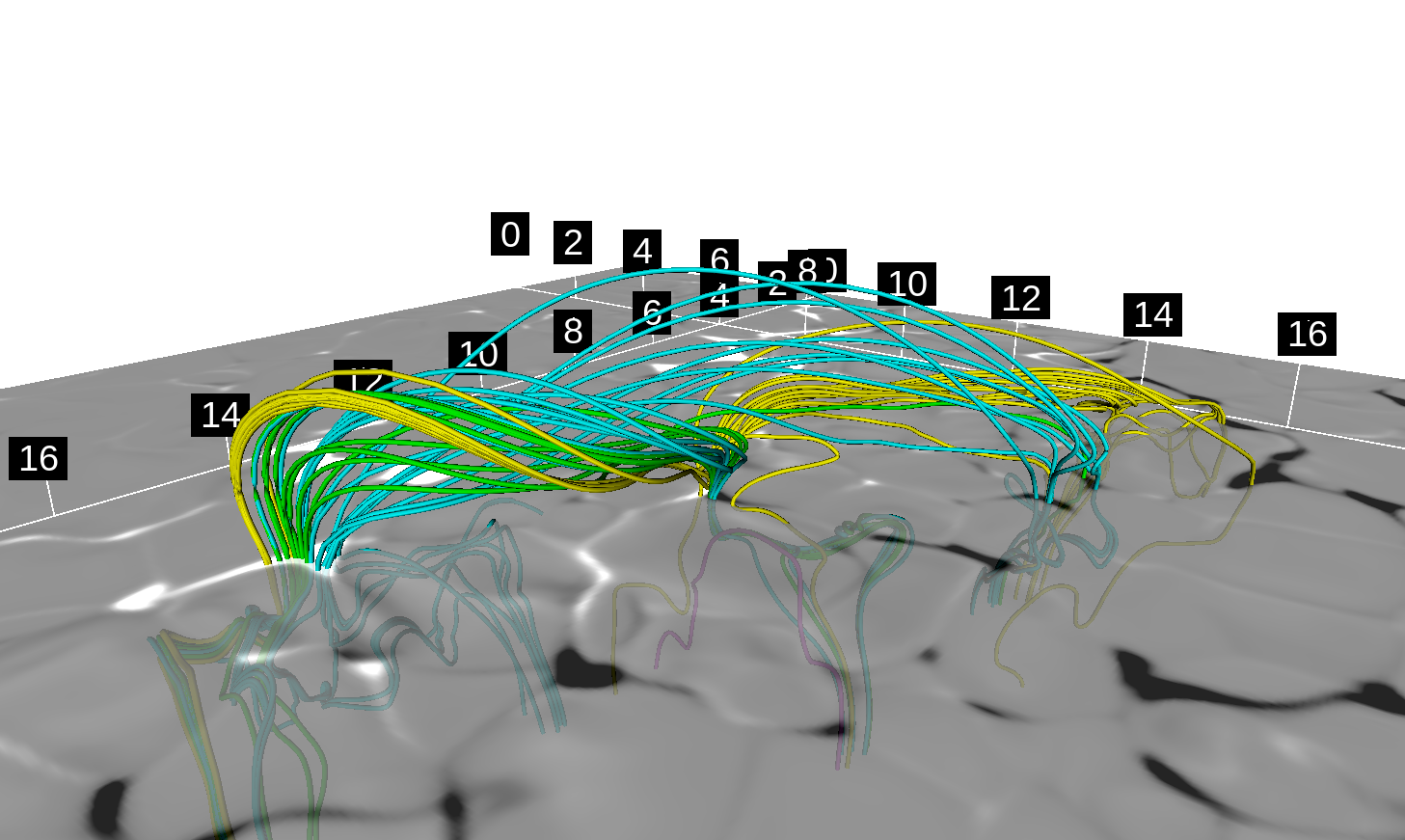}\label{fig_fr2_3}}%
    \hfill\!%
    \subfloat{\includegraphics[height=\szH\textheight,trim= 6.9cm 0 7.8cm 0, clip, valign=t]{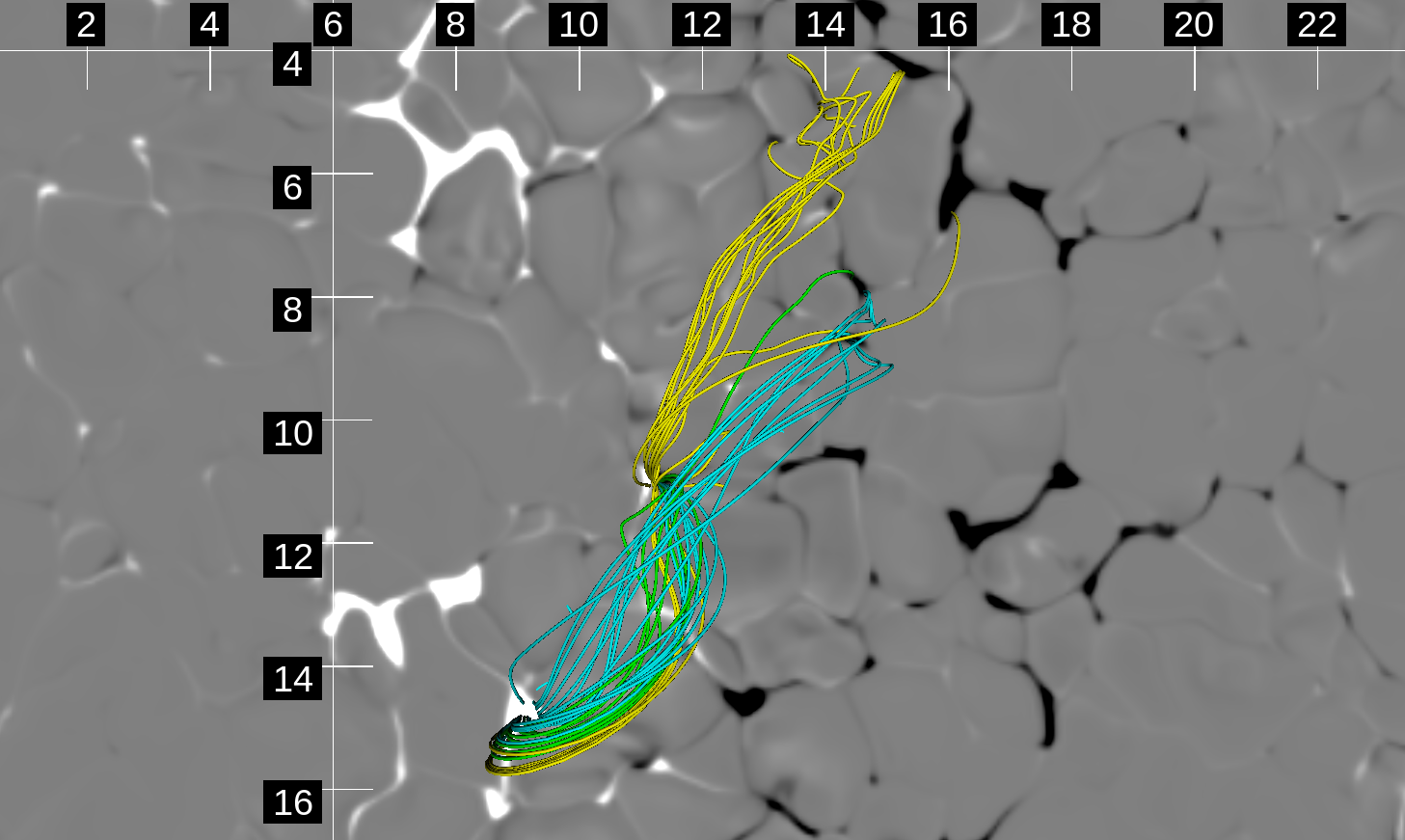}}
    \hfill\!%
    \subfloat{\includegraphics[height=\szH\textheight,trim= 0 2mm 0 2mm, clip,valign=t]{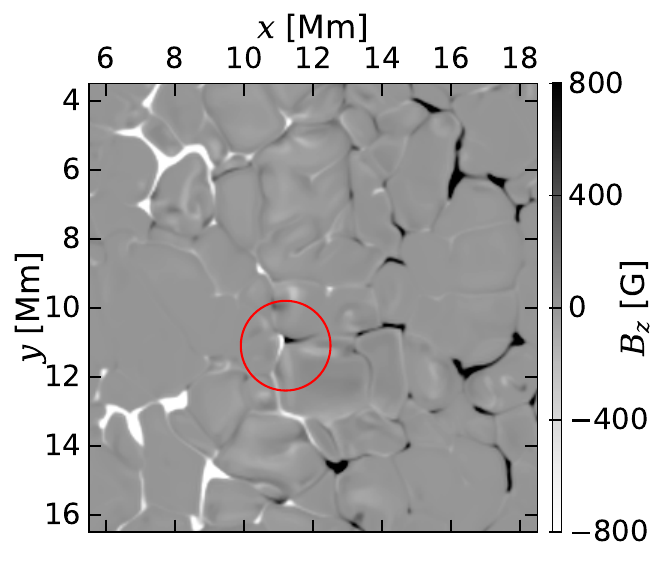}}\\

    \setcounter{subfigure}{3}
    \subfloat[Snapshot 2000, ${t=\text{333m20s}}$]{\includegraphics[height=\sza\textheight,trim= 2.5cm 2cm 15mm 0, clip, valign=t]{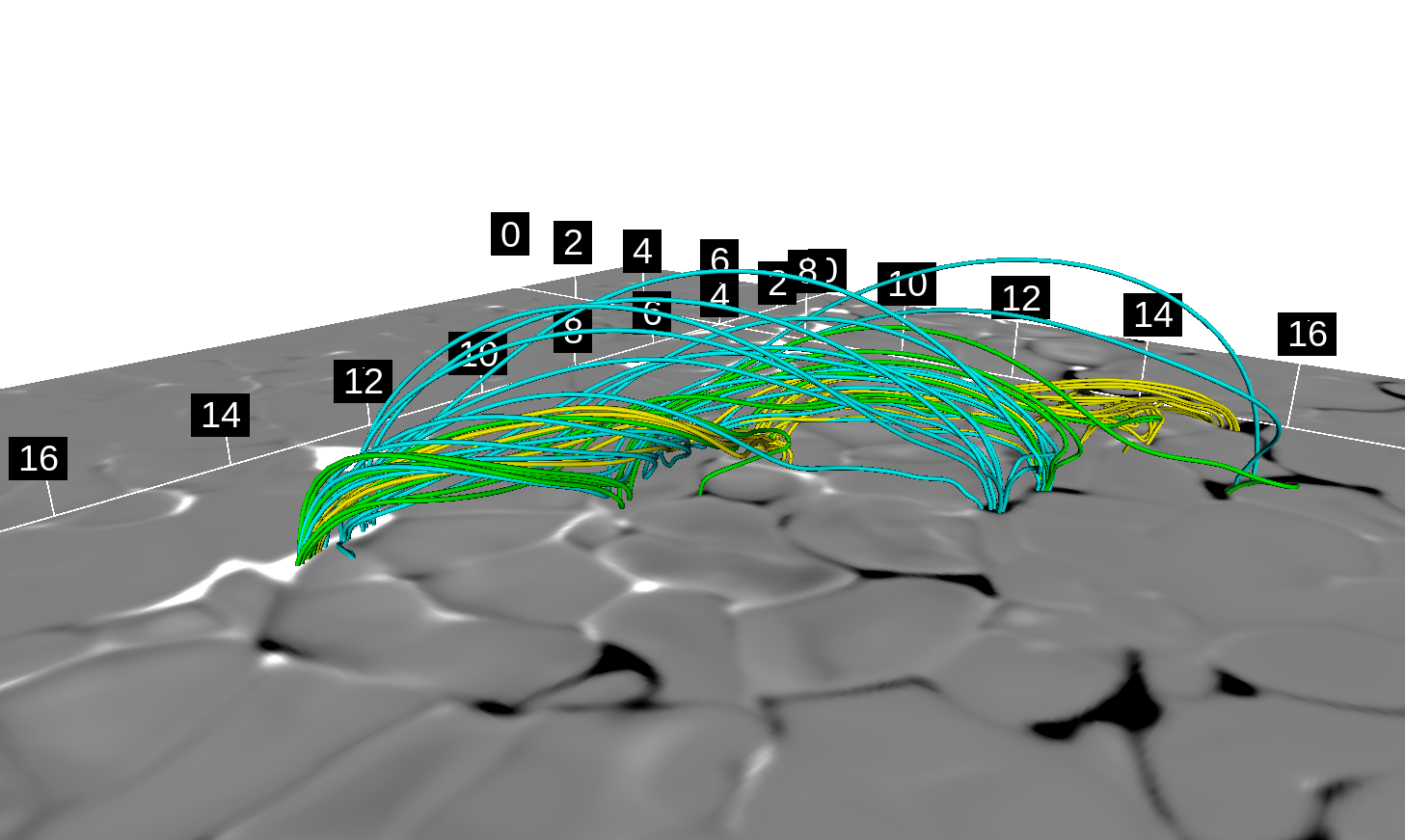}\label{fig_fr2_4}}%
    \hfill\!%
    \subfloat{\includegraphics[height=\szH\textheight,trim= 6.9cm 0 7.8cm 0, clip, valign=t]{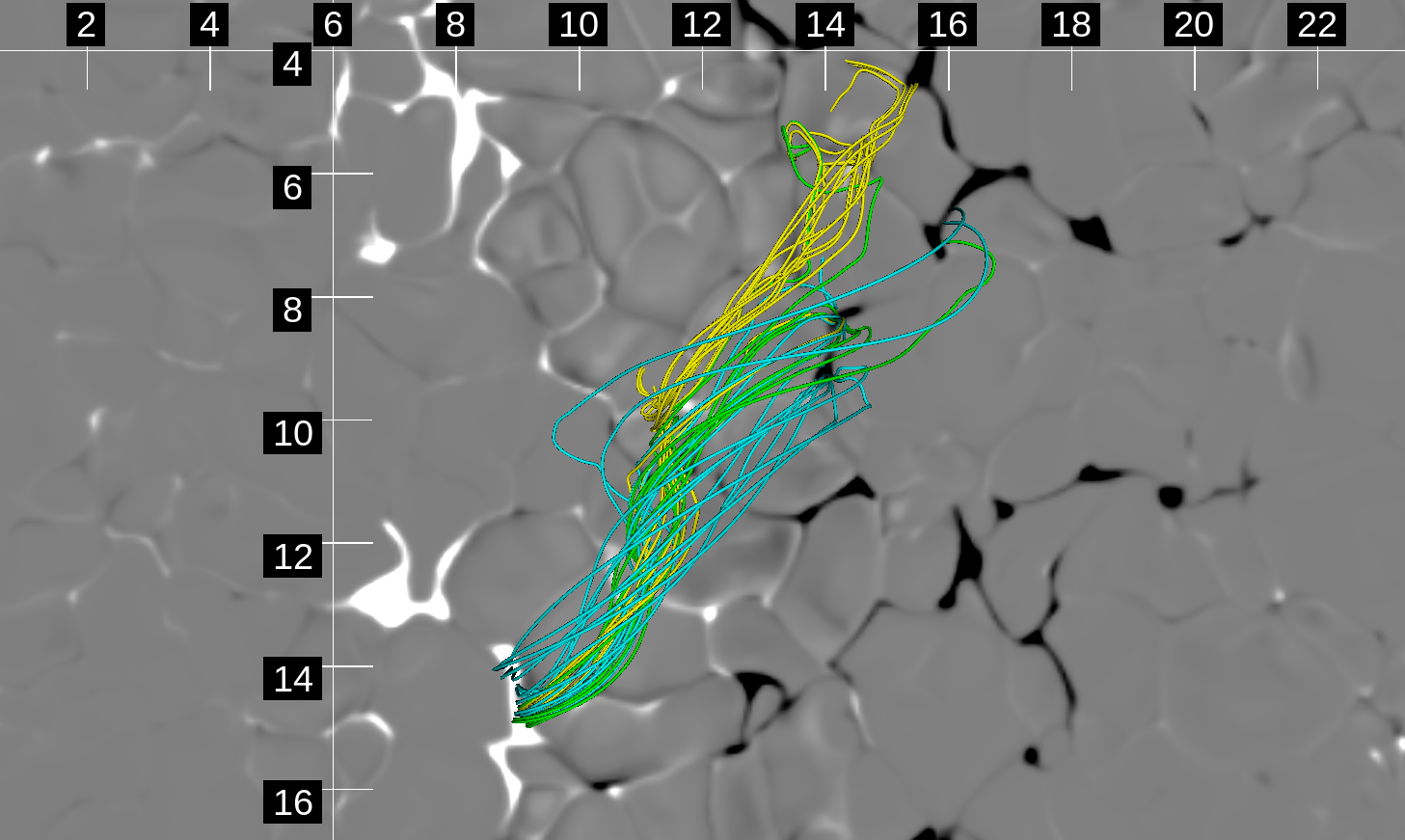}}
    \hfill\!%
    \subfloat{\includegraphics[height=\szH\textheight,trim= 0 2mm 0 2mm, clip,valign=t]{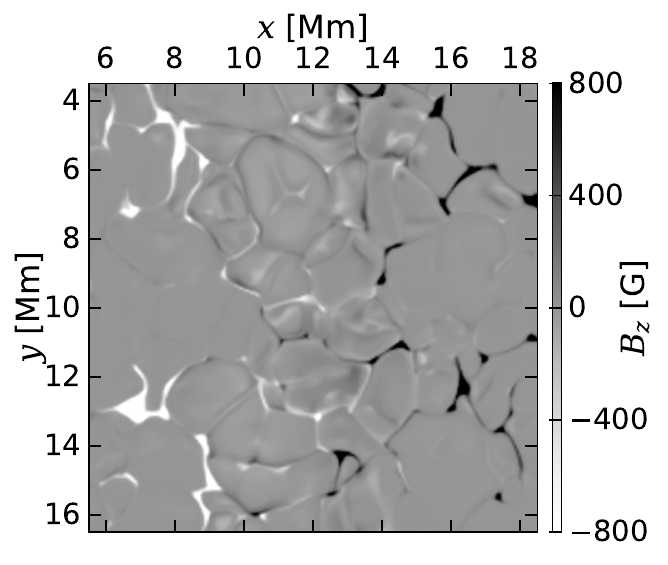}}\\
    
    \caption{
        TPTC reconnection in the \bifrost{} simulation. In this figure, time advances downwards, unlike in the previous figures, while all three panels per subfigure row show different views of the same snapshot.
        The left panels show four families of 3D field lines of interest in different colors (cyan, green, magenta, yellow), tracked by corks, above a 2D magnetogram. The magnetogram in the left panel in (c) is transparent by choice to show the magenta field line below the surface.
        The middle panels show the same, but directly from above. 
        The right panels show only the 2D magnetogram, directly from above, where the cancellation of interest occurs within the red circle. There is no red circle in (d) as the cancellation then has completed.
        }
    \label{fig_fr2}
\end{figure*}

\subsection{Slipping reconnection}
\label{sec_res_fr1a}

Possible examples of slipping reconnection are shown in Fig.~\ref{fig_fr1a}. 
The colored field lines were selected at the time shown in Fig.~\ref{fig_fr1a_5}, as it was close to the end of the slipping process we were observing. That made it ideal for finding field lines that were or had been slipping, whereupon we studied them backwards in time to see how they arrived at such a configuration. The chosen field lines had large values of the expression in Eq.~\eqref{eq_joverb}, i.e. they were related to strong and narrow current layers, and their footpoints were in the canceling white polarity within the red circle. We chose three separate groups of field lines that passed through the cross section of the resulting flux rope in physically separated areas, to see how they wrapped around each other. The field lines were tracked with corks chosen between the middle of the field lines and their footpoints in the black polarity at this time.

All the field lines remain rooted in the black pole (B1) at approximately ${(x,y)=(\SI{15}{\mega\meter},\SI{6}{\mega\meter})}$ throughout this process and beyond. The footpoints in the white polarity, however, are gradually slipping to the same white pole (W1) at ${(x,y)=(\SI{12.1}{\mega\meter},\SI{10.6}{\mega\meter})}$ within the red circle in Fig.~\ref{fig_fr1a_6}. First the green field lines slip, then the magenta, then the cyan. Each family spend ${6\!-\!\SI{10}{\minute}}$ on this slipping, in which the photospheric footpoints in the white polarity move approximately $\SI{4}{\mega\meter}$.
Hence, the average field line footpoint slipping velocities are ${7\!-\!11~\si{\kilo\meter/\second}}$. The maximum slipping velocity observed in this Fig., which has $\SI{2}{\min}$ cadence or more between the panels, is approximately $\SI{25}{\kilo\meter/\second}$. The peak velocities would be higher. This matches slipping velocities observed during a solar flare, ranging from $\SI{4.5}{\kilo\meter/\second}$ to $\SI{136}{\kilo\meter/\second}$~\citep{dudik2014_slipping}. The slipping of these footpoints stops after the last shown snapshot. Later, these field lines remain twisted around each other, in approximate this position, as a flux rope that later becomes more twisted.

During the slipping reconnection, there is convergence and eventually cancellation between the black pole and W1 within the red circle in Fig.~\ref{fig_fr1a_1}. This convergence can be forcing the slipping reconnection through a quasi-separatrix layer that resides higher up in the corona.
Since the reconnection does not occur in the photospheric layer, this process is different from the process proposed by~\citet{van_ballegooijen_formation_1989}. Nevertheless, it still contributes to the generation and twisting of the flux rope.

\subsection{U-loop emergence}
\label{sec_res_fr1b}

An example of U-loop emergence is shown in Fig.~\ref{fig_fr1b}. This is detectable as flux cancellation in the magnetogram.
The green field lines were selected to go down into the intermediate black pole. They were tracked by corks selected along the longest visible part of the field lines pre-emergence, 50~s before ${t_b}$, close to the intermediate white pole. The pale yellow field lines represent approximately the core of the flux rope. 

The entire area of interest within the red circle is convected to the left, toward the white polarity footpoints of the flux rope. The black polarity cancels mostly against the white polarity to the left of it in Fig.~\ref{fig_fr1b_1}. As the U-loop starts to emerge in Fig.~\ref{fig_fr1b_2}, and the field lines to a lesser extent are weighed down by the denser plasma, the field lines curl around the flux rope in Fig.~\ref{fig_fr1b_3}, increasing its overall twisted flux. That this U-loop emergence led to increased twist was dependent on the field line connectivity and orientation relative to the flux rope.

Judging from the magnetogram alone, this was a candidate for reconnection \`{a} la~\citet{van_ballegooijen_formation_1989}. That could have been the case if the long green field lines ending in B1 instead would have been emerging from the white polarity to the left of the black pole in focus, instead of the white polarity to the right of it. It was necessary to visualize the 3D field lines and track them with corks to deduce what happened in this process.
Even though it is not a reconnection event, this flux emergence is still a flux cancellation event that adds twist to the flux rope.

\subsection{$\Omega$-loop submergence independent of the flux rope}
\label{sec_res_fr1c}

An example of $\Omega$-loop submergence is shown in Fig.~\ref{fig_fr1c}. This is detectable as flux cancellation in the magnetogram. The three sets of colored field lines were tracked by corks selected at time $t_c$, corresponding to Fig.~\ref{fig_fr1c_2}. The green and cyan field lines were selected to go down through different areas of their shared black pole. The corresponding corks were selected closer to their footpoints in this black pole, about 20\% to 40\% along the part of the field lines visible above the photosphere. The magenta field lines were selected to come up through the white pole that later cancels. The corresponding corks were selected close to the middle of the part of the field lines visible above the photosphere. A few of the cyan field lines follow the magenta lines into the $\Omega$-loop at $t_c$. Similarly, beyond the $\Omega$-loop (meaning at lower $x$), most of the magenta field lines, a few above and many below the photosphere, go in the same direction as the cyan field lines toward their footpoints in the white polarity.

Most of the area of interest, within the red circle, is convected horizontally toward larger $y$ (downwards in the 2D magnetogram). The white pole within the red circle in Fig.~\ref{fig_fr1c_2} cancels evenly against the black pole at larger $y$ and the one at larger $x$, before reaching the state in Fig.~\ref{fig_fr1c_3}. The magenta field lines are convected downwards along $z$, into the convection zone, between these two snapshots. The previously photospheric $\Omega$-loop is submerged in Fig.~\ref{fig_fr1c_3}. This submerging and related flux cancellation is not linked to reconnection for most of the magenta field lines. This can be seen by how the magenta field lines are connected beyond the, now, submerged $\Omega$-loop (meaning at lower $x$). Most of the magenta field lines are still going in the same direction as in Fig.~\ref{fig_fr1c_2}, meaning along the cyan ones that are visible above the photosphere toward their footpoints in the white polarity. 
A single magenta field line follows the green field lines above the photosphere. This reconnection happens below the photosphere, in the high plasma-$\beta$ regime. In other words, it is not the kind of coronal reconnection leading to a coronal flux rope that we are looking for. This cancellation does not contribute to the twist of the flux rope.

Judging from the magnetogram alone, this cancellation between the footpoints of the flux rope was also a candidate for reconnection \`{a} la~\citet{van_ballegooijen_formation_1989}. However, like the event in Sect.~\ref{sec_res_fr1b}, when visualizing the field lines in 3D and tracking them with corks, it was clearly not the case. Since the green and cyan field lines both enter the black pole in the same direction, they cannot reconnect to make a longer field line twisted around the pale yellow flux rope. We note not this event partially overlaps in time with the U-loop emergence discussed in Sect.~\ref{sec_res_fr1b}, without them seeming to strongly affect each other.

The cancellation discussed in this section is an $\Omega$-loop submergence, but it happens just after an $\Omega$-loop emergence. The white pole is not visible in Fig.~\ref{fig_fr1c_1}, because the $\Omega$-loop is still below the photosphere. The loop is convected upwards, through the photosphere, until $t_c$, before it is going down again. Such a peak-a-boo behavior of flux appearance and cancellation is an example that the cancellation is not always related to reconnection and twisting of the flux rope. 
In comparison to other types of simulations described in the introduction, this mechanism could not have happened naturally on its own, meaning without prescription, without modeling the convection zone.

\subsection{Thick-photosphere tether-cutting (TPTC)}
\label{sec_res_tptc}

An example of thick-photosphere tether-cutting (TPTC) reconnection is shown in Fig.~\ref{fig_fr2}. It is detectable as flux cancellation in the magnetogram. There is no clear PIL between the white and black polarities, a trait it has in common with cancellation events often observed. That is also part of what separates this simulation from the other cancellation simulations that were mentioned in the introduction. Furthermore, the cancellation is more extended in both spatial and temporal dimensions than the events previously discussed, making an equivalent event on the Sun more easily detectable.
The overall geometry of the field lines involved is similar to that in the tether-cutting model for solar eruptions~\citep{Moore_2001}. However, the altitude of the reconnection site is located neither in the corona as during a flare, nor in the transition region as in models of eruptive flux ropes formed by flux emergence ~\citep{Manchester_2004, archontis_flux_2008}, but instead the reconnection occurs within the photosphere, as in the original flux-cancellation model by~\citet{van_ballegooijen_formation_1989}.

All four families of colored field lines, displayed in Fig.~\ref{fig_fr2}, were chosen to have large values of the expression in Eq.~\eqref{eq_joverb} at ${t=\text{291m40s}}$, corresponding to Fig.~\ref{fig_fr2_1}.
The yellow field lines correspond to the flux rope we have been presenting the gradual twisting of so far. Its footpoints are W1 and B1, and therefore we denote it W1B1. The footpoints are marked in Fig.~\ref{fig_fr2_1}. The corks were selected between W1 and the middle of the field lines. As mentioned in Sect.~\ref{sec_res_full}, here we show how this flux rope W1B1 reconnects with W2B2, the green untwisted field lines in Fig.~\ref{fig_fr2_1}. The corresponding corks were selected closer to B2 than W2. The corks of the magenta field lines, W1B2, were selected in the middle between the two footpoints. The corks of the cyan field lines, W2B1, were selected approximately above B2.

The photospheric area of interest, within the gradually shrinking red circles, display significant flux cancellation over a much more extended time than the events previously discussed. The first and last snapshots displayed are separated by over $\SI{40}{\minute}$. Within this time, W1 and B2 gradually approach each other through Fig.~\ref{fig_fr2_2} and Fig.~\ref{fig_fr2_3}, before having canceled in Fig.~\ref{fig_fr2_4}. 
However, this non-line-tied simulation highlights that this flux cancellation is associated with a TPTC reconnection, including an X-crossing. The reconnection does not take place exactly in the ${z=0}$ plane where we visualize the flux cancellation, but over an extended volume within the photosphere, mainly within ${z\in[0,-1]~\si{\mega\meter}}$, which we call the thick photosphere. There, the plasma~${\beta=P_p/P_B}$ is still high, but decreasing, and the temperature is close to aa minimum, as seen from Fig.~\ref{fig_ramp_1d}.

Following the flux cancellation, W1B2, the magenta $\Omega$-loop, is submerged and transported by solar convection down and out through the bottom of the simulation box, which is at $\SI{2.5}{\mega\meter}$ below the photosphere.
In Fig.~\ref{fig_fr2_3}, there is only one cork still within the simulation box, hence the single magenta field line below the photosphere. As W1B2 (magenta) is submerged, W1B1 (yellow) and W2B2 (green) reconnect in such a way that the flux rope becomes longer, stretching from W2 to B1. In Fig.~\ref{fig_fr2_3}, the green and yellow field lines still provide the generated flux rope with feet in W1 and B2. In Fig.~\ref{fig_fr2_4}, after the complete cancellation of these poles, the feet are gone and the green and yellow field lines are twisted around each other, with a maximum photospheric footpoint separation of $\SI{12}{\mega\meter}$. The flux rope extends about ${\SI{2}{\mega\meter}}$ into the solar atmosphere. The black pole B1, which have been quite steady so far, now starts to break up.

As for the previously discussed events, one could not have concluded from the magnetograms alone that this cancellation was reconnection \`{a} la~\citet{van_ballegooijen_formation_1989}. The 3D field lines were necessary to analyze the snapshots individually, while the corks used to track the field lines over time were necessary to analyze the dynamics. 
This analysis confirmed the picture from \citet{van_ballegooijen_formation_1989} for a realistic treatment of convection, including the occurrence of the submergence of the small reconnected $\Omega$-loop. 
Additionally, our analysis revealed that there is a small but nonzero difference in altitude within the photosphere between where the magnetic fields are measured and where reconnection occurs (i.e. up to ${\SI{1}{\mega\meter}}$). Hence, the traditional picture of reconnection had to be expanded to allow reconnection to start before the cancellation between opposite flux concentrations, since reconnection is actually driven by the converging motions of the polarities toward one another, not the cancellation itself.

\section{Discussion}
\label{sec_disc}

A key motivation for this study was to see if flux ropes actually can form above the photosphere from flux cancellation, as has been shown in line-tied simulations with prescribed driving at the photospheric boundary. Such prescribed driving has the benefit that one can control everything, but one has to know how the Sun works for that control to represent reality. In other words, all ropes that have come out of such simulations could have been artifacts of the prescribed driving. This is just one of the many logical arguments against these simulations. However, everyone does it -- one of the authors has run many -- but it does have its assumptions that have not been treated before. 
A key contribution of the present study is that it shows that flux ropes in fact can be generated also without these assumptions. Hence, the simplifying assumption of line tying is not a prerequisite for qualitatively forming flux ropes in the atmosphere. However, the line tying assumption still has other problems like the decaying signed flux discussed in the introduction.

The present study took on an exploratory approach in simulating the formation of flux ropes through flux cancellation without the assumption of line-tied footpoints at the photosphere. This methodology for forming a flux rope has been proven to work with the following set of parameters for the LFFF: ${n_h=30}$, ${\alphan^2=0.5}$, and ${\max[B_z(z=0)]=\SI{50}{\gauss}}$. Only one set of parameters has been tested because of the numerical cost and complexity of running such a simulation, as expanded upon in App.~\ref{app_Simulation}. The simulation did produce a flux rope as hypothesized, through multiple distinguishable flux cancellation events along a non-smooth PIL, driven by self-consistent convection.

By ramping up the LFFF in the QS simulation, sheared arcades appear in the \bifrost{} simulation. That is similar to the configuration after steps (a) and (b) in the theory by~\citet[see Fig.~1]{van_ballegooijen_formation_1989}. The next step of the theory, and several other simulations discussed in the introduction, is to further shear the arcades by a prescribed convection of the plasma and field line footpoints toward or along the PIL, until the arcades reconnect. In \bifrost{}, as in the Sun, the convection is not prescribed, but stochastic and self-consistently driven by the convection zone. This causes the stochastic patterns exemplified by the magnetograms in Fig.~\ref{fig_ramp_z0}. That is the most significant difference of the idealized theory and simulations to the more realistic simulation presented in this paper.

The formation of the flux rope in the simulation bears some resemblance to observations referred to in the introduction. The flux rope forms along an irregular PIL, separating two areas of patchy polarities~\citep{Schrijver_cme_2011,wang_2013}. Throughout the simulation, multiple flux cancellation events, separated in time and space, contribute to the formation of the flux rope ~\citep{wang_2007,wang_2013}. The resulting flux rope is gradually more aligned along the irregular PIL. However, the resulting flux rope is rather short and low, possibly due to the limited shear of the initial arcades or size of the simulation box. Hence, it does not warrant quantitative comparison to observed flux ropes.

Now, that the process of forming flux ropes through cancellation of sheared coronal arcades has been established with this simulation, it would of interest to test other field configurations. Using an LFFF with a high number of harmonics ${(n_h)}$ caused the field to be centered around ${x=L_x/2}$ as intended, but it also caused the shearing angle to be smaller close to the photosphere than higher up in the corona, as visible in Fig.~\ref{fig_frT_1}. This is opposite to the configuration in~\citet{van_ballegooijen_formation_1989}, where there are several sheared arcades in parallel over an even more sheared field line. It is also opposite to the observations by~\citet{Schmieder_1996}, which show that the shear is concentrated at lower heights, close to the PIL. 
Similar setups to the one presented, with different parameters for the LFFF, are expected to lead to qualitatively similar results, albeit with quantitative differences -- e.g., a stronger shear would likely cause a longer flux rope. Hence, instead of inserting an LFFF, one could add a nonlinear force-free field with a shear distribution more in line with observations. Such a simulation would be more quantitatively comparable to observations.

The work presented in this paper focus on the cancellation process in a dispersed flux concentration environment, which is the case for decaying active regions. Flux cancellation can also happen between emerging sunspots. That was modeled in~\citet{rempel_2023}, in which they formed a flux rope and an eruption. This simulation was run with \muram{}, which also can simulate atmospheric dynamic driven by stochastic and self-consistent flows in the convection zone.
In order to build the full picture of flux rope formation in both situations, it may be interesting to perform a similar kind of detailed analysis, as presented here, also on the reconnection process in the setting of colliding sunspots.

\section{Conclusion}

In this paper, we have shown and analyzed the stepwise formation and twisting of a flux rope in a self-consistent radiative MHD simulation including a free convection zone. 
Other simulation studies on flux rope formation often limit themselves to simulating the atmosphere, keeping the field lines line-tied to the photospheric boundary where there is prescribed convection or diffusion to drive the cancellation, or driving it by flux-emergence through a stratified convection zone and atmosphere with reconnection happening in the transition region, often leading to eruptions. 
A key aspect of the present study, in comparison, is the modeling of the stochastic convection of an unstable convection zone. This extension was necessary to include the complexities of the real Sun, as well as being able to understand what happens below the photosphere post-cancellation and post-submergence of $\Omega$-loops. The simulation was run with the state-of-the-art parallel code, \bifrost{}. To enable the tracking of field lines without following the prescribed convection of their footpoints at the boundary, we have used Lagrangian markers, denoted \corks{}, to accurately track the motion of selected field lines with time.

Within the simulation, which extends $\SI{28}{\mega\meter}$ into the corona, an LFFF with sheared arcades was slowly introduced into a relaxed QS simulation. After the careful introduction of the LFFF, the magnetic field at the photosphere was mainly bipolar on the large scale, but patchy and without a straight PIL on the small scale. This is qualitatively the same as observed PILs on the Sun. 
Following this initial setup, the stochastic self-consistent plasma flows in the convection zone formed a flux rope above the photosphere, through several events connected to observable flux cancellation in the photosphere. Over the course of about $\SI{2.5}{\hour}$ of solar time, a flux rope was formed with a maximum horizontal footpoint separation of approximately $\SI{12}{\mega\meter}$ and extending ${\SI{2}{\mega\meter}}$ into the solar atmosphere (Fig.~\ref{fig_fr2_4}).

The gradual formation of this flux rope included multiple events, separated in time and space. A few representative events with different atmospheric responses have been studied in more detail. A slipping reconnection event forms the initial flux rope. This event might be driven by observed flux cancellation at the ${z=0}$ plane, but the reconnection occurs higher up in the atmosphere. A U-loop emergence later contributes to the twist of this flux rope. However, the flux cancellation observed as the loop emerges is not linked to reconnection.
An $\Omega$-loop submergence occurs shortly after, between the photospheric footpoints of the flux rope. This submergence, observable as flux cancellation, was found to occur shortly after the emergence of the same field lines. Nevertheless, this cancellation event was found to not directly contribute to the twist of the flux rope, even though it occurred in close vicinity to the flux rope. This analysis raises a caution for not mistaking correlation for causation when analyzing observations.

The most significant event analyzed in this study is the thick-photosphere tether-cutting event (TPTC, see Sect.~\ref{sec_res_tptc}) that combined the already formed flux rope with additional untwisted arcades. Together, they formed an even longer flux rope, as well as some shorter field lines arcades that was later submerged. This event is the realistic counterpart to the more ideal and symmetrical reconnection proposed by~\citet{van_ballegooijen_formation_1989}. It is also similar to the observed reconnection of two loops to a longer loop in~\citet{wang_2013}.
However, it is the first time such an event has been simulated in a thick photosphere in a convectively-driven MHD simulation with a gradually-forming stable flux-rope. We find that the reconnection occurs within an extended volume of the thick photosphere, hence the TPTC name. The reconnection starts and lasts during the convergence of two eventually canceling polarities. In other words, the flux cancellation is clearly related to the reconnection, and not only appearing in more artificial models. By simulating the convection zone, we also follow the shorter field lines, which were formed during the reconnection, down below the photosphere after the submergence. This is how such a TPTC event occurs on the Sun.

A flux rope did form in this simulation, in spite of the fragmented nature of the large-scale PIL we introduced in the middle of the simulation box. 
That an atmospheric flux rope was formed is the same end result as in the planar line-tied simulations, the non-convective flux-emergence simulations, and the theoretical illustration by~\citet{van_ballegooijen_formation_1989}. However, these past models all have smooth elongated PILs, along which reconnection happens all at once, to form the flux rope. But many observed solar prominences are not formed above smooth PILs. Hence, a question we wanted to answer with this work was whether the Sun could form flux ropes above patchy canceling PILs. It can.

\begin{acknowledgements}
      We would like to thank 
      M.~Szydlarski and M.~Carlsson for help with setting up and running \bifrost{} simulations,
      M.~Druett for help tracking the field lines with \corks{} in \bifrost{}, 
      as well as A.~Prasad for visualization with \texttt{Vapor}.
      
    This research has been supported by the European Research Council through the Synergy Grant number 810218 (“The Whole Sun”, ERC-2018-SyG) and by the Research Council of Norway through its Centres of Excellence scheme, project number 262622 (Rosseland Centre for Solar Physics -- RoCS). 
    The simulations were performed on resources provided by Sigma2 – the National Infrastructure for High Performance Computing and Data Storage in Norway. 
    This work was supported by the Action Th\'ematique Soleil-Terre (ATST) of CNRS/INSU PN Astro, co-funded by CNES, CEA and ONERA.

    Reproduced with permission from Astronomy \& Astrophysics, \textcopyright{} ESO.
      
\end{acknowledgements}


\bibliographystyle{aa} 
\bibliography{TOP_ref} 


\begin{appendix} 

\section{Relaxing a reconnecting simulation}

\label{app_Simulation}

Preparing, running, and analyzing the simulation presented in this paper caused unforeseen challenges. 
Numerical experimentation was, therefore, necessary to get the simulation to relax to a realistic model of the process under study. In this appendix, we add explanation for some of the modifications done to the simulation, described in Sect.~\ref{sec_met_QS}, to combat these unforeseen challenges. While not critical for the physical conclusions drawn in this paper, it could be relevant for someone attempting to run a similar simulation in the future.


At first, the flux feeding through the lower boundary was turned off, to not affect the magnetic field topology and reconnection we later wanted to study. After this modification, the simulation would need time to relax to the new BCs. The spatial resolution was reduced at the same time to conserve numerical resources, using the \texttt{backstaff} tool. Unexpectedly, the modified resolution caused enhanced p-mode waves through the box originating at the lower boundary. Furthermore, these waves were not fully canceled at the original coronal boundary at ${\SI{-14.3}{\mega\meter}}$, but were in part reflected back into the box after about ${\SI{10}{\minute}}$, stirring the interior of the simulation box.

To minimize the reflections from the coronal boundary, it was attempted -- mid-simulation, in the copied state of the QS simulation -- to change the vertical BCs. Instead of extrapolating into the ghost zones and using the standard time-derivative equations, we tested characteristic BCs. However, that led to nonphysical point heating in the coronal boundary, likely as a consequence of quickly changing the boundary equations. The combined impact of the waves and their reflection is believed to have caused heating, possibly due to reconnection, high up in the corona in early tests of this simulation.

To rather minimize the push causing the wave at the lower boundary in the first place, we ended up rather modifying that BC instead. The simulation drives the average mass and energy densities over time to keep them close to a set of desired values. By changing the resolution, the depth of the lower boundary was slightly shifted, making the set of desired values inappropriate for the new depth. Hence, the desired values for these parameters at the lower boundary was reset to their new calculated averages just after modifying the resolution. By starting at the new averages, the driving of the problematic waves was minimal, alleviating this problem.

By later introducing the LFFF, a process described in Sect.~\ref{sec_exp_lfff}, more waves occurred within the box as the simulation was again trying to relax to a self-consistent configuration including this new field. When introducing the LFFF all at once, it created strong waves, which stirred and mixed the interior of the simulation box. By rather introducing the extra field in 120 smaller insertions over ${\SI{20}{\minute}}$, the simulation box had time to readjust, greatly reducing the stirring of the simulation box interior. 

When the waves generated by ramping the LFFF reached the coronal boundary, which in part reflected them, it would sometimes cause numerical instabilities. It was found that the magnetic field of the added LFFF, when convected into the coronal boundary, again caused isolated heating, which the wave method in the Spitzer module was unable to convect away correctly, even at very short time steps. This problem caused further studies into the Spitzer module~\citet{cherry_2024,furuseth_2024}. These studies found that the Courant condition for the wave method in fact gives a lower limit on the time step, below which, the method becomes unstable. This is opposite most explicit methods, which typically has an upper limit for the time step.

The challenges near the coronal boundary, both when relaxing the QS simulation and when ramping the LFFF, were alleviated by extending the corona almost twice as far away from the photosphere. This was done by adding horizontally stratified layers. However, as a consequence of adding stratified layers, the simulation required to be relaxed for longer before ramping the LFFF. These simulations with an extended corona, typically reached an approximately steady state after about 700 snapshots, or about \SI{2}{\hour}. We relaxed the simulation for almost one more hour after this before starting to ramp the LFFF.

With the presented openness and transparency regarding this effort, it may seem that the gradual twisting of a flux rope was a stroke of luck, finally occurring after hundreds of attempts. That was not the case. Similar flux ropes occurred in several earlier simulations that had, e.g., nonphysical strong waves through the domain due to the change of the resolution, nonphysical strong reconnection at the coronal boundary due to the BC, and a lower resolution of ${\Delta x=\SI{125}{\kilo\meter}}$.

\section{Tracking an evolving flux rope in 3D}

\label{app_tracking}

The flux rope was easy to visualize in single snapshots with both home-made numerical analysis tools and more professional tools like \vapor{}~\citep{vapor2019}. However, displaying how a flux rope evolves over time in 3D is challenging, particularly in this kind of simulation where the motion of the field lines are not controlled by the convection at the boundary (photosphere), but rather by self-consistent plasma convection. 

We tracked the motion of the field lines with \corks{}, point particles that were convected with the plasma. They worked very well to track the field lines over shorter periods of time, but over longer periods, they could be convected far away from the area of interest. That made it impossible to, e.g., track the full history of the field lines over the $\SI{2.5}{\hour}$ of solar time shown in Fig.~\ref{fig_frT}. Therefore, those figures were not made with \corks{}. The slightly more than $\SI{40}{\minute}$ shown in Fig.~\ref{fig_fr2} seemed to be borderline, in our simulation, indicated by the magenta field lines convected down and out of the simulation box. The other colored families of field lines in this figure were also challenging to illustrate, as the corks were convected far along the field lines during that time interval. This challenge should be kept in mind when planning a similar numerical experiment in the future.

\end{appendix}
\end{document}